\newif\iffinal
\newif\ifdraft

\finaltrue
\iffinal
\draftfalse
\else
\drafttrue
\fi

\ifdraft
\documentclass[12pt]{scrartcl}
\else
\documentclass[11pt]{scrartcl}
\usepackage[letterpaper,margin=1in]{geometry}
\fi

\usepackage{amsmath}
\usepackage[ruled,noline,noend]{algorithm2e} % no vertical lines, no ends globally
\usepackage{amssymb,mathtools,etoolbox,forloop,booktabs,tabularx}
\usepackage{amsthm}
\usepackage{enumitem}
\usepackage{dsfont}
\usepackage[kerning]{microtype}
\usepackage{comment,cprotect}
\usepackage[svgnames]{xcolor}
\usepackage{soul}\setul{0.5ex}{0.3ex}\setulcolor{blue}
\usepackage[%hidelinks
colorlinks,linkcolor=NavyBlue,citecolor=DarkGreen]{hyperref}
\usepackage[nameinlink,capitalize]{cleveref}
\ifdraft
\usepackage[left,mathlines]{lineno}
\fi
\usepackage{bm,nicefrac,csquotes}

\newtheorem{theorem}{Theorem}
\newtheorem{theorem*}{Theorem}
\newtheorem{lemma}{Lemma}
\newtheorem{claim}{Claim}

\newtheorem*{claim*}{Claim}
\newtheorem{fact}{Fact}
\newtheorem{invariant}{Invariant}
\theoremstyle{definition}
\newtheorem{observation}{Observation}
\newtheorem{corollary}{Corollary}
\newtheorem{definition}{Definition}
\newtheorem*{buildingblock*}{Building Block}
\newtheorem*{subroutine*}{Subroutine}
\usepackage{tikz}\usetikzlibrary{calc}

\def\tO{\tilde{O}}

\newcounter{reqs}
\newcounter{i}
\def\req #1{\begingroup\refstepcounter{reqs}\hyperref[requirementlist]{\color{red} \ul{(\thereqs)~requires #1}}%
    \global\csdef{reqs\thereqs}{#1}\expandafter\label\expandafter{\thereqs labelreqs}\endgroup}

\ifdraft
\usepackage{scrlayer-scrpage}
\ihead{\Acrobatmenu{GoBack}{\fbox{\texttt{Back}}}~~\hyperref[current]{\fbox{\texttt{Current}}}}

\else

\fi

\makeatletter
\if@cref@capitalise
\crefname{property}{Property}{Properties}
\crefname{case}{Case}{Cases}
\crefname{step}{Step}{Steps}
\crefname{requirement}{Req.}{Reqs}
\else
\crefname{property}{property}{properties}
\crefname{case}{case}{cases}
\crefname{step}{step}{steps}
\crefname{requirement}{req.}{reqs}
\fi
\makeatother
\crefname{step}{Step}{Steps}
\crefname{case}{Case}{Cases}
\crefname{requirement}{Req.}{Reqs}
\Crefname{property}{Property}{Properties}
\makeatother

\newcommand{\volF}{\operatorname{vol}_F}
\def\capc{\operatorname{cap}}
\newcommand\restr[2]{{% we make the whole thing an ordinary symbol
  \left.\kern-\nulldelimiterspace% automatically resize the bar with \right
  #1 % the function
  \vphantom{\big|} % pretend it's a little taller at normal size
  \right|_{#2} % this is the delimiter
  }}

\makeatletter
\renewcommand\paragraph{\@startsection{paragraph}{4}{\z@}%
  {-0.5ex \@plus -1ex \@minus -0.2ex}%
  {-2ex}%
  {\raggedsection\normalfont\sectfont\nobreak\size@paragraph}%
}
\makeatother

\newcommand{\magicpart}{\hyperref[bb:part]{\textnormal{\textsc{PartitionCluster}}}}
\newcommand{\algTWT} {\hyperref[bb:algTWT]{\textnormal{\textsc{TwoWayTrim}}}}
\newcommand{\algSC}[0]{\hyperref[bb:algSC]{\textnormal{\textsc{SparsestCutApx}}}}
\newcommand{\algFC}[0]{\hyperref[bb:algFC]{\textnormal{\textsc{FairCutFlow}}}}
\newcommand{\algCH}[0]{\hyperref[bb:algCH]{\textnormal{\textsc{ConstructHierarchy}}}}

\newcommand{\zero}[0]{zero-$A_t$ block }
\newcommand{\id}[0]{id-$A_t$ block }

\def\P{{\mathcal{P}}}
\def\X{{\mathcal{X}}}
\def\Y{{\mathcal{Y}}}
\def\C{{\mathcal{C} }}

\makeatletter
\def\deg{\def\argsub{}\def\argsup{}\@ifnextchar_{\collectsub}{\@deg}}%
\def\@deg{\@ifnextchar^{\collectsup}{\@@deg}}%
\def\@@deg{\operatorname{deg}_{\argsub}^{\argsup}\@ifnextchar|{\!}{}}%
\def\collectsub_#1{\def\argsub{#1}\@deg}
\def\collectsup^#1{\def\argsup{#1}\@@deg}
\makeatother

\def\con{\mathrm{con}}
\def\parent{\text{par}}
\def\v #1{\bm{#1}}
\def\1{\mathds{1}}
\def\diag{\operatorname{diag}}
\def\Tr{\operatorname{Tr}}
\def\partialc #1{\operatorname{deg}_{#1}}
\def\pp #1{\partial #1}
\def\q{q^*}
\def\t{\tau^*}
\newcommand{\bal}{\beta^*}
\newcommand{\E}{\mathbb{E}}
\newcommand{\mb}[1]{\v{#1}}
\newcommand{\vu}[0]{{V \setminus U}}
\newcommand{\sa}[0]{{S \setminus U}}

\def\Tpart{T_{\operatorname{part}}}
\def\Tfc{T_{\operatorname{fc}}}
\def\Ttwt{T_{\operatorname{twt}}}
\def\Tsc{T_{\operatorname{sc}}}

\begin{document}
\setcounter{page}{-1}
\ifdraft\linenumbers\fi

\title{An Improved Quality Hierarchical Congestion Approximator in Near-Linear Time}

%\affil[1,2]{Institute of Science and Technology Austria (ISTA), Klosterneuburg, Austria

\author{Monika Henzinger\thanks{ISTA, Klosterneuburg, Austria}, Robin Münk\thanks{Technical University of Munich, Germany}, Harald Räcke\thanks{Technical University of Munich, Germany}}
\date{}
\maketitle
\thispagestyle{empty}

\begin{abstract}
A single-commodity congestion approximator for a graph is a compact data structure that
approximately predicts the edge congestion required to route any set of single-commodity flow demands in a network.
A hierarchical congestion approximator (HCA) consists of a laminar family of cuts in the graph and has numerous applications in approximating cut and flow problems in graphs, designing efficient routing schemes, and managing distributed networks.

There is a tradeoff between the running time for computing an HCA and its approximation quality.
The best polynomial-time construction in an $n$-node graph gives an HCA with approximation quality $O(\log^{1.5}n \log \log n)$. Among near-linear time algorithms, the best previous result achieves approximation quality $O(\log^4 n)$.
We improve upon the latter result by giving the first near-linear time algorithm for computing an HCA with approximation quality $O(\log^2 n \log \log n)$.
Additionally, our algorithm can be implemented in the parallel setting with polylogarithmic span and near-linear work, achieving the same approximation quality. This improves upon the best previous such algorithm, which has an $O(\log^9n)$ approximation quality. 
We also present a lower bound of $\Omega(\log n)$
for the approximation guarantee of hierarchical congestion approximators.

Crucial for achieving a near-linear running time is a new partitioning routine that, unlike previous such routines, manages to avoid recursing on large subgraphs. To achieve the improved approximation quality, we introduce the new concept of border routability of a cut and provide an improved sparsest cut oracle for general vertex weights.

\end{abstract}
\clearpage

\thispagestyle{empty}
\tableofcontents
\clearpage

\setcounter{page}{1}

%%%%%%%%%%%%%%%%%%%%%%%%%%%%%%%%%%%%%%%%%%%%%%%%%%%%%%%%
%%%%%%%%%%%%%%%%%%% Introduction %%%%%%%%%%%%%%%%%%%%%%%
%%%%%%%%%%%%%%%%%%%%%%%%%%%%%%%%%%%%%%%%%%%%%%%%%%%%%%%%

\section{Introduction}\nocite{ADK25}
A congestion approximator for a graph is a compact data structure that
approximately predicts the edge congestion required to route a given set of
flow demands in a network.

Typically, it consists of a collection of cuts. Each cut provides a lower bound
on the congestion needed to realize a particular demand: namely, the total
demand crossing the cut divided by the cut’s capacity. The prediction of the
approximator for a given demand vector is then the maximum of these lower
bounds over all cuts in the collection. 

By the classical max-flow–min-cut theorem \cite{FF56}, an approximator that
includes all $2^n$ possible cuts would yield exact predictions for
single-commodity flows. Surprisingly, Räcke~\cite{Rae02} showed that already a
\emph{linear number of cuts} suffice to approximate the required congestion for
all demands within a factor of $O(\log^3n)$. Importantly, this guarantee
extends to multicommodity flow demands, demonstrating that the flow-cut
structure of an undirected graph can be captured efficiently using only a small
collection of cuts.

Räcke’s original result was non-constructive, i.e., it established the
existence of a congestion approximator but did not provide an efficient way
to compute it. Independently, Bienkowski, Korzeniowski, and Räcke~\cite{BKR03} and
Harrelson, Hildrum, and Rao~\cite{HHR03} provided the first polynomial-time
constructions of congestion approximators. Their algorithms achieved
approximation guarantees of $O(\log^4 n)$ and $O(\log^2 n \log \log n)$,
respectively.
Subsequently, Räcke and Shah~\cite{RS14} constructed a congestion approximator
for single-commodity flows, obtaining an existential guarantee of 
$O(\log n \log \log n)$ and a polynomial-time construction with guarantee 
$O(\log^{1.5} n \log \log n)$.

A key feature of all the above constructions is that the collection of cuts
forms a \emph{laminar family}. We refer to a congestion approximator with this
property as a \emph{hierarchical congestion approximator}. Intuitively, this
means that the cut structure of the graph is approximated by a hierarchical decomposition that can be represented by a single tree, a
property that plays a central role in many applications, including
oblivious routing, approximation algorithms for sparsest cut and multicommodity
flow, network design problems, and routing schemes in distributed
networks~\cite{AGMM09,BFK+11,CKS04,EKLN07,KKM12,KPS11}.

As many of these applications require working with large graphs, the
mere polynomial-time guarantees of previous constructions posed a significant
obstacle. Räcke, Shah, and Täubig~\cite{RST14} addressed this challenge by
giving an almost-linear-time algorithm that produces a hierarchical
congestion approximator with an approximation guarantee of $O(\log^4 n)$
for multicommodity flows.
Interestingly, this result relied on a technique of Sherman~\cite{She13} to
compute approximate maximum single-commodity flows in almost-linear time, 
given access to a congestion approximator. The quality of this approximator 
directly affected the running time of his method. Thus, improving Sherman's 
algorithm to almost-linear time would have improved the RST14 result to 
almost-linear time, and vice versa. This circular dependency was ultimately 
resolved by Peng~\cite{Pen16}, who showed that both problems can be solved
in nearly linear time. 

More recently, Li, Rao, and Wang \cite{LRW25} proposed a bottom-up approach to
constructing congestion approximators, which contrasts structurally with the
recursive, top-down techniques used in earlier work. Unlike Peng’s method,
their algorithm does not rely on heavy recursion, leading to smaller
logarithmic factors in the running time, albeit with a weaker approximation
guarantee of $O(\log^{10}n)$.

In this paper, we substantially improve the approximation guarantee for
near-linear-time constructions of hierarchical congestion approximators. We
show how to compute a congestion approximator for single-commodity flows with
approximation guarantee $O(\log^2 n \log \log n)$ in nearly-linear time. This
improves upon RST14 by almost two logarithmic factors and nearly matches the
best known guarantee of $O(\log^{1.5} n \log \log n)$ for single-commodity
congestion approximators, which takes time $\tO(n^2)$.
Formally, we show the following theorem.
\begin{theorem}\label{thm:main:normal}
Given an undirected graph with $n$ vertices, there is a near-linear
time algorithm that computes a hierarchical congestion approximator that
w.h.p.\ has approximation guarantee $O(\log^2n\log\log n)$ for single-commodity flows.
\end{theorem}

Note that a single-commodity hierarchical congestion approximator is equivalent
to a \emph{tree cut sparsifier}. Formally, a tree cut sparsifier $T$ of a graph
$G=(V,E)$ with quality $q \geq 1$ is a weighted tree whose leaves correspond to
the vertices of $G$ and for any pair $A,B$ of disjoint subsets of $V$,
$1/q \cdot \operatorname{mincut}_T(A,B) \geq \operatorname{mincut}_G(A,B) \geq
\operatorname{mincut}_T(A,B)$, where, for any graph $H$,
$\operatorname{mincut}_H(A, B)$ denotes the value of a minimum cut separating
$A$ from $B$ in $H$.
The stronger notion of \emph{tree flow sparsifier} is equivalent to a
hierarchical congestion approximator for multi-commodity flows. For
completeness, we give a lower bound of $\Omega(\log{n})$ for the approximation
quality of any hierarchical congestion approximator
% for multi-commodity flows 
in \cref{sec:lowerbound} based on a reduction from oblivous routing.

By the definition of the flow-cut gap $\lambda$ for concurrent multi-commodity flows\footnote{%
Define for a multi-commodity flow demand $d$ and a cut $C\subseteq V$,
$d(C,V\setminus C)=\sum_{(x,y)\in C\times V\setminus C}d(x,y)$, and
$\Phi(d) := \max_S d(S,V\setminus S)/ \capc(S, V \setminus S)$. Then the
flow-cut gap of $G$ w.r.t.\ $d$ is the ratio
$\operatorname{opt}_G(d)/\Phi(d)$, where $\operatorname{opt}_G(d)$ is the
optimum congestion for routing $d$ in $G$. The \emph{flow-cut gap} $\lambda$ of
$G$ is
the maximum of this ratio over all demands $d$. If the maximum is just taken
over product multicommodity flow demands, then the corresponding flow-cut gap
is denoted by $\lambda^*$.
}, a result with approximation guarantee $\alpha$ for single-commodity flows implies a result for
multi-commodity flow with approximation guarantee $\lambda\alpha$.
It is known that $\lambda=O(\log n)$
in any graph, while some graph classes have a smaller gap.
This leads to to the following corollary.

\begin{corollary}\label{thm:main:normal-multi}
Given an undirected graph with $n$ vertices, there is a near-linear
time algorithm that computes a hierarchical congestion approximator that
w.h.p.\ has approximation guarantee $O(\log^3 n\log\log n)$ for multi-commodity flows.
\end{corollary}

Note that all previous near-linear time constructions for congestion
approximators that obtain a polylogarithmic guarantee only give a correct
solution with high probability.
Our algorithm can be easily parallelized.
\begin{theorem}\label{thm:main:parallel}
Given an undirected graph with $m$ edges and $n$ vertices, there is a
parallel algorithm that requires $O(m\operatorname{polylog}n)$ work, has
$O(\operatorname{polylog} n)$ span, and constructs a hierarchical congestion
approximator that w.h.p.\ has approximation guarantee $O(\log^2n\log\log n)$.
\end{theorem}
The previous best parallel construction for congestion approximators is due to
Agarwal et al.~\cite{AKL+24} who obtain an approximation guarantee of
$O(\log^9n)$ with $O(m\operatorname{polylog} n)$ work and
$O(\operatorname{polylog} n)$ depth.

\makeatletter
\def\@fnsymbol#1{*}
\renewcommand*{\thefootnote}{\fnsymbol{footnote}}
\makeatother
\begin{table}[t]
\centering
\small
\renewcommand{\arraystretch}{1.15}
\setlength{\tabcolsep}{6pt}
\begin{tabular}{cccc c}
\toprule
\textbf{Approximation} & \textbf{Type} & \textbf{Running Time} & \textbf{Notes} & \textbf{Work} \\
\midrule
$O(\log^3 n)$ & multi  & --- & --- & \cite{Rae02} \\
$O(\log^4 n)$ & multi  & $\mathrm{poly}(n)$ & --- & \cite{BKR03} \\
$O(\log^2 n \log \log n)$ & multi  & $\mathrm{poly}(n)$ & --- & \cite{HHR03} \\
$O(\log n \log \log n)$ & single & --- & --- & \cite{RS14}\footnotemark\\
$O(\log^{1.5} n \log \log n)$ & single & $\mathrm{poly}(n)$ & --- & \cite{RS14}\footnotemark \\
$O(\log^4 n)$ & multi  & $O(m^{1+o(1)})$ & w.h.p. & \cite{RST14} \\
$O(\log^4 n)$ & multi  & $\tO(m)$ & w.h.p. & \cite{Pen16} \\
$O(\log^{10} n)$ & single  & $\tO(m)$ & bottom up, w.h.p. & \cite{LRW25} \\
$O(n^{o(1)})$ & multi  & update time: $n^{o(1)}$ & dynamic, w.h.p. & \cite{GRST21} \\
$O(\log^2 n \log \log n)$ & single & $\tO(m)$ & w.h.p. & \textbf{this paper}\footnotemark \\
$O(\log^9 n)$ & multi & work: $\tO(m)$, span: $\tO(\log^c n)$ & parallel, w.h.p. & \cite{AKL+24} \\
$O(\log^2 n \log \log n)$ & single & work: $\tO(m)$, span: $\tO(\log^c n)$ &
                                                                             parallel, w.h.p. & \textbf{this paper}\footnotemark \\
$\Omega(\log n)$ & single & --- & --- & \cref{obs:lowerbound}\\
\bottomrule
\end{tabular}
\caption{Comparison of hierarchical congestion approximator results. A result
  with guarantee $\alpha$ for single-commodity flows implies a result for
multi-commodity flow with guarantee $\lambda\alpha$, where $\lambda=O(\log n)$
is the \emph{flow-cut gap} of the graph. For the marked results ($*$) this
step loses only a factor of $\lambda^*$ where $\lambda^*$ is the flow-cut gap w.r.t.\
product multicommodity flows. The latter e.g. is constant for graphs excluding
a fixed minor.~\cite{KPR93}}
\label{tab:congestion-approximators}
\end{table}

\paragraph{Very Recent Development.} 
Our algorithm relies on a generalized sparsest cut oracle whose running time for our application is linear in the number of (parallel) edges in an unweighted graph. 
In independent work, very recently, Agassy, Dorfman and Kaplan~\cite{ADK25} published a paper on arXiv that also presents a generalized sparsest cut oracle.
However, they do not use their result for obtaining a congestion approximator and offer no guarantees for the parallel setting.
In our application, their algorithm takes time linear in the number of edges in a capacitated graph (not the sum of the edge weights).
As the other parts of our algorithm work in capacitated graphs, we obtain the following corollary for the sequential setting.

\begin{corollary}\label{crl:main-wgt}
Given an undirected, capacitated graph with $n$ vertices, there is an algorithm that takes near-linear time (in the number of edges) and computes a hierarchical congestion approximator that w.h.p.\ has approximation guarantee $O(\log^2 n\log\log n)$ for single-commodity flows.
\end{corollary}

\subsection{Further Related Work}
Apart from the work on hierarchical congestion approximators there is also the
work on congestion approximators where the family of cuts does not form a
laminar family. Räcke~\cite{Rae08} introduced the concept of approximating the
cut-structure of a graph not by a single tree as in a hierarchical congestion
approximator but by a probability distribution over trees. These trees can be
constructed in polynomial time and the collection of all cuts over all trees
can be interpreted as a congestion approximator. This gives a congestion
approximator with approximation guarantee of $O(\log n)$.

Madry~\cite{Mad10} approximated the graph instead by a probability distribution
over $j$-trees, which are trees with a small number of additional edges. His
construction has a worse approximation guarantee of $O(n^{o(1)})$ but allows
to sample a $j$-tree from the distribution in almost linear time. This gave
almost linear time approximation algorithms for many applications before RST14
gave the first congestion approximator in nearly linear time. This technique
is also very important in the area of dynamic graph algorithms.

For many applications approximating the graph by a probability distribution of
trees as in \cite{Rae08} or by a distribution over $j$-trees is sufficient. 
However, some applications do require a single tree~\cite{KKM12,BFK+11}.

\paragraph{Connections to Expander Decompositions}
An \emph{expander decomposition} of an undirected graph $G = (V, E)$ is a 
partition of its vertex set into disjoint pieces $V_1, \dots, V_k$ such that 
each piece induces a subgraph $G[V_i]$ with conductance at least $\phi$. 
The goal is to minimize the number of edges that go between different pieces 
(we say such an edge is \emph{cut}). 

It is known that any expander decomposition may need to cut 
$\phi m \log n$ edges in the worst case. 
In the following, we say that an expander decomposition has 
\emph{gap}~$g$ if it cuts at most $g \phi m$ edges.
There are close connections between the problem of finding good hierarchical
congestion approximators and that of computing expander decompositions with
small gap.

In one direction, Goranci et al.~\cite{GRST21} showed that an algorithm for
expander decomposition can be used to compute a hierarchical congestion
approximator with quality $n^{o(1)}$ in almost-linear time. This result allowed
them to obtain dynamic congestion approximators with the same $n^{o(1)}$
quality.

Another relationship between near linear time constructions for congestion
approximators and expander decompositions is in terms of techniques. Efficient
constructions for both problems usually rely on variants of the Cut-Matching
game as introduced by~\cite{KRV06}. We generalize a Cut-Matching game that
was presented by Agassy et al.~\cite{ADK22} with the application of computing an expander
decomposition. They obtained the currently best gap of $O(\log^2n)$ for this
problem.

\section{Technical Overview}
\subsection{Basic Notation and Definitions}

\paragraph{Graphs.} We are given an undirected graph $G=(V,E)$ with $|V| = n$ and
$|E|=m$ and parallel edges allowed. Without loss of generality we assume throughout
the paper that the graph is connected. Given an edge $(u,v)$, $\capc(u,v)$ denotes the number of parallel edges between $u$ and $v$. For $S \subseteq V$ we use $G[S]$ to
denote the subgraph induced by the vertices in $S$. A \emph{cut} usually refers
to an edge set $E(S,V \setminus S)$ for a vertex set $S \subset V$, but, by
abuse of notation, we sometimes also refer to the set $S$ itself as a cut. For
$F \subseteq E$ we use $\deg_F(v)$ to denote the sum of the capacities of the
edges of $F$ that are incident to $v$. If $F=E$, then we use $\deg(v)$ instead
of $\deg_E(v)$.

For a set $C \subseteq V$, let the \emph{boundary} or \emph{border} $\pp{C}$
denote the set of edges with exactly one endpoint in $C$ and for any set
$F \subseteq E$, and $v \in V$ let $\partialc{F}(v)$ be the capacity of the
edges of $F$ that are incident to $v$. For a collection of (not necessarily disjoint)
subsets ${\mathcal C}=\{C_1,\dots,C_k\}$ we use $\partial{\mathcal C}$ as a
shorthand for $\bigcup_{C\in{\mathcal C}}{\partial C}$. Note that $\partialc{\pp{C}}(v) = \partialc{\pp{\C}}(v).$

\paragraph{Partition.} A \emph{partition $\X$ of a subset $S$ of  $V$} is a family of disjoint non-empty subsets
$X_1,\dots,X_k$ such that $\bigcup_k X_i = S$. 
Note that $\X = \{S\}$ means that the partition contains exactly one set, namely $S$, i.e., it is a trivial partition.

For a partition $\X$ we use $\pp{\X}$ to denote the set of edges such that for each edge exactly one endpoint belongs to one set $A$ of $\X$  and the other endpoint does not belong to $A$. Given two partitions $\X$ and
$\X'$ of two disjoint vertex sets $X$ and $X'$ we use $\X \uplus \X'$ to denote
the set-union of the family of subsets $\X$ and with the family of subsets of
$\X'$, ie. $\X \uplus \X' = \{C | C \in \X \text{ or } C \in \X'\}$.

\paragraph{Vertex weight functions.} We call a function
$x:V \rightarrow {\mathbb R}$ a \emph{(vertex) weight function} for the vertices
of $V$. Let $S \subseteq V$. Then $x(S) = \sum_{v \in S} x(v)$. For example,
$\partialc{F}(S) = \sum_{v \in S} \partialc{F}(v)$.

We use
$x|_S: V \rightarrow {\mathbb R}$ to denote the function where $x|_S(v) = x(v)$
for $v \in S$ and $x|_S(v) = 0$ for $v \not\in S$. Given two weight functions
$x$ and $x'$ we use $x \le x'$ when $x(v) \le x'(v)$ for all $v \in V$ and use
$|x|$ to be the function where $|x|(v) = |x(v)|$ for all $v \in V$.

\paragraph{Flows.} A \emph{demand} or \emph{demand function} is a vertex weight
function $d$ such that $\sum_{v \in V} d(v) = 0.$ { A flow
  $f:V \times V \rightarrow \mathbb{R}$ satisfies $f(u,v) = -f(v,u)$ for all
  $u,v \in V$ and $f(u,v) =0$ for all pairs $(u,v) \not\in E$. A flow is
  \emph{feasible} if $\capc(u,v)\ge |f(u,v)|$ for all $(u,v) \in E$. If
  $f(u,v) >0$ then we say that \emph{flow mass is routed from $u$ to $v$}. For
  each vertex $u \in V$ the \emph{net-flow out of $u$} is
  $f(u) = \sum_{v \in V} f(u,v)$, i.e., the total flow mass routed from $u$ to
  its neighbors in $G$ minus the total flow mass routed from its neighbors to
  $u$.}
The \emph{net-flow into $u$} is $-f(u)$. 

A flow \emph{routes a demand}
$d$ if the net-flow out of each vertex $v$ is $d(v) \in \mathbb R$ (which might be negative in which case the net-flow into $v$ is positive). For
$S \subseteq V$, a flow within an induced sub-graph $G[S]$ routes
$d:V\rightarrow\mathbb{R}$ if the net-flow out of each vertex $v$ is $d(v)$ and $d(v) = 0$ for all $v \not\in S$.

Given two non-negative vertex weight functions $\mb{s},\mb{t}: V\rightarrow\mathbb{R}$ with $\sum_v \v{s}(v) = \sum_v \v{t}(v)$, we say that a flow $f$ is
an $(\mb{s},\mb{t})$-flow if it routes the demand $\mb{s}-\mb{t}$. If $|f(u,v)| \le c \cdot \capc(u,v)$ for all $u,v \in V$, then $f$ has \emph {congestion $c$}. If $c=1$, then $f$ is feasible. 

\paragraph{Congestion Approximator.} Let $\alpha \ge 1$. An
\emph{$\alpha$-congestion approximator $\C$} for a graph $G=(V,E)$ with edge
capacities $\capc$ is a collection of subsets of $V$ such that for any demand
$d$ satisfying for all $C \in \C$, $|d(C)| \le \partialc{\pp{\C}}(C)$, it is guaranteed that there exists a flow in $G$ routing $d$ with congestion $\alpha$.  Thus, for an $\alpha$-congestion approximator $\C$,  the existence of a flow routing $d$ can be checked
simply by checking this condition for all $C \in \C$. We refer to $\alpha$ as
the \emph{approximation guarantee} of the congestion approximator. Note that if the
condition is violated for at least one set $C \in \C$ then we know that no such
flow with congestion 1 exists, but we do not know that no flow with congestion
$\alpha>1$ exists.

\paragraph{Expansion.}
We call $G=(V,E)$ \emph{$\v{\pi}$-expanding with quality $q>0$}, if every set
$X\subseteq V$ with $\v{\pi}(X)\le\v{\pi}(V \setminus X)$ fulfills
$\capc(X,V-X)\ge q\cdot\v{\pi}(X)$. If this holds for $q=1$, we just call the
graph \emph{$\v{\pi}$-expanding.} Note that the larger $q$, the stricter the
requirement. For $S \subset V$, we say that \emph{$G[S]$ is $\v{\pi}$-expanding
  with quality $q$} if it is $\v{\pi}|_S$-expanding with quality $q$. If,
however, a cut $X$ with $\v{\pi}(X)\le\v{\pi}(V \setminus X)$ exists such that
$\capc(X,V \setminus X) < q \cdot \v{\pi}(X)$, we say that $X$ is {\em
  $q \cdot \v{\pi}$-sparse}. There is an equivalent characterization of
expansion using flows. The following lemma is due to a simple application
of the max-flow min-cut theorem.
\begin{lemma}\label{lem:cutflow}
A graph $G=(V,E)$ is $\v{\pi}$-expanding with quality $q$ iff every
demand function $\v{d}$, with $|\v{d}|\le\v{\pi}$ can be routed with congestion
at most $1/q$.
\end{lemma}

\paragraph{Hierarchical Decomposition.}
A (partial) \emph{hierarchical decomposition} of $G$
is a sequence $\P=(\P_1,\dots,\P_L)$ of distinct partitions
such that
\begin{itemize}
\item $\P_1$ is the partition $\{V\}$ with a single cluster, and
\item for every $i>1$, $\P_{i}$ is a \emph{refinement} of $\P_{i-1}$, i.e., for
$X\in \P_{i}$ there exists a cluster $P\in\P_{i-1}$ with $X\subseteq P$. We
call $P$ the \emph{parent cluster} of $X$ and denote it by $\parent_\P(X)$.
\end{itemize}

We refer to the unique cluster $V\in\P_1$ as the \emph{root-cluster} of $\P$ and use the \emph{height}  $L$ to denote the number of partitions in the hierarchical decomposition.
We define $\parent_\P(V)=V$. If $\P_L=\{\{v\}\mid v\in V\}$ is the partition
into singletons, we call $\P$ \emph{complete}.

A hierarchical decomposition can be represented in a natural way by a tree
structure $\mathcal T$ where each cluster $X$ in a partition is represented by
a node in the tree (and by abuse of notation we also call that node $X$) and
$X$ is a child of $\parent(X)$. Thus all nodes that belong to the same
partition are on the same level of $\mathcal T$.

\paragraph{\mathversion{bold}$\gamma$-Border Routability.\mathversion{normal}}

While constructing the hierarchy we will use the concept of border routability.
Given a set $C\subseteq V$ and a subset $U \subseteq C$,  
we say that a cut $E(U,C\setminus U)$ is \emph{$\gamma$-border-routable 
through  $U$}, if for any non-negative weight function $\v{s}$ with $|\v{s}| \le \deg_{E(U,C \setminus U)}$ there exists a non-negative vertex weight $\v{t}$ with $|\v{t}| \le\deg_{E(U,V \setminus C)}/\gamma$ and an $(\v{s},\v{t})$-flow in $G[U]$ with
congestion $2$.

\subsection{Our Contribution}
Nearly all constructions for hierarchical congestion approximators
(e.g.~\cite{Rae02,HHR03,BKR03,RS14}) proceed in a top down manner. The graph is
partitioned recursively into smaller and smaller pieces until the remaining
pieces are just singletons. This gives a hierarchical decomposition $\P$ of the
graph. If within this decomposition every cluster fulfills some expansion
properties w.r.t.\ its sub-clusters then one obtains a good congestion
approximator. This general scheme has already been used in the result by
Räcke~\cite{Rae02} that showed the existence of good congestion approximators.
The expansion property that was used in this paper (translated into our
notation) was that a non-leaf cluster $C\in\P_i$ must be
$\alpha\cdot\deg_{\partial\P_{i+1}}$-expanding. Such a hierarchy then results
in a congestion approximator with guarantee $L/\alpha$, where $L$ is the number
of levels in $\P$.

With this in mind the na\"ive approach for constructing a congestion
approximator is to design a partitioning routine that partitions a sub-cluster
such that it expands well w.r.t.\ its sub-clusters, and then to apply this
partitioning routine recursively. However, this approach has severe limitations
when it comes to obtaining fast running times. Already at a single level the
partitioning of a cluster $C$ into sub-clusters such that the above expansion
property is fulfilled is challenging. The original polynomial time construction
(\cite{BKR03} and \cite{HHR03}) maintain a partition of the current cluster
that has to be partitioned, and keep modifying this partition until the
expansion properties are fulfilled. They show that the number of edges between
clusters of the partition monotonously decreases, and therefore the
construction terminates in polynomial time. Clearly, this step is a first
bottleneck for obtaining fast algorithms.

\subsubsection{Improved Sparse Cut Oracle}\label{sec:impr-sc}
Räcke et al.~\cite{RST14} found a solution to this problem by adapting the
Cut-Matching game framework of Khandekar, Rao, and Vazirani~\cite{KRV06} and
combining it with a fast algorithm for single-commodity maxflow due to
Sherman~\cite{She13}. This technique has later been refined by Saranurak and
Wang~\cite{SW19} and applied to the problem of finding expander decompositions
in graphs.

The Cut-Matching game framework can be viewed as an efficient implementation of
a sparse cut oracle. Given a sparsity parameter $\phi$, an \emph{approximate
sparse cut oracle} either declares that the graph has expansion at least $\phi$
or returns a cut with sparsity at most $\alpha\phi$ for some $\alpha\ge
1$.
One can obtain a good expander decomposition by repeatedly partitioning the
graph along (approximate) sparse cuts until all pieces have expansion at least
$\phi$. However, doing so with just a sparsest cut oracle might be very very
slow as in every step a small piece might be cut from the graph, leading to
very little progress.

Saranurak and Wang strengthened the Cut-Matching game framework so that it
gives a sparse cut oracle that guarantees to find a \enquote{balanced} sparse
cut if one exist. More precisely, if the oracle returns a sparse cut that only
contains a very small subset of the vertices, then it guarantees that the
larger side is \enquote{nearly expanding}. They showed that from this \enquote{nearly
  expanding} property they could obtain a large subset that is a
proper $\phi$-expander, and that guarantees that they always make good progress
towards finding an expander decomposition.

The first ingredient for our improved construction algorithm for congestion
approximators is therefore an improved sparse cut oracle.

\begin{buildingblock*}[\algSC$(G,\v{\pi},\phi)$]\label{current}
Given a graph $G=(V,E)$, a non-negative (vertex) weight function $\v{\pi}$ and a
sparsity parameter $\phi \in (0,1)$, the algorithm \algSC{} computes a (potentially
empty) set $R \subseteq V$, with  $\v{\pi}(R) \leq \v{\pi}(V\setminus R)$ such that
    \begin{enumerate}
    \item $R$ is $\phi$-sparse w.r.t.\ $\v{\pi}$, i.e.,
    $\capc(R, V \setminus R) \leq\phi \v{\pi}(R)$; and %\label[property]{prop:SC:sparse}
    \item if $R$ is very imbalanced, i.e.,
    $\v{\pi}(R) < \bal\v{\pi}(V)$, then $G$ is
    $ (\phi/\q) \cdot \v{\pi}|_{V\setminus R}$ expanding with high
    probability. 
    \end{enumerate}
\end{buildingblock*}
\noindent
We show how to implement this sub-routine with $\q=O(\log n)$ and
$\bal=\Omega(1/\log n)$. Observe that if $R=\emptyset$ the first condition is
always fulfilled. One can view a traditional sparse cut oracle as an
oracle that only fulfills the second condition if the returned set is
$R=\emptyset$; then the oracle certifies that $G$ is $\phi/\q$-expanding
w.r.t.\ $\v{\pi}$.

Saranurak and Wang~\cite{SW19} used a similar oracle with weight function
$\v{\pi}=\v{\1}$ in the definition of sparsity and $\q=O(\log^2 n)$. Agassy et al.~\cite{ADK22}
showed how to improve to $\q=O(\log n)$ by basing their oracle on the improved
Cut-Matching game due to Orecchia, Schulman, Vazirani and
Vishnoi~\cite{OSVV08}. Their oracle uses weight function $\v{\pi}=\deg_V$
(conductance). For our application it is crucial to allow an arbitrary non-negative weight
function, and we show how to obtain this oracle by adapting the techniques used
by Agassy et al. 

Note that the expansion criterion that we use in the imbalanced case above is
different from the concept of a \emph{near-expander} as introduced
by~\cite{SW19} (and also used by~\cite{ADK22}). In our terminology a set
$S\subseteq V$ is a near-expander in $G$ if $\tilde{G}$ is $\deg|_S$-expanding,
where $\tilde{G}$ is the graph obtained from $G$ by contracting the vertices in
$V\setminus S$ into a single vertex. Saranurak and Wang~\cite{SW19} and Agassy
et al.~\cite{ADK22} require that in the imbalanced case the set $V\setminus R$
is a near-expander in $G$. 
Note that $G$ being $\deg|_S$-expanding
implies that $S$ is a near-expander in $G$, but not vice versa~\footnote{Our definition is in our opinion more intuitive: For example, assume that $k$ is an integral divisor of $n$ and consider the extreme example where $G$ is a set of $k$ disjoint stars, each with $n/k$ nodes. Then the set $S$ of the $k$ centers of the star is a near-expander, even though it it not connected. However, $G$ is not $\deg|_S$-expanding as the stars in this example are disconnected.}.

\paragraph{Very Recent Independent Work.}
As mentioned before, very recently, Agassy, Dorfman and Kaplan~\cite{ADK25} published a paper on arXiv, that also presents a generalized sparsest cut oracle.
While our presented algorithm only handles integral, nonnegative vertex weights, they handle arbitrary, nonnegative vertex weights and their running time is $\tO(m \operatorname{polylog}(W))$, where $W$ is the ratio of largest to smallest vertex weight.
Thus, for the unweighted setting, this matches our running time and approximation quality.

\subsubsection{Handling Bad Child Events}
The above sparse cut oracle is the first main ingredient that we need,
to implement a partitioning routine with which we can recursively construct the
hierarchy. However there is another more severe obstacle for obtaining fast
algorithms for hierarchical congestion approximators, namely \emph{bad child
  events}. Suppose that we start our recursive decomposition by applying our
partitioning routine first to $G[V]$, then to its sub-clusters, and so on. It
may happen that we arrive at a cluster $C$ for which it is impossible to find a
sub-clustering $\X$ so that $G[C]$ is $\alpha\deg_{\partial \X}$-expanding with a
reasonable large value for $\alpha$. 
This problem already existed in the
original polynomial-time constructions and was solved in different ways:
Bienkowski et al.~\cite{BKR03} used a partitioning routine that not only
ensured that a cluster $C$ is expanding w.r.t.\ its sub-clusters but also
guaranteed that no bad child events could possibly occur. Harrelson et
al.~\cite{HHR03} introduced the concept of a so-called \emph{bad child event:} If a cluster
occurred that could not properly be partitioned, the cluster was split and both
parts were re-introduced as children of the parent cluster (thereby altering
the partition of the parent cluster). The latter approach guaranteed the better
approximation guarantee but also substantially increases the complexity of
the algorithm\,---\,making it more difficult to obtain a variant that runs in
nearly linear time.

Because of the difficulty Räcke et al.~\cite{RST14} used a completely different
top-down approach for constructing their hierarchy, which lead to an
approximation guarantee of $O(\log^4n)$ and takes $\Omega(n^2)$ time. 

In this paper we show how to obtain a partitioning routine, called
\magicpart{}, that can deal with bad child events and still run in near-linear
time. Basically, if a cluster $C$ does not sufficiently expand, then it finds a sparse cut such that one
side of the cut is border-routable and the cut is either ``balanced'' or the other side of the cut
is sufficiently expanding. More formally it guarantees the following properties.

\begin{subroutine*} [\magicpart{}$(G,C,\X,\phi)$]
We are given a graph $G=(V,E)$, a subset $C\subseteq V$, a partition $\X$
of $C$ with $z:=\max_{X\in\X}|X|$, and an expansion
parameter $\phi$ with $0<\phi\le 1/4$. The procedure \magicpart{} returns
a (possibly empty) subset $U \subset C$ with $|U| \le |C|/2$ and a new
partition $\Y$ of $C$ with 
$U\in\Y$ (if $U\neq\emptyset$) and $|Y|\le\max\{z,|C|/2\}$ for all $Y\in\Y$. Furthermore,
\begin{enumerate}
  \item the cut $E(U,C\setminus U)$ is $1/\phi$-border routable through $U$
    with congestion $2$, w.h.p.
      \item \textbf{either}
 $\deg_{\partial \Y}(U) \ge \Omega(1/\log n) \cdot \deg_{\partial \Y}(C)$ and
      $\deg_{\partial\Y}(C)\le \deg_{\partial\X}(C)+2\capc(U,C\setminus
      U)$

      \noindent
      \textbf{or} $G[C \setminus U]$ is $\deg_{\partial \mathcal{Y}}$-expanding
      with quality $\phi/(500\q)$, w.h.p. 
  \end{enumerate}
\end{subroutine*}
If the partitioning routine returns a non-empty set $U$ we call $U$ a \emph{bad
child}. The first condition implies that the capacity of the cut
$E(U,C\setminus U)$ is fairly small: at most $\phi\cdot\deg_{\partial C}(U)$.
Otherwise, it would clearly not be possible to send a flow where each source
edge in $E(U,C\setminus U)$ sends $1$ and each vertex receives at most
$\phi\cdot\deg_{\partial C}(U)$ as required by the border-routability
condition. If now $\deg_{\partial C}(U)\le \deg_{\partial C}(C\setminus U)$
holds then this cut certifies that $C$ is at most
$\phi\deg_{\partial C}$-expanding, which means that any partition $\X$ of
$C$ would be $\deg_{\partial\X}$-expanding with quality at most $\phi$.
Depending on the value of $\phi$ this would be problematic for constructing the
hierarchy. Note that it may happen that the routine returns a $U$ that actually
is not an obstacle for further partitioning as described above (e.g.\ it may be
that $\deg_{\partial C}(C\setminus U)\ll\deg_{\partial C}(U)$).
Nevertheless, we still refer to $U$ as a bad child.

The border-routability condition is essential to guarantee that after introducing
possibly several bad child events at a parent cluster $P$ the
$\deg_\Y$-expansion of $P$ does not degrade too much due to the change in its
sub-clustering $\Y$ within the hierarchical partition. More specifically we
show that the quality of expansion decreases by at most a factor of $2e$.
How is this achieved? Suppose a bad child event is introduced at the parent $P$
because a sub-cluster $C$ of $P$ is partitioned into $C$ and $C\setminus U$.
This changes the expansion guarantee due to the new inter cluster edges
$E(U,C\setminus U)$. The border-routability guarantees that demand at these
edges can be cheaply routed to the edge set $E(U,V\setminus C)$, i.e., the
border edges of $C$ incident to $U$. In particular these edges only receive
very little demand. If we can route all demand on newly introduced edges 
to \enquote{old} edges (inter-cluster edges that existed before any bad child
event) with small congestion and an \enquote{old} edge does not receive
too much additional demand a good expansion is guaranteed. 

The main challenge is to prove that border-routability is sufficient to
preserve good expansion even after an arbitrary sequence of bad child events.
Each such event introduces new inter-cluster edges, and we must show that their
demand can always be routed to \enquote{old} edges with bounded congestion. This
suffices to guarantee that the $\deg_\Y$-expansion of $P$ degrades by at most a
constant factor.

The second condition of the sub-routine \magicpart{} is the crucial difference to the
construction in~\cite{HHR03} and, in combination with the improved sparse cut
oracle, allows for a linear-time algorithm. The construction in~\cite{HHR03}
only guarantees that you \emph{either} find a bad child, \emph{or} you are able to completely
partition $C$ with a partitioning $\X$ so that $C$ is $\alpha\deg_{\partial \X}$-expanding
for a large enough value of $\alpha$. Our routine also \emph{gives some expansion
guarantee for the sub-clustering when there is a bad child}. In particular in the
\enquote{or}-case it says that $G[C\setminus U]$ has sufficient expansion
w.r.t.\ its sub-clustering (or the cut fulfills some other useful properties such as being balanced - see below). 
Hence, we can introduce the clusters $U$ and
$C\setminus U$ as new children at the parent of $C$ (replacing $C$) and at the
same time introduce the sub-clustering for $C\setminus U$. Then one only has to
recurse on the smaller sub-cluster $U$. 
We refer to this case as an \emph{imbalanced} bad child; otherwise we call
the child \emph{balanced}.

\subsubsection{Tracking Progress}
Designing the above interface and efficiently implementing  the algorithm
\magicpart{} forms the central component of our technical contribution. The
main challenge lies in defining an appropriate measure of progress that ensure
that the overall running time is near-linear, when using \magicpart{} to
construct a congestion approximator. For this we need to ensure
\begin{enumerate}
\item progress \emph{during} \magicpart{}, so that the running time of one call
is small;\label[property]{progress:iteration}
\item progress between different calls to \magicpart{}, so that
the running time of repeated/recursive calls generated by bad child events is
small;\label[property]{progress:badchild}
\item progress in the final result to guarantee that the constructed hierarchy
has logarithmic height.\label[property]{progress:hierarchy}
\end{enumerate}
\Cref{progress:iteration} is obtained by using $\deg_{\partial \X}(C)$ as a progress
measure, where $\X$ is the sub-clustering of $C$. When the procedure \emph{first}
starts on partitioning a sub-cluster, $\X$ is chosen as the partition into singletons. 
Then $\X$ is changed during perhaps several iterations. In every non-terminating
iteration it is guaranteed that $\deg_{\partial \X}(C)$ decreases by a reasonable
amount. This makes sure that \emph{one} call to \magicpart{} terminates quickly.

For the progress in \Cref{progress:badchild} there are two cases: a
\emph{balanced bad child} or an \emph{imbalanced bad child}. The second case is
straightforward: we only recurse on $U$ as $C\setminus U$ is expanding w.r.t.\
its sub-clustering. As $|U|\le|C|/2$ there is enough progress. However, for the
\emph{balanced} case we were unable to obtain a progress in terms of
cardinality: we have to recurse on both sides and one side could have nearly all the
vertices.

Instead, we measure the progress again in terms of a reduction in
$\deg_{\partial \X}(C)$. For this it is crucial that during a recursive/repeated call
(a call caused by a bad child event), the \magicpart{} routine does not again
start with a clustering into singletons but it continues to work on the
sub-clustering from the previous run that is given as a parameter. However,
$\deg_{\partial \X}(C)$ is not necessarily monotone. It might increase during a
terminating iteration of \magicpart{}. 

The property in the \enquote{or}-case (\emph{balanced bad child}) first of all
guarantees that during the call to \magicpart{} $\deg_{\partial \X}(C)$ did not increase
by too much. Only by $2\capc(U,C\setminus U)$. Together with the fact that
$\capc(U,C\setminus U)$ is a sparse cut (i.e.,
$\deg_{\Y}(U)\gg \capc(U,C\setminus U)$) and
$\deg_{\partial \Y}(U) \ge \Omega(1/\log n) \cdot \deg_{\partial \Y}(C)$ we get
the desired progress for the cluster $C\setminus U$: 
$\deg_{\Y}(C\setminus U)\le (1-\Omega(1/\log n))\deg_{\partial \X}(C)$.

The progress over the hierarchy (\Cref{progress:hierarchy}) is guaranteed by
\magicpart{} because the implementation only changes the clustering $\X$ by
so-called \emph{fuse}-operations. The operation $\X - T$ returns the partition
$\{A \setminus T\mid A \in \X , A \setminus T \ne \emptyset \}$; and a
\emph{fuse operation} changes $\X$ to $(\X-T)\cup \{T\}$. We only apply
fuse-operations for sets $T$ with $T\le |C|/2$. This has the effect that in the
final hierarchy a cluster can be at most half the size of its grand-parent.
Note that we cannot guarantee that a cluster has at most half the size of its parent:
Intuitively, when cluster is created it is at most half the size of its parent at this time. However, its parent can change (and decrease in size), while grandparents never change.

\subsubsection{Implementing \magicpart}
In a first step the routine computes a sparsest cut w.r.t.\ the weight function
$\deg_{\partial \X}$ using \algSC{}. If the returned cut $R$ is empty, we have the desired
expansion and can return an empty bad child.

Otherwise, there are several cases. One case is that we identify a set $T$ that
a) contains at most half the vertices; b) has a logarithmic fraction of the
overall weight, and c) is sparse (i.e.,
$\capc(T,C\setminus T)\lesssim\phi\bm{\pi}(T)$). If now most of the weight of $T$
does \emph{not} lie on the boundary we can fuse $T$ in $\X$ and substantially
decrease $\deg_{\partial \X}(C)$ (importantly we guarantee that we only fuse small
  sets). If on the other hand most of $T's$ weight is on its boundary it must mean
that the edges in $E(T,V\setminus C)$ contribute most of this weight as
otherwise the cut $(C\setminus T,T)$ wouldn't be sparse. In this case we want
to return $T$ as a balanced bad child. However, for this we need to ensure
border-routability. In order to obtain this we compute a fair
cut~\cite{LNPS23,LL25} $T'$ between edges in $E(T,C\setminus T)$ and the border
edges $E(T,V\setminus C)$. Then we return this set $T'$. This set still
contains a lot of weight and fulfills all properties required in the
\enquote{either}-case of \magicpart{}.

How do we find a suitable set $T$? If the sparsest cut $R$ returned from
\algSC{} contains a reasonable fraction of the weight $\deg_{\partial \X}(C)$ we can
essentially use this set (we take the one from $R$ and $C\setminus R$ that has
smaller cardinality). Otherwise, the guarantees from \algSC{} give us that
$G[C]$ is $\deg_{\partial \X}|_{C\setminus R}$-expanding. Now, we actually would like
to return $R$ as an imbalanced bad child. However, we have to 
ensure that
\begin{itemize}
\item $G[C\setminus R]$ is $\deg_{\partial \X}|_{C\setminus R}$-expanding (instead
of $G[C]$ being $\deg_{\partial \X}|_{C\setminus R}$-expanding)
\item the cut edges $E(R,C\setminus R)$ are border-routable through $R$
\end{itemize}
The property that $G[C]$ is $\deg_{\partial \X}|_{C\setminus R}$-expanding is similar to
the notion of a near expander as introduced by Saranurak and Wang~\cite{SW19} (as discussed before).
Expander trimming~\cite{SW19} also works for our notion. This would find a set
$A\subseteq C\setminus R$ that is properly $\deg_{\partial \X}$ expanding. However, we
crucially need the border-routability property. Therefore, we introduce a
sub-routine \algTWT{} that finds a subset $A'$ that is $\deg_{\partial \X}$-expanding
\emph{and} for which $E(A',C\setminus A')$ is border-routable through
$C\setminus A'$. This is implemented by two trimming operations (using suitable fair $(\v{s},\v{t})$-flows): we first trim
to obtain a set $A$ as above; then we compute a fair cut between the edges
$E(A,C\setminus A)$ and $E(C\setminus A,V\setminus C)$ in order to guarantee
border-routability; this fair cut is then added to $A$ to form $A'$.
Importantly, we show that we can implement the second trim operation so that it
does not destroy the expansion guarantee that we obtained from the first step.

\subsubsection{Usage of Fair Cuts}
We generalize the notion of $s,t$ fair cuts to the setting where $\v{s}$ and $\v{t}$ are vertex weightings. We show two use cases of this generalization can be used with a suiatable choice for $\v{s}$ and $\v{t}$ to compute sparse cuts in the cut matching game and for guaranteeing border routability in partition cluster.

%%%%%%%%%%%%%%%%%%%%%%%%%%%%%%%%%%%%%%%%%%%%%%%%%%%%%%%%
%%%%%%%%%%%%%%%%%%% Preliminaries %%%%%%%%%%%%%%%%%%%%%%
%%%%%%%%%%%%%%%%%%%%%%%%%%%%%%%%%%%%%%%%%%%%%%%%%%%%%%%%

\section{Building Blocks}\label{sec:buildingblocks}

We next present a list of algorithmic problems and solutions that are used as
subroutines by our algorithm in order to build the desired hierarchical
decomposition. As these subroutines are crucial for our algorithm we call \emph{building blocks.}
\begin{definition}[Fair Cut/Flow Pair]\label{def:fair-cut}
\label{bb:faircutdef}
Let $G=(V,E)$ be a graph with integral edge capacities $\capc$ and let
$\v{s}$ and $\v{t}$ be two non-negative vertex weight functions. For
any parameter $\alpha \geq 1$, we say that a set $U \subseteq V$ and a feasible
flow $f$ is an $\alpha$-fair ($\v{s}$,$\v{t}$)-cut/flow pair $(U,f)$
if
\begin{enumerate}
    \item net sources do not send too much:\hfill\\
    for each vertex $v$ with $\v{s}(v)-\v{t}(v)\ge0$: $0\le f(v)\le \v{s}(v)-\v{t}(v)$\label[property]{prop:fc:net-s}

    \item net targets do not absorb too much:\hfill\\
     for each vertex $v$ with $\v{s}(v)-\v{t}(v)\le0$: $0\ge f(v)\ge \v{s}(v)-\v{t}(v)$\label[property]{prop:fc:net-t}

    \item net sources in $V\setminus U$ are nearly saturated:\hfill\\
     for each vertex $v$ with $\v{s}(v)-\v{t}(v)\ge0$, $v \in V \setminus U$: $f(v)\ge (\v{s}(v)-\v{t}(v))/\alpha$
     \label[property]{prop:fair-cut:generate}

    \item net targets in $U$ are nearly saturated:\hfill\\
     for each vertex $v$ with $\v{s}(v)-\v{t}(v)\le0$, $v \in U$: $f(v)\le (\v{s}(v)-\v{t}(v))/\alpha$
     \label[property]{prop:fair-cut:absorb}

    \item edges from $U$ to $V \setminus U$ are nearly saturated:\hfill\\
      each edge $\{u,v\} \in E(U, V \setminus U)$  with $u\in U$ and
      $v \in V \setminus U$ sends at least $\capc(u, v)/\alpha$
      flow in the direction from $u$ to $v$. In particular, no flow is sent in
      the reverse direction on these
      edges. \label[property]{prop:fair-cut:cross-cut}
    \end{enumerate}
 An \emph{$\alpha$-fair $(\v{s},\v{t})$-cut $U$ in $V$} is a vertex set for
    which a flow $f$ exists such that $(U,f)$ is an
    $\alpha$-fair ($\v{s}$,$\v{t}$)-cut/flow pair in $V$.   
\end{definition}
Note that this definition is not symmetric in the sense that the fact that
$(U,f)$ is an $\alpha$-fair ($\v{s}$,$\v{t}$)-cut/flow pair in $V$
does not imply that this also holds for $(V \setminus U,f)$. It does, however,
imply that $(V \setminus U, -f)$ is an $\alpha$-fair ($\v{t}$,$\v{s}$)-cut/flow pair in $V$. 
Our first building block computes an $\alpha$-fair ($\v{t}$,$\v{s}$)-cut/flow pair in $V$,

\begin{buildingblock*}[\algFC$(G, \v{s}, \v{t}, \alpha)$]\label{bb:algFC}
Given a graph $G = (V, E)$ with integral edge capacities
$\capc$, two non-negative vertex weight functions $\v{s}, \v{t}$, and a
parameter $\alpha \geq 1$, the algorithm \algFC{} outputs an $\alpha$-fair
$(\v{s}, \v{t})$-cut/flow pair $(U, f)$.
\end{buildingblock*}

For many applications of~\algFC{}, the flow is not needed
explicitly and just the cut would be sufficient alongside the promise of \emph{existence} of a flow. However, the implementation we
use (\cref{lemma:fair-cut}) computes both simultaneously, so we define it here to
output both as there is no additional overhead for providing an explicit flow. 
Specifically, in \cref{sec:fair-cut} we show how to obtain the following lemma with a simple reduction to known results.

\begin{theorem}\label{lemma:fair-cut}
For a graph $G=(V,E)$ with $m = \deg(V)$ and $\alpha > 1$ our algorithm \algFC{} has running time $\Tfc(m,\alpha) = \tO(m/(\alpha-1))$.
\end{theorem}

Our next building block computes an approximately sparsest cut, as defined below. Its output guarantees use parameters $\q \geq 1$ and $\bal \in (0, 1/2]$, which correspond to the achieved quality of expansion and balance guarantee, respectively.

\begin{buildingblock*}[\algSC$(G,\v{\pi},\phi)$]\label{bb:algSC}
Given a graph $G=(V,E)$, an integral,
non-negative vertex weight function $\v{\pi}$ and a sparsity parameter
$\phi \in (0,1)$, the algorithm \algSC{} computes a (potentially empty) set
$R \subseteq V$, with $\v{\pi}(R) \leq \v{\pi}(V\setminus R)$ such that
    \begin{enumerate}
    \item $R$ is $\phi$-sparse w.r.t.\ $\v{\pi}$, i.e.,
    $\capc(R, V \setminus R) \leq\phi \v{\pi}(R)$; and \label[property]{prop:SC:sparse}
    \item if $R$ is very imbalanced, i.e.,
    $\v{\pi}(R) < \bal\v{\pi}(V)$, then $G$ is
    $ (\phi/\q) \cdot \v{\pi}|_{V\setminus R}$ expanding with high
    probability. \label[property]{prop:SC:exp}
    \end{enumerate}
\end{buildingblock*}

In \cref{sec:spx} we show our implementation of \algSC{} and prove the following theorem. The algorithm is based on a general adaptation of the cut-matching game, allowing for weights on the vertices (\cref{sec:cmgame}), as is needed here.

\begin{theorem}\label{lemma:spx}
Given a graph $G=(V,E)$ with $n = |V|$ and $m=\deg(V)$ and an algorithm for
\algFC{} that runs in time $\Tfc$, we can implement \algSC{} with parameters
$\q = O(\log{\bm{\pi}(V)})$ and $\bal =  1/(2 \log{\bm{\pi}(V)})$ in running time $\Tsc=O \big(\log^4(\bm{\pi}(V)) \cdot (\Tfc(m,3/2) + \bm{\pi}(V) \log{(\bm{\pi}(V))} + m \log{n})\big)$. 
\end{theorem}

While the theorem is more general, for all our purposes, we will have
$\bm{\pi}(V) = O(m)$. Hence, throughout the paper we use the resulting values
$\q = O(\log{n})$ and $\bal = \Omega(1/\log{n})$ as the quality and balance
guarantee of our implementation of \algSC{}. For technical reasons, we will
further assume that $\q \geq (\log{n})/125$. 
We also use the value
$\t:=\min\{\frac{1}{440\q},\, \bal\}$. In our implementation, we thus have
$\t = \Omega(1/\log{n})$. Together with our implementation of \algFC{}
(\cref{lemma:fair-cut}), we get the following corollary.

\begin{corollary}
  If $\bm{\pi}(V) = O(m)$, then \algSC{} can be implemented with $\q = O(\log{n})$ and $\bal = \Omega(1/\log{n})$ in time $O(\log^5{n} \cdot (\Tfc(m,3/2)) + m ) = \tO(m)$.
\end{corollary}

%%%%%%%%%%%%%%%%%%%%%%%%%%%%%%%%%%%%%%%%%%%%%%%%%%%%%%%%
%%%%%%%%%%%%%% Congestion Approximator %%%%%%%%%%%%%%%%%
%%%%%%%%%%%%%%%%%%%%%%%%%%%%%%%%%%%%%%%%%%%%%%%%%%%%%%%%

\section{Congestion Approximator}
Our goal is to design an algorithm that constructs a hierarchical decomposition that fulfills certain properties and will then show that this naturally give a congestion approximator.
More specifically, we call every set that belong to a partition of the constructed (partial) hierarchical decomposition a \emph{cluster}.
Our goal is to design a hierarchical decomposition in which every
  cluster fulfills certain expansion properties w.r.t.\ its partitioning into
  sub-clusters. Formally, we define for a hierarchical decomposition
  $\P=(\P_1,\dots,\P_L)$ the function $f_\P$ for a level-$i$ cluster $X$ in the hierarchy as
\[
  f_\P(X):=\left\{
  \begin{array}{ll}
       1                                         & i=1\\
       3\log\log n\log_2({2|\parent_\P(X)|}/{|X|}) & \text{otherwise}
  \end{array}\right.
\] 
that defines an \emph{expansion bound} for every cluster. We say that a
non-leaf cluster $X\in\P_i$, $i<L$ is \emph{$\gamma$-well expanding} for some $1\ge \gamma >0$ if $G[X]$ is at least
$\deg_{\partial \P_{i+1}}$-expanding with quality $\gamma /f_\P(X)$.
Our hierarchy construction algorithm \algCH{} will build a hierarchy so that every non-leaf cluster is $\gamma$-well expanding with $\gamma =\Theta(1/\log n)$.

Our goal is to find a complete hierarchical decomposition
$\P=(\P_1,\dots,\P_L)$ such that (A) each non-leaf cluster is $\gamma$-well expanding and
(B) $L= O(\log n)$. The next theorem states that the set of clusters in such a decomposition is a ``high-quality'' congestion approximator.

\begin{theorem}\label{thm:main}
A complete hierarchical decomposition of logarithmic height in which each
non-leaf cluster is $\gamma$-well-expanding gives a congestion
approximator with approximation guarantee
$6\log\log n(L+\log n)/\gamma=O(\log^2 n\log\log n)$.
\end{theorem}

\begin{proof}
Throughout this proof we use $X_v^{(i)}\in\P_i$ to denote the level-$i$ cluster
that contains $v$ within the partition $\P_i, i\ge1$ from the hierarchy.

Suppose we are given a demand vector $d:V\rightarrow \mathbb{R}$ such that
$|d(X)|\le\partialc{\partial\P_i}(X)$ holds for every $X\in \P_i$ and every 
$i\in\{1,\dots,L\}$. We have to show that we can route $d$ in $G$ with congestion at
most $6\log\log n(L+\log n)/\gamma$.
This shows that $\P$ is a $6\log\log n(L+\log n)/\gamma$-congestion approximator. 

We next explain how we route $d$. The basic idea is to
route $d$ level by level. To do so we need to define a suitable demand $d_i$ for each level $i$.

For level 1, we define $d_1(v) = 0$ for every $v \in V$.
For $i>1$ we define the \emph{level-$i$ demand} $d_i(v)$ of a vertex $v$ as 
$$
  d_i(v)=\frac{\partialc{\partial\P_i}(v)}{\partialc{\partial\P_i}(X_v^{(i)})}d(X_v^{(i)})\enspace.
$$
Recall that wlog the graph is connected and, thus, $\partial\P_i(X)\neq 0$ for each $X \in \P_i$.
Further observe that $d_i(X) = \sum_{v \in X} d_i(v) = \sum_{v\in
  X}\frac{\partialc{\partial\P_i}(v)}{\partialc{\partial\P_i}(X)}d(X) = d(X)$
for each $X\in\P_i$. From this we get
$\sum_v d_i(v)=\sum_{X\in\P_i}d_i(X)=\sum_{X\in\P_i}d(X)=\sum_{v}d(v)=0$, i.e., $d_i$ is a proper
demand vector.

Note that if the hierarchical decomposition is complete, i.e., $\P_L$ consists of only singleton clusters, then the definition of $d_i$ implies that $d_L = d$.
Next we assume we have  for
$i\in\{1,\dots,L-1\}$ a flow that routes demand $d_{i+1}-d_{i}$. 
Summing all these flows gives a flow that routes demand $d_L-d_1=d$.
Thus, we are left with giving a flow that routes the demand $d_{i+1}-d_i$. 

Again, we partition the demand into several sub-demands, one for each cluster $X\in\P_i$. Formally, the \emph{subflow for cluster} $X\in\P_i$ routes demand vector $d_{i+1}|_X-d_i|_X$ \emph{inside} 
$G[X]$. Note that $d_{i+1}|_X-d_i|_X$ is indeed a proper demand vector for $G[X]$ as $\P_{i+1}$ is a refinement of $\P_i$ and, thus, 
$d_{i+1}(X)= \sum_{Z \in \P_{i+1}, Z \cap X \ne \emptyset} d_{i+1}(Z) = \sum_{Z \in \P_{i+1}, Z \cap X \ne \emptyset} d(Z) = d(X) = d_i(X)$ holds for any cluster $X \in \P_i$ which implies $\sum_{v \in X} (d_{i+1}|_X-d_i|_X)(v) = 0$

Let $S\subseteq X$ be any subset in $X$ and assume wlog that
$\partialc{\partial\P_{i+1}}(S)\le \partialc{\partial\P_{i+1}}(X\setminus S)$
inside $G[X]$. We show next that the total demand that has to cross $S$ in
$G[X]$ is at most $2 f_\P(X)\operatorname{cap}(S,X\setminus S)/\gamma$. By the
maximum flow-minimum cut theorem~\cite{FF56} it follows that there exists a
(single-commodity) flow that routes  $d_{i+1}|_X-d_i|_X$ \emph{inside} 
$G[X]$ with congestion at most $2
f_\P(X)/\gamma$. As the sets $X \in \P_{i}$ are disjoint, it follows that
$d_{i+1}-d_i$ can be routed in $G$ with congestion at most $2
f_\P(X)/\gamma$. Summing up the flows $d_{i+1}-d_i$ for $i \in \{1, \dots, L-1\}$ will then show that $d$ can be routed in $G$
with congestion at most $2 (L-1)
f_\P(X)/\gamma$

We still have to show  that the total demand that has to cross $S$ in
$G[X]$ is at most $2 f_\P(X)\operatorname{cap}(S,X\setminus S)/\gamma$. 
For $i>1$ this demand is 
\begin{equation*}
\begin{split}
\Big|\sum_{v\in S}d_{i+1}(v)-\sum_{v\in S}d_i(v)\Big|
   &\le \sum_{v\in S}|d_{i+1}(v)|+\sum_{v\in S}|d_i(v)|\\
   &= \sum_{v\in S}\frac{\partialc{\partial\P_{i+1}}(v)}{\partialc{\partial\P_{i+1}}(X^{(i+1)}_v)}|d(X^{(i+1)}_v)|+
        \sum_{v\in S}\frac{\partialc{\partial\P_{i}}(v)}{\partialc{\partial\P_{i}}(X_v^{(i)})}|d(X_v^{(i)})|\\
   &\le\partialc{\partial\P_{i+1}}(S)+\partialc{\partial\P_{i}}(S)\le
     2\partialc{\partial\P_{i+1}}(S)\\
   &\le2 f_\P(X)\operatorname{cap}(S,X\setminus S)/\gamma\enspace,
\end{split}
\end{equation*}
where the third inequality uses the fact that
$\partialc{\partial\P_j}(Y)\ge |d(Y|)$ holds for any level $j$ cluster $Y$, and
the fifth inequality uses that $G[X]$ is $\partialc{\partial\P_{i+1}}$-expanding
with quality $\gamma/f_\P(X)$. 

For $i=1$ the only cluster is $V$. The demand in the flow $d_{2}-d_1=d_{2}$
that has to cross the cut $S$ is
\begin{equation*}
\begin{split}
\Big|\sum_{v\in S}d_{2}(v)\Big|
   &\le \sum_{v\in S}|d_{2}(v)|
   = \sum_{v\in S}\frac{\partialc{\partial\P_{2}}(v)}{\partialc{\partial\P_{2}}(X^{(2)}_v)}|d(X^{(2)}_v)|\\
   &\le\partialc{\partial\P_{2}}(S)
   \le2f_\P(V)\operatorname{cap}(S,V \setminus S)/\gamma\enspace.
\end{split}
\end{equation*}
This means that the subflow problem for any cluster $X$ in the hierarchical
decomposition can be solved with congestion $2f_\P(X)/\gamma $. 

Now fix an edge
$e=(u,v)$ and let $k$ denote the largest level in the hierarchy such that
$u$ and $v$ are contained in the same $k$-level cluster. Let $X_i$,
$i\in\{1,\dots,k\}$ denote the $i$-level cluster that contains $e$. The
congestion of $e$ due to all flow problems is at most 
$
2\sum_{i=1}^kf_\P(X_i)/\gamma 
\le6\log\log n(1+\sum_{i=2}^k(1+\log|\parent_\P(X_i)|-\log|X_i|))/\gamma 
=6\log\log n(k+\log |\parent_\P(X_1)|-\log|X_k|)/\gamma 
\le 6\log\log n(L+\log |V|-\log |X_k|)/\gamma \le 6\log\log
n(L+\log n)/\gamma$.
\end{proof}

%%%%%%%%%%%%%%%%%%%%%%%%%%%%%%%%%%%%%%%%%%%%%%%%%%%%%%%%
%%%%%%%%%%%%%%% Construction Algorithm %%%%%%%%%%%%%%%%%
%%%%%%%%%%%%%%%%%%%%%%%%%%%%%%%%%%%%%%%%%%%%%%%%%%%%%%%%
\section{Hierarchy Construction Algorithm}
\label{sec:hierarchy}\label{subsec:ConstructingHierarchy}

In this section we give an algorithm that efficiently constructs a
$\gamma$-well expanding hierarchical decomposition   $\P=(\P_1,\dots,\P_L)$.
To build the hierarchy $\P=(\P_1,\dots,\P_L)$ we need the following subroutine,
called \magicpart{}, whose implementation is presented in \Cref{sec:cluster}.
Here we
describe how \magicpart{} is used to efficiently construct the hierarchy.

\begin{subroutine*} [\magicpart{}$(G,C,\X,\phi)$]\label{bb:part}
We are given a graph $G=(V,E)$, a subset $C\subseteq V$, a partition $\X$
of $C$ with $z:=\max_{X\in\X}|X|$, and an expansion
parameter $\phi$ with $0<\phi\le 1/4$. The procedure \magicpart{} returns
a (possibly empty) subset $U \subset C$ with $|U| \le |C|/2$ and a new
partition $\Y$ of $C$ with
$U\in\Y$ (if $U\neq\emptyset$) and $|Y|\le\max\{z,|C|/2\}$ for all $Y\in\Y$. Furthermore,
\begin{enumerate}
  \item the cut $E(U,C\setminus U)$ is $1/\phi$-border routable through $U$
    with congestion $2$, w.h.p.\ \label[property]{prop:partition:border-routable}
      \item \textbf{either}
 $\deg_{\partial \Y}(U) \ge (\t/20) \cdot \deg_{\partial \Y}(C)$ and
      $\deg_{\partial\Y}(C)\le \deg_{\partial\X}(C)+2\capc(U,C\setminus
      U)$

      \noindent
      \textbf{or} $G[C \setminus U]$ is $\deg_{\partial \mathcal{Y}}$-expanding
      with quality $\phi/(500\q)$, w.h.p.\label[property]{prop:partition:expansion}
  \end{enumerate}
\end{subroutine*}
We now show how \magicpart{} can be used to efficiently build a complete
hierarchical decomposition, where each non-leaf cluster is $\gamma$-well
expanding. Starting with an initial partition that consists only of the set
$V$, we apply \magicpart{} repeatedly to the leaf clusters of the current
partial hierarchical decomposition until each leaf cluster consists of a
singleton vertex. The following observation shows that in the first call to
\magicpart{} no bad child event occurs, i.e., the returned set $U$ is the empty
set.
\begin{observation}\label{obs:part-no-bdry}
    Let $\X$ be the partition of $V$ into singletons.
   Executing \magicpart{}$(G, V, \X, \phi)$ returns an empty set $U$ together with a partition $\Y$ and $G$ is $\deg_{\pp{\Y}}$-expanding with quality $\frac{1}{500}\phi/\q$.
\end{observation}
This follows since $V$ has no border edges, which implies that no non-trivial
$\phi$-border-routable set $U \subset V$ can exist. Consequently, it follows that $U$ is empty. Additionally, the output guarantees of \magicpart{}$(G, V, \X, \phi)$ imply that $G[V]$
is $\deg_{\pp{\Y}}$-expanding with quality
$\frac{1}{500}\phi/\q$ as it is not possible that $0 = \deg_{\pp{\Y}}(U) \ge \t/20 \cdot \deg_{\pp{\Y}}(C)>0.$ 

We next give the details of \algCH{}.

{
\makeatletter
\@beginparpenalty=10000
\makeatother

\pagebreak[1]
\bigskip\par\noindent
\textbf{Algorithm} \algCH{($G$)}\label{bb:algCH}
\begin{itemize}
\item \textbf{Construct levels 1 and 2.}~~Let $\P$ be the (partial)
hierarchical decomposition consisting only of one level, $\P_1$, that
contains only the root-cluster $V$, i.e., $\P_1=\{V\}$. Let $\X$ be the partition of
$V$ into singletons and set $\phi \gets 1/f_\P(V)$. Call
\magicpart{}$(G, V, \X, \phi)$ to obtain a set $U$ and a partition $\Y$
of $V$. Note that then $G[V]$ is $\deg_{\partial \Y}$-expanding with quality
$\frac{1}{500} \phi / \q = 2e\gamma/f_\P(V)$, as $U$ must be empty (see
\cref{obs:part-no-bdry}). We set $\P_1 = \{V\}$ and $\P_2 = \Y$. Then
$\P=(\P_1,\P_2)$ is a partial hierarchical decomposition with height $L=2$.

\item
\textbf{Construct further levels.}~~
\begin{enumerate}
\item Assume $\P =(\P_1 , \ldots, \P_L)$ is the partial hierarchical
decomposition constructed so far. If every cluster is a singleton cluster, return $\P$ and terminate the algorithm. Otherwise, mark every non-singleton cluster
$C \in \P_L$ as unprocessed and initialize $\X_C$ as the partition of
singletons for each such $C$. 
\label[step]{nextlevel}
  
\item \textbf{While} there is an unprocessed cluster $C \in \P_L$, process $C$
by calling \magicpart{}$(G, C, \X_C,1/f_\P(C))$ to obtain a set $U \subset C$,
$|U| \le |C|/2$, and a partition $\Y$ of $C$.\label[step]{step:two}
\begin{itemize}
\item If $U = \emptyset$, update $\X_S \gets \Y$, mark $C$ as processed, and
end this iteration of the while loop, i.e. go to the beginning of
\cref{step:two}.
  \item \emph{Bad child event.} Otherwise, split $C$ in $\P_L$ into two parts by removing $C$
  from $\P_L$ and inserting $U$ and $C \setminus U$. Note that $U \in \Y$, so by
  removing $U$ from $\Y$ we obtain a partition $\Y'$ of $C \setminus U$.  Set $\X_{C \setminus U} \gets \Y'$ and $\X_U \gets \{U \}$
  \footnote{
  This operation can be seen as a fuse operation (see~\cref{sec:cluster}), namely $\X_U = (\X_U - U)\cup \{U\}$}.

  \item If $\deg_{\partial \Y}(U) \ge \t/20 \cdot \deg_{\partial \Y}(C)$ and
      $\deg_{\partial\Y}(C)\le \deg_{\partial\X}(C)+2\capc(U,C\setminus
      U)$
  (\enquote{either}-case of \cref{prop:partition:expansion}), mark both new clusters $U$ and
  $C \setminus U$ as unprocessed.
    \item Otherwise, mark $U$ as unprocessed and $C \setminus U$ as processed.
  \end{itemize}

  \item Once all clusters are processed, set
  $\P_{L+1} = \bigcup_{C\in\P_L} \X_C$, add $\mathcal{P}_{L+1}$ to
  $\mathcal{P}$, and increment $L$.  
\end{enumerate}
\end{itemize}
}

  \begin{lemma}
  Algorithm \algCH{} constructs a complete hierarchical decomposition where
  every non-leaf cluster is $\gamma$-well-expanding with
  $\gamma=1/1000e\q = \Theta(1/\log n)$.
  \label{lem:wellexpanding}
  \end{lemma}
  \begin{proof}
  To prove the lemma we need to show that every non-leaf cluster $C$ in the
  hierarchy is at least $\deg_{\pp{\P_{i+1}}}$-expanding with quality
  $\gamma /f_\P(C)$, where $i$ is the level of $C$. Fix a non-leaf cluster $C$
  on some level $i$ in the final hierarchy. We first show that $C$ is
  $\deg_{\pp{\P_{i+1}}}$-expanding with quality $2e\gamma /f_\P(C)$,
  where $e$ is the Euler's constant, just after the level $i+1$ has been
  constructed. However, this is not sufficient. During the construction of
  level $i+2$ bad child events may occur, which means that clusters within
  $\P_{i+1}$ are further subdivided. This in turn may worsen the expansion
  property of the level-$i$ cluster $C$, where this subdivisions happened. We
  will show that after the construction of level $i+2$ is finished, $C$ still
  is $\deg_{\pp{\P_{i+1}}}$-expanding with quality $\gamma/f_\P(C)$
  (i.e., the expansion is only a constant factor less). The further
  construction of the hierarchy does not change $\P_{i+1}$ anymore, and, hence,
  $C$ has the desired expansion in the end. As this holds for any $C$ the lemma
  follows.

   \paragraph{\mathversion{bold}Step 1: Analysis of the expansion of level $i$ clusters after the
    construction of level $i+1$.\mathversion{normal}}%
  \mathversion{normal} To construct $\P_{i+1}$ Algorithm \algCH{} calls the
  subroutine \magicpart{} on every non-processed clusters until no such
  clusters are left and then the union of the partition $\X_C$ of each
  processed cluster $C$ forms the new partition $\P_{i+1}$. Thus, to show that
  each cluster is $\deg_{\pp{\P_{i+1}}}$-expanding with quality
  $2e\gamma/f_\P(C)$ just after level $i+1$ has been constructed, it
  suffices that each processed cluster $C$ is $\deg_{\pp{\X_C}}$-expanding with
  quality $2e\gamma/f_\P(C)$ when it is marked as being processed.

  To process an unprocessed cluster $C$, algorithm \algCH{} calls
  \magicpart{}$(G, C, \X_C, 1/f_\P(C))$ which returns a set $U$ and a new
  partition $\Y$. \algCH{} then proceeds as follows. It makes sure that the
  cluster $C \setminus U$ is part of $\P_i$ either because $U=\emptyset$ and
  $C$ already belongs to $\P_i$ or because $U \ne \emptyset$ and $C$ is removed
  from $\P_i$ and $C \setminus U$ (and also $U$) are added to $\P_i$. The
  cluster $U$ is marked as unprocessed. The cluster $C \setminus U$ is marked
  as processed if (i) either $U=\emptyset$ or (ii) the conditions of the
  \enquote{either}-case of \cref{prop:partition:expansion} of \magicpart{} do
  not hold (i.e., $\deg_{\pp{\Y}}(U) < \t/20 \cdot \deg_{\pp{\Y}}(C)$ or
  $\deg_{\pp{\Y}}(C) > \deg_{\pp{\X}}(C) + 2\capc(U, C \setminus U)$.
  
  In both cases the partition
  $\X_{C\setminus U}$ returned by \magicpart{} fulfills the first part of 
  \cref{prop:partition:expansion} of \magicpart{}, i.e., $G[C\setminus U]$ is
  $\deg_{\pp{\X_{C\setminus U}}}$-expanding with quality 
  $1/(500\q f_\P(C))$.
  By setting $\gamma = 1/(1000e\q)$ it follows
  that this quality equals $2e\gamma /f_\P(C)$.

  The construction of level $i+1$ only ends once all clusters are marked as
  processed and, thus, each cluster $C$ in $\P_i$ is
  $\deg_{\pp{\X_C}}$-expanding with quality $2e \gamma/f_\P(C)$. Note
  that once all clusters in $\P_i$ are marked as processed it follows that each
  cluster $C$ in $\P_i$ is $\deg_{\pp{\P_{i+1}}}$-expanding with quality
  $2e \gamma/f_\P(C)$ as
  $\deg_{\pp{\X_C}}|_C = \deg_{\pp{\P_{i+1}}}|_C$.

   \paragraph{\mathversion{bold}Step 2: Analysis of the expansion of level $i$ clusters after the
    construction of level $i+2$.\mathversion{normal}}%
   Fix a level-$i$ cluster $C$ after level $i+1$ of the
  hierarchy has been constructed, and let $\X$ denote the partition of $C$ at
  this time. During the construction of level $i+2$ the partition of $C$ may
  change due to bad child events. Let $\Y$ denote the partition of $C$ after
  the construction of level $i+2$ has finished. Observe that $\Y$ is a
  refinement of $\X$. We need to show that $C$ is
  $\deg_{\pp{\P_{i+1}}}$-expanding with quality $\gamma/f_\P(C)$ when
  the construction of level $i+2$ is complete. This is equivalent to showing
  that $C$ is $\deg_{\pp{\Y}}$-expanding with this quality.

  \def\mis{{\color{red}\text{\bfseries ??}}} Let $\mb{d}$ be an arbitrary demand
  vector in $C$ with $|\mb{d}|\le\deg_{\pp{\Y}}|_C$. We show that we can route $\mb{d}$ with
  congestion at most $f_\P(C)/\gamma $  inside $G[C]$, which implies the desired expansion guarantee
  for $C$.
  The idea is to (a) first route $\v{d}$ from $\deg_{\pp{\Y}}$ to
  $\deg_{\pp{\X}}$ with small congestion (using border-routability) and call the resulting flow the  \emph{to-border flow for $C$} and (b)
  then apply the fact that we know from above how to route any demand $\v{d'}$
  with $|\v{d'}| \le \deg_{\pp{\X}}$ with small congestion.

  More formally, in order to route $\v{d}$ we will (a) route a
  to-border $(\mb{d},\mb{t})$-flow for some suitable demand $\mb{t}$ with
  $|\mb{t}|\le\alpha\deg_{\pp{\X}}|_C$ for some constant $\alpha\le e$ with
  congestion $\con \le 1/(2 \gamma)$. Then (b) we are left with routing demand
  $\mb{t}$ inside $G[C]$, but we know already that we can do that with small
  congestion: We are guaranteed that $G[C]$ is
  $2e \gamma/f_\P(C)\cdot\deg_{\pp{\X}}$-expanding, and, thus, by
  \cref{lem:cutflow}, we can route any demand $\mb{d'}$ with
  $|\mb{d'}|\le\alpha\deg_{\pp{\X}}|_C$ with congestion
  $ \alpha f_\P(C)/(2e\gamma )\le f_\P(C)/(2\gamma )$ inside $G[C]$ for
  $\alpha \le e$. The sum of the two flows then has congestion at most
  $f_\P(C)/\gamma$ and routes $\v{d}$.

  \textbf{To-border flow for $C$.}
  It remains to show that for any demand vector $\mb{d}$ with
  $|\mb{d}|\le\deg_{\pp{\Y}}|_C$ there exists a demand vector $\mb{t}$ with
  $|\mb{t}|\le\alpha\deg_{\pp{\X}}|_C$ for some suitable constant $\alpha\le e$
  such that we can route a to-border $(\mb{d},\mb{t})$-flow with small congestion $\con$.
  
  Specifically, we need to upper bound $\con$ by $4\alpha \log n$. This gives
  the desired bound as $\con\le 4\alpha\log n\le f_\P(C)/(2\gamma)$, as
  $4\alpha \log n \le 4e \log n \le \log\log n/(2\gamma)$ which holds as
  $\gamma = 1/(1000 e \q) \le \log \log n/(8e \log n)$ as
  $\q \ge (\log n)/125$.
  
  %For an edge with both endpoints in $C$, but each one belonging to a different child of $C$ $\con$ is upper bounded by $\alpha$. %Thus summed over all clusters $C$ to which an edge of $G$ belongs in the final hierarchy it follows that the congestion of all the to-border flows of the edge is at most $4 \alpha \log n$.}

  We need to find a demand $\mb{t}$ and route the $(\mb{d},\mb{t})$-flow $f$
  inside $C$ with congestion at most $4\alpha \log n$. As $\mb{d}$ is a demand
  and the flow $f$ routes the demand $\v{d}-\v{t}$, it follows that $\v{t}$ is
  a demand. Thus, it is sufficient to show that there exists a \emph{vertex
    weight vector} $\v{t}$ such there exists a $(\mb{d},\mb{t})$-flow $f$
  inside $C$ with congestion at most $4\alpha \log n$.
  
  To do we will show for every cluster $X\subseteq C$ with $X \in\X$ that there
  exists a vertex weight vector $\v{t}|_X$ with
  $|\v{t}|_X| \le \alpha \deg_{\pp{\X}}|_X$ so that we can route the flow
  $(\v{d}|_X,\v{t}|_X)$ in $G[X]$ with congestion at most $4\alpha \log n$.
  Note that $\v{d}$ equals $\sum_{x \in \X} \v{d}|_X$. Let
  $\v{t} := \sum_{x \in \X} \v{t}|_X$. Note that $\v{t}$ is a vertex weight
  vector with $|\mb{t}|\le\alpha\deg_{\pp{\X}}|_C$ and the combination of the
  resulting the $(\mb{d}|_X,\mb{t}|_X)$-flows over all clusters $X \in \X$
  routes exactly the $(\mb{d},\mb{t})$-flow inside $C$.
  
  Fix a cluster $X\in\X$. During the construction of level $i+2$, $X$ might be
  repeatedly partitioned due to bad child events. We represent this partitioning proces by  a binary tree
  $T$: the root vertex is $X$, the
  internal vertices are intermediate clusters created during the partitioning
  process, and the leaf vertices are the child-clusters of $X$ in $\Y$, and, thus, in the final hierarchical decomposition. Thus
  nodes in $T$ are always clusters that are created during the partitioning
  process, ie during the construction of level $i+2$ in \algCH{}. Only the root and the 
  leaves of $T$ exist when the construction of level $i+2$ ends.
  At every internal node of $T$ a flow is constructed between
  its two children and internal to one child. The to-border flow for $C$ will be the \emph{sum} of all these flows.
  
  Every non-leaf
  cluster $S$ has two children: one bad child $U$ and $S\setminus U$ (we assume
  wlog.\ that $U$ is the left child in the binary tree). We say an internal
  node has at left-depth $\ell$ in $T$, if the number of left edges on its path
  to the root is $\ell$. The root $X$ has thus left-depth 0. Let $h$ denote the
  left height of $T$, i.e., the maximum left depth of any node in $T$.
  \begin{claim}
  A node $S$ of $T$ at left-depth $\ell$ has $|S|\le |X|/2^{\ell}$ and $T$ has left height at most $\log n$.
  \end{claim}
  \begin{proof}This follows because a bad child $U$ of a parent $P$ has
  $|U|\le|P|/2$, thus the
  left-depth of any cluster can be at most $\log |X| \le \log n$.
  \end{proof}

  Now by restricting $\v{d}|_X$ to a node $S$ of $T$ we receive $\v{d}|_S := (\v{d}|_X)|_S$.
  \begin{claim}\label{cla:routable}
  For a node $U$ of $T$ that is a bad child at left depth $\ell$ 
  for any $\mb{s}$ with $|\mb{s}|\le\deg_{E(U,S\setminus U)}|_U$ 
  there exists a $\mb{t}$ with
  $|\mb{t}|\le \deg_{\partial S}|_U/(3\ell\log\log n)$ such that we can route the 
 $(\mb{s}$,$\mb{t})$-flow
  \textbf{in} $G[U]$ with congestion $2$.
  \end{claim}
  \begin{proof}
  During the construction of the hierarchy we obtain the bad child $U$ of a
  cluster $S$ by calling the subroutine \magicpart{}$(G,S, \X_\S, 1/f_\P(S))$ on
  some cluster $S$ with $\phi=1/f_\P(S)$ and we are guaranteed that the edge
  set $E(U,S\setminus U)$ is $\frac{1}{\phi}$-border routable, i.e.,
  $f_\P(S)$-border routable in $S$, through $U$ with congestion 2. Since
  $X = \parent_{\P_{i+1}}{(S)}$ in the hierarchical decomposition when
  \magicpart{} is called, it follows that
  $\log(|\parent_{\P_{i+1}}{(S)}|/|S|) \ge \ell$. Thus, it holds that
  $f_\P(S)\ge 3\log\log n\log(2|X|/|S|)\ge 3\ell\log\log n$. Thus,
  $f_\P(S)$-border routability implies $3\ell \log \log n$-border routability.
    Now the definition of border routability guarantees that for any $\mb{s}$ with $|\mb{s}|\le\deg_{E(U,S\setminus U)}|_U$ 
  there exists a $\mb{t}$ with
  $|\mb{t}|\le \deg_{E(U,V\setminus S)}|_U/(3\ell\log\log n) = \deg_{\partial S}|_U/(3\ell\log\log n)$ such that we can route the 
 $(\mb{s}$,$\mb{t})$-flow in $G[U]$ with congestion $2$.
  \end{proof}

  \noindent
  Now we are ready to give a construction for the vertex weight vector  $\mb{t}|_X$ and explain how to route the $(\mb{d}|_X,\mb{t}|_X)$-flow inside cluster $X$ with congestion at most $4\alpha \log n$.

  Define for a left depth $\ell\in\{0,\dots,h\}$
  $$\alpha_\ell=\prod_{j=h-1}^\ell\Big(1+\frac{1}{(j+1)\log\log
    n}\Big)=\Big(1+\frac{1}{(\ell+1)\log\log n}\Big)\alpha_{\ell+1}\enspace,$$
  and $\alpha_{h} = 1$. Note that $\alpha_{\ell} > \alpha_{\ell + 1} \ge 1$ for $0 \le \ell < h$ and
  \begin{equation*}
  \begin{split}
  \alpha :=\alpha_0 & =\prod_{j=h-1}^0\Big(1+\tfrac{1}{(j+1)\log\log n}\Big)
        \le \exp\Big(\tfrac{1}{\log\log n}{\textstyle\sum_{j=1}^h}\tfrac{1}{j}\Big)\le
        e\enspace,
  \end{split}
  \end{equation*}
  where we used $h\le\log_2n$.

 We prove via induction over $T$ (from leaves to the root $X$) the following \textbf{inductive claim:} {\em For any node $S$ of $T$ at left depth $\ell$
 there exists a vertex weight vector $\mb{t}|_S$ with $|t_S| \le \alpha_\ell\deg_{\partial S}|_S$ and 
  we can route the flow
  $(\mb{d}|_S, \mb{t}|_S)$ inside $S$ such that the congestion is 0, if $S$ is a leaf cluster, and the congestion is $\alpha_{\ell}$ on the edges of $E(U,S\setminus U)$, and $4\alpha_\ell(h-\ell)$ for edges in $G[U]$ and $G[\sa]$,  where $U$ and $\sa$ are the children of $S$ in $T$ with $U$ being the bad child.
  }
  
  Note that this implies that for the root $X$ (which has left depth 0)
  there exists a $(\mb{d}|_X,\mb{t}|_X)$-flow inside cluster $X$ for some vertex weight vector $\mb{t}|_X$ with $|\mb{t}|_X| \le \alpha \deg_{\pp{\X}}|_X$ with congestion at most $\alpha h \le \alpha \log n$.

  For a leaf cluster $S$ (which means $S\in\Y$) we have
  $|\v{d}|_S|\le\deg_{\pp{\Y}}|_S=\deg_{\partial S}|_S\le\alpha_h\deg_{\partial S}|_S$, so we set $ \mb{t}|_S=  \mb{d}|_S$, i.e.,  we fulfill the condition without any routing, i.e., congestion 0.

  Now suppose that we have
  a non-leaf cluster $S$ at left depth $\ell<h$ that has a bad child $U$. Further,
  assume that we already found vertex weight vectors $\v{t'}|_U$ and $\v{t'}|_\sa$ and a $(\mb{d}|_U, \mb{t'}|_U)$-flow with congestion $\alpha_{\ell+1} (h-\ell -1) $ and a 
  $(\mb{d}|_{S\setminus U}, \mb{t'}|_{S\setminus U})$-flow with congestion $\alpha_{\ell} (h-\ell) $
  so that  $|\mb{t'}_U| \le \alpha_{\ell+1}\deg_{\partial U}$ and 
  $|\mb{t'}_\sa| \le \alpha_{\ell}\deg_{\partial (S\setminus U)}$
  (observe that the left depth of $U$ is $\ell+1$).

  We first route a $(\mb{t'}|_\sa,\mb{t^*})$-flow at vertices from $S\setminus U$ that are incident to $U$.
  For this, a vertex $v\in S\setminus U$ with $\mb{t'}|_\sa(v) \ge 0$ sends
  $\alpha_\ell\deg_{E(U,S\setminus U)}(v)$ flow along its incident
  $E(U,S\setminus U)$-edges to $U$ ($v$ may send less if it is running out of ``supply'') and if $\mb{t'}|_\sa (v)< 0$ $v$ receives  $\alpha_\ell\deg_{E(U,S\setminus U)}(v)$ flow ($v$ may receive less flow if its ``deficit'' becomes zero) across each edge of 
  $E(U,S\setminus U)$ that is incident to $v$.
  More formally, for $v \in \sa$, we set $\mb{t^*}(v) = \max(0, \mb{t'}|_\sa(v) - \alpha_\ell\deg_{E(U,S\setminus U)}(v))$  if $\mb{t'}|_\sa(v) \ge 0$ and $\mb{t^*}(v) = \min(0,\mb{t'}|_\sa(v) + \alpha_\ell\deg_{E(U,S\setminus U)}(v))$ otherwise.
  Define $\v{t^*}(v)$ for $v \in U$ correspondingly.
  Note that this can be done with congestion $\alpha_{\ell}$ and that for $v \in \sa$,
   $|\v{t^*}(v)| \le \alpha_\ell\deg_{\partial (S\setminus U)}(v)-\alpha_\ell\deg_{E(U,S\setminus
    U)}(v)=\alpha_\ell\deg_{\partial S}(v)$.

\noindent    
   After this step  $\v{t^*}(v)$ at a vertex  $v \in U$ fulfills
\begin{equation*}
\begin{split}
  |\v{t^*}(v)| 
   &\le |\v{t'}|_U|(v) + \alpha_{\ell}\deg_{E(U,S\setminus U)}(v) 
   \le  \alpha_{\ell+1}\deg_{\partial U}(v)+\alpha_{\ell}\deg_{E(U,S\setminus U)}(v)\\
   &=\alpha_{\ell+1}\deg_{\partial S}(v)+(\alpha_{\ell}+\alpha_{\ell+1})\deg_{E(U,S\setminus U)}(v)\enspace.
\end{split}
\end{equation*}

  In a second step we want to reduce  $|\v{t^*}(v)|$ for each $v \in U$ to at most $\alpha_{\ell+1}\deg_{\partial S}(v)$. Recall that $U$ is a bad child. Thus, we use \cref{cla:routable} that shows for every 
   $\mb{\tilde s}$ with $|\mb{\tilde s}| \le (\alpha_{\ell}+\alpha_{\ell+1})\deg_{E(U,S\setminus U)}$ 
  the existence of a vertex weight $\mb{\tilde t}$ with $|\mb{\tilde t}| \le (\alpha_{\ell}+\alpha_{\ell+1}) \deg_{\partial S}|_U/(3(\ell+1)\log\log n) 
      = (\alpha_{\ell}+\alpha_{\ell+1}) \deg_{\partial S}/(3(\ell+1)\log\log n)$ and a $(\mb{\tilde s},\mb{\tilde t})$-flow in $G[U]$
 with congestion $2(\alpha_{\ell}+\alpha_{\ell+1}) \le 4 \alpha_{\ell}$. 
 Thus, we apply the claim with $\v{\tilde s}(v) := \v{t^*}(v) - \alpha_{\ell+1}\deg_{\pp{S}}(v)$ if $\v{t^*}(v) >\alpha_{\ell+1}\deg_{\pp{S}}(v)$, $\v{\tilde s}(v) := \v{t^*}(v) + \alpha_{\ell+1}\deg_{\pp{S}}(v)$ if $\v{t^*}(v)< -\alpha_{\ell+1}\deg_{\pp{S}}(v)$, and $\v{\tilde s}=0$ otherwise.
 Note that $|\mb{\tilde s}| \le \max(0, |\v{t^*}| - \alpha_{\ell+1}\deg_{\pp{S}}) \le (\alpha_{\ell}+\alpha_{\ell+1})\deg_{E(U,S\setminus U)}$, i.e., it fulfills the requirements of \cref{cla:routable}.
  Finally, we set $\mb{t}|_S(v) = \mb{t^*}(v)$ for $v \in \sa$ and $\mb{t}|_S(v) = \mb{t^*}(v) -\mb{\tilde s}(v) + \mb{\tilde t}(v)$ for $v \in U$.
 
  Thus, $|\v{t}|_S|(v)$ at any vertex $v \in U$ is at most 

  \begin{equation*}
  \begin{split}      
  |\v{t}|_S|(v)
    &\le | \mb{t^*}(v) -\mb{\tilde s}(v)| + |\mb{\tilde t}(v)|\\
    &\le  \Big(\alpha_{\ell+1}+\frac{\alpha_{\ell+1}+\alpha_{\ell}}{{3(\ell+1)\log\log
      n}}\Big) \deg_{\partial S}(v)\\
    &\le \Big(\alpha_{\ell+1}+\frac{\alpha_{\ell+1}}{(\ell+1)\log\log
      n}\Big) \deg_{\partial S}(v)
  = \alpha_\ell \deg_{\partial S}(v)\\
   \end{split} 
  \end{equation*}
  where we used $\alpha_\ell\le2\alpha_{\ell+1}$, and for a vertex $v \in \sa$
  it is at most $\alpha_\ell\deg_{\partial S}(v)$.
 
  Finally we analyze the congestion. The edges in $E(U,\sa)$ have no congestion
  in the recursive flows $(\mb{d}|_U, \mb{t'}|_U)$-flow and a
  $(\mb{d}|_{S\setminus U}, \mb{t'}|_{S\setminus U})$-flow and receive
  congestion $\alpha_\ell$ in step one. By the inductive assumption the edges
  in $G[U]$ have congestion of $4 \alpha_{\ell+1}(h-(\ell+1))$ and receive a
  congestion of $4\alpha_\ell$ in step two. Thus their total congestion is
  $4 \alpha_{\ell}(h-\ell).$ By the inductive assumption the edges in $G[\sa]$
  have congestion of $4 \alpha_{\ell}(h-\ell)$ and receive no additional
  congestion in either step one or step two. Thus their congestion is
  still $4 \alpha_{\ell}(h-\ell).$ It follows that we can route the flow $(\v{d}|_S, \v{t}|_S)$ inside $S$ with congestion $4 \alpha_{\ell}(h-\ell)$.
\end{proof}
  %%%%%%%%%%%%%%%%%%%%%%%%%%%%%%%%%%%%%%%%%%%%%%%%%

  \begin{lemma}\label{lem:levels}
  The complete hierarchical decomposition constructed by Algorithm \algCH{} has height $L=O(\log n)$.
  \end{lemma}
  \begin{proof}
  We show that any cluster $C$ in the final hierarchy has size at most $|P|/2$,
  where $P$ is the grandparent of $C$. 
  
  At the start of the while loop a cluster $S$ on level $L$ is chosen. Let $P$
  denote the parent of $S$ in the hierarchy at this time and observe that $P$
  is also contained in the final hierarchy while $S$ might not be due to bad child
  events that subdivide $S$. 

  The call to \magicpart{}$(G, S, \X_S,1/f_\P(S))$ at the beginning of the while
  loop guarantees that the bad child $U$ that is returned has size at most
  $|S|/2$ and all clusters in the returned partition $\Y$ have size at most
  $|S|/2$. Any generated sub-cluster in this or the following iterations on
  clusters that are subsets of $C$ are a set or a subset of the partition $\Y$
  and will therefore have size at most $|S|/2$. Such a sub-cluster $C$ is then
  inserted as a child of $S$ or as a child of a part of $S$ that is obtained
  via a bad child event. In either case the size of $C$ is at most
  $|S|/2\le|P|/2$, where $P$ is the parent of $S$, and, hence, the grandparent
  of $C$.
  \end{proof}

  \begin{lemma}\label{lem:runningtimeCH}
  The time for executing the while loop is
  $O(\log(m)/\t\cdot\Tpart(m))$, where
  $\Tpart(m)=\Omega(m)$ is the running time for the
  subroutine \magicpart{}.
  \end{lemma}
  \begin{proof}
  Fix a leaf cluster $C$ at the start of the while loop. We analyze how much
  work the algorithm performs on sub-clusters of $C$ until all
  sub-clusters of $C$ are marked as processed.

  The following claim shows that a bad child event $U$ on some subcluster
  $S\subseteq C$ makes progress in the sense
  that an \emph{unprocessed} new cluster resulting from the bad child event
  is substantially \enquote{smaller} than $S$.
  \begin{claim}
  Suppose a call to \magicpart$(G,S,\X,\phi)$ on some subcluster $S$  
  returns with a bad child $U$. Then the 
  resulting sub-clusters
  $U$ and $S\setminus U$ fulfill
  \begin{enumerate}
  \item $|U| \le |S|/2$\label[property]{propU}
  \item
  $\deg_{\partial\Y}(S\setminus U) \le (1-\t/80)\deg_{\partial\X}(S)$
  if $S\setminus U$ is declared as unprocessed by Algorithm \algCH{}, where
  $\Y$ denotes the new partition of $S$.\label[property]{propSmU}
  \end{enumerate}
  \end{claim}
  \begin{proof}
  \cref{propU} is directly guaranteed by the output of \magicpart{}. For
  \cref{propSmU} observe that $S\setminus U$ is only declared
  \emph{unprocessed} if we are in the \enquote{either}-case of
  \cref{prop:partition:expansion} in \magicpart{}, i.e., if
  $\deg_{\partial \Y}(U) \ge \t/20 \cdot \deg_{\partial \Y}(S)$ and
  $\deg_{\partial\Y}(S)\le \deg_{\partial\X}(S)+2\capc(U,S\setminus U)$. This gives
  \begin{equation*}
  \begin{split}
  \deg_{\partial\Y}(S\setminus U)
  &=   \deg_{\partial\Y}(S)-\deg_\Y(U)\\
  &\le \deg_{\partial\X}(S)+2\capc(U,S\setminus U)-\deg_{\partial\Y}(U)\\
  &\le \deg_{\partial\X}(S)+2\phi\deg_{\partial\Y}(U)-\deg_{\partial\Y}(U)\\
  &\le \deg_{\partial\X}(S)+(2\phi-1)\t/20 \cdot \deg_{\partial\Y}(S)\\
  &\le \deg_{\partial\X}(S)-\t/40 \cdot \deg_{\partial\Y}(S)\\
  \end{split}
  \end{equation*}
  The second inequality holds because \cref{prop:partition:border-routable} of
  \magicpart{} implies that $\capc(U,S\setminus U)\le\phi\capc(U,V\setminus
    S)\le\phi\deg_{\partial\Y}(U)$ and the final inequality uses $\phi\le 1/4$.
    If now  $\deg_{\partial\Y}(S)\le \deg_{\partial\X}(S)/2$ we clearly have
    $\deg_{\partial\Y}(S\setminus
    U)\le\deg_{\partial\Y}(S)\le\deg_{\partial\X}(S)/2$. Otherwise, we obtain
    $\deg_{\partial\Y}(S\setminus U)\le (1-\t/80)\deg_{\partial\X}(S)$ by plugging
    $\deg_{\partial\Y}(S)\ge \deg_{\partial\X}(S)/2$ into the above inequality.
    The claim follows.
  \end{proof}

  From the above claim it directly follows that a vertex from $C$ can be
  contained at most
  $\log_2|C|+\log_{1/(1-\t/80)}\deg(C)=O(\log(\deg(C))/\t)$ times in a
  sub-cluster $S\subseteq C$ during the while loop. Consequently, the total
  work performed for calls to \magicpart{} during the while loop is at most
  $O(\log(m)/\t\cdot\sum_C \Tpart(m_C))\le O(\log(m)/\t\cdot\Tpart(m))$,
  where $\Tpart(m)$ is the time required for Algorithm \magicpart{} on a
  cluster with volume $m$ (the inequality uses the fact that $\Tpart(m)$ grows
  at least linearly with the volume). The remaining cost of inserting a new
  cluster into the partition, or comparing the size of clusters is just linear
  in $m$, and, hence dominated by the cost for calling $\Tpart(m)$.
\end{proof}

\begin{theorem}
  Suppose we are given an undirected graph $G=(V,E)$ with volume $m=\deg(V)$, and an algorithm
  for \magicpart{} with running time $\Tpart$.
  Then we can construct a hierarchical
  congestion approximator with quality $O(\q\log n\log\log n)$ in time
  $O(\log^2(m)/\t\cdot \Tpart(m)$. Here $\q = O( \log n)$
  and $\t\le1/160$ are the quality- and  balance-guarantee, respectively, of the
  \algSC{}-routine used by \magicpart{}.
\end{theorem}

\begin{proof}
\cref{lem:wellexpanding} shows that the hierarchy constructed by algorithm
\algCH{} is $\gamma$-well-expanding with $\gamma=\Theta(1/\log n)$.
\cref{lem:levels} shows that our hierarchical decomposition has height
$L = O(\log n)$. Hence, the quality of the congestion approximator follows by
\cref{thm:main}. The running time for constructing one level of the hierarchy
is $O(\log(m)/\t\cdot\Tpart(m))$ due to \cref{lem:runningtimeCH}. Since the
number of levels is logarithmic due to \cref{lem:levels} the theorem follows.
\end{proof}

\noindent
Plugging in the result for our building blocks from \cref{sec:buildingblocks} we get
the following corollary.
\begin{corollary}
With the implementations of the building blocks in 
\cref{sec:buildingblocks} we obtain a running time of
$O(\log^2(m)/\t\cdot\Tpart(m))=
O(\log^3(m)\cdot\Tpart(m))=
O(\log^{10} \cdot \Tfc(m, 2))=\tO(m)$.
\end{corollary}

\section{Partitioning a Cluster}\label{sec:cluster}

\makeatletter
\newcommand{\iflabelexists}[1]{%
  \expandafter\ifx\csname havelabel:#1\endcsname\relax%    
    \expandafter\@secondoftwo%
  \else%
    \expandafter\@firstoftwo%
  \fi%
}

\newcommand{\nlabel}[2][]{%
  \iflabelexists{#2}{%
  }{%
    \global\expandafter\def\csname havelabel:#2\endcsname{}%
    \label[#1]{#2}%
  }%
}
\makeatother

\def\RoutineTWT{
\begin{subroutine*}[\algTWT$(G, C, R, \v{\pi}, \phi)$]\nlabel{bb:algTWT}
We are given a graph $G=(V,E)$, two subsets $R,C$ with $R\subset C\subseteq V$,
a non-negative vertex weighting $\v{\pi}$, and a parameter $\phi > 0$ such that
\begin{enumerate}[label=(\alph*)]
  \item $\capc(R, C \setminus R) \leq \phi \v{\pi}(R)$, and
  \nlabel[requirement]{req:twt:Rsparse}   \item $G[C]$ is $\delta\cdot\phi\v{\pi}|_{C \setminus R}$-expanding for some
  $\delta \in (0, 1]$.\nlabel[requirement]{req:twt:exp}
\end{enumerate}
The procedure \algTWT{} outputs a
three-partition of $C$ into $(A,B,U)$ such that
\begin{enumerate}[label=(\alph*)]
  \item $A \subseteq C \setminus R$ and $\capc(A,C\setminus A)\le 2\capc(R,C\setminus
  R)$\nlabel[property]{prop:twt:Acap}
  \item $\v{\pi}(B\cup U)\le (11/\delta)\cdot\v{\pi}(R)$\nlabel[property]{prop:twt:Avol} 
  \item $E(U,C\setminus U)$ is $1/\phi$-border routable through $U$ with
  congestion $2$\nlabel[property]{prop:twt:border} 
  \item $G[A\cup B]$ is
  $\frac{1}{25}\cdot \delta \phi \, \big(\deg_{E(B,V\setminus B)}|_{A \cup B} +
  \v{\pi}|_A \big)$-expanding\nlabel[property]{prop:twt:expanding} 
\end{enumerate}
\end{subroutine*}
}

In this section we present the algorithm \magicpart{} that we use to compute a
new partition $\Y$ of a cluster $C$ when given a partition $\X$ and an
expansion parameter $\phi$. Recall its definition at the beginning
of~\Cref{subsec:ConstructingHierarchy}. It first uses \algSC{} with a vertex
weight that depends on $\X$ to find a $(\phi/20) \cdot \deg_{\partial \X}$-sparse cut in
$C$. It then uses \algTWT{} to ``slightly shift'' the cut. The resulting cut is
border-routable through one side, called $U$, of the cut. Ideally the other side
of the cut would be expanding with a suitable parameter and the algorithm can
terminate. However, we can only show that the ideal case happens under certain
conditions. If they do not hold, then we either (i) modify the original
partition $\X$ so that $\deg_{\pp{\X}}(C)$ decreases by a multiplicative factor
and repeat the algorithm from the beginning or (ii) we ``trim'' one of the
sides of the original cut, finding a border-routable set $U$ and modifying the
partition $\X$ such that the second condition in
\cref{prop:partition:expansion} of \magicpart{} holds.

Achieving this crucially relies on the following subroutine, whose algorithm we
will present and analyze in \cref{subsec:twt}. \RoutineTWT

Let us define the following notation. Suppose we are given a cluster $S$, a
partition $\X$ of $S$, and a subset $T\subseteq S$. The operation $\X - T$
returns the partition
$\{A \setminus T\mid A \in \X , A \setminus T \ne \emptyset \}$ of
$S \setminus T$. When modifying a partition $\X$ of $C$ we use the operation
$\X\leftarrow(\X - T)\cup\{T\}$, for which we say that we \emph{fuse the set
  $T$ in $\X$}, i.e., we first remove $T$ from every set in the partition and
then we add $T$ itself as set to the partition.
\begin{claim}\label{claim:fuse}
    For any $B \subseteq C$ if $\Y = (\X-B)\cup \{B\}$, it holds that
    $\deg_{\pp{\Y}} (C) \le \deg_{\pp{\X}}(C) - \deg_{\pp{\X}}(B) + 2\capc(B, C\setminus B) + \capc(B, V\setminus C).$
\end{claim}
\begin{proof}
    The fuse operation joins all the vertices in $B$ into one set. As a result, edges with both endpoints inside $B$ do not belong to $\Y$, leading to a reduction of $\deg_{\pp{\X}}(B)$. However, all the edges on the boundary of $B$ are added to $\Y$. This leads to an increase of 
    $2\capc(B, C\setminus B) + \capc(B, V\setminus C),$ as
    the edges in $E(B,C \setminus B)$ have both endpoints inside $C$ and, thus, contribute twice to   $\deg_{\pp{\Y}} (C)$ (if there weren't already there),
    while the edges in $E(B, V \setminus C)$ have only one endpoint inside $C$ and, thus, contribute only once to  $\deg_{\pp{\Y}} (C)$.
\end{proof}
Thus, if $\deg_{\pp{X}}(B)$ is large in comparison to $2\capc(B, C\setminus B) + \capc(B, V\setminus C)$ (i.e., a large part of $\deg_{\pp{X}}(B)$ is ``inside'' $B$), the fuse operation ``decreases volume'', i.e., $\deg_{\pp{\Y}} (C) < \deg_{\pp{\X}}(C)$. Otherwise, the ``volume increase'' is limited, i.e., $\deg_{\pp{\Y}} (C) \le \deg_{\pp{\X}}(C) + 2\capc(B,C\setminus B)$.

Next we present algorithm \magicpart{} whose correctness we prove afterwards.

\bigskip
\noindent
Algorithm \magicpart$(G,C,\X,\phi)$\label{bb:algpart}

\begin{enumerate}
   \item Set $\v{\pi} \gets \deg_{\partial \X}$ and
         $R \leftarrow $ \algSC{}$(G[C], \v{\pi}, \phi/20)$ 
         \label[step]{step:sparsecut}

         \begin{enumerate}
         \item  If $R$ has small $\v{\pi}$ volume $\big[\v{\pi}(R)\le\t\cdot\v{\pi}(C)\big]$ then\hfill\\
             $(A,B,U) \leftarrow$ \algTWT{}$(G,C,R,\v{\pi},\phi)$.
              If $|A|\ge |C|/2$, perform two fuse operations, namely set
               $\Y\leftarrow \big(\X-B \big) \cup \{B\}$ and 
                $\Y\leftarrow \big(\Y- U \big) \cup \{U\}$
             and
             return $\Y$ and $U$. Otherwise, set $T\leftarrow A$ and go to
             \cref{step:processT}.\label[case]{case:unbalanced}\label[case]{case:assignTa}\label[case]{case:twtcondition}
              
         \item If $R$ does not have small $\v{\pi}$ volume $\big[\v{\pi}(R)\ge\t\cdot\v{\pi}(C)\big]$, then \hfill\\
             set $T$ to be the set out of $\{R, C\setminus R\}$ that has the
             smaller cardinality and go to \cref{step:processT}.\label[case]{case:assignTb}
         \end{enumerate}

   \item
   \label[step]{step:processT}
   
         \begin{enumerate}
         \item If $\deg_{\pp{C}}(T)\le \bm{\pi}(T)/2$, then fuse $T$
         in $\X$, i.e., $\X\leftarrow (\X-T)\cup \{T\}$ and go to
         \cref{step:sparsecut}.\label[case]{case:repeat}

         \item Otherwise,  ``trim'' $T$  and return a balanced bad child $T'$ as follows:
             Formally, let $\v{s} = \deg_{E(T, C\setminus T)}|_T$ and
             $\v{t} = \smash{\frac{1}{2}\phi \deg_{\partial C}|_T}$. We compute
             a $2$-fair cut $X \subseteq T$ in $G[T]$ using \algFC{}$(G[T], \v{s},\v{t},2)$ and let $T'=T\setminus X$. We then
             return $T'$ together with the partition that is generated by fusing $T'$ in $\X$, namely $\Y\leftarrow(\X - T')
             \cup \{T'\}$.\label[case]{case:balanced}
         \end{enumerate}
\end{enumerate}
If the algorithm returns a non-empty cluster together with $\Y$, we call this cluster a \emph{bad child}.
We call each execution of \cref{step:sparsecut} and \cref{step:processT} an \emph{iteration}.
Note that the procedure terminates in~\cref{case:assignTa} and~\cref{case:balanced}, and
only starts a new iteration if it reaches
\cref{case:repeat}. As we will show below, in the latter case, the remaining $\v{\pi}$ volume decreases
sufficiently so that this cannot happen too often.

The correctness proof now proceeds as follows: We first show
in~\Cref{claim:condition} that the conditions for \algTWT{} are fulfilled whp
whenever it is called in \cref{case:twtcondition}. Then we show that $T$
fulfills a certain set of conditions at the beginning of \cref{step:processT}.
This is
then sufficient to show correctness, i.e., that
\cref{prop:partition:border-routable} and \cref{prop:partition:expansion} of
\magicpart{} are fulfilled at termination.

\begin{claim}\label{claim:condition}
     In \cref{case:twtcondition} of the algorithm the conditions for executing
     \algTWT{} are fulfilled for $\delta = 1/(20\q)$ with high probability.
\end{claim}
\begin{proof}
\cref{req:twt:Rsparse} of \algTWT{} is fulfilled
because $R$ is a $\phi/20$-sparse cut returned from the call to \algSC{}.
By \cref{prop:SC:exp} of \algSC{} we further have that $G[C]$ is
$\phi/(20\q) \, \bm{\pi}|_{V \setminus R}$-expanding with high probability, since $\bm{\pi}(R) \leq \bal \bm{\pi}(C)$ when we reach \cref{case:twtcondition}. This gives \cref{req:twt:exp} of \algTWT{} with $\delta = 1/(20\q)$.
\end{proof}

\begin{invariant}\label{claim:T-props}
At the beginning of \cref{step:processT} the set $T$
fulfills\label{cla:Tproperties}
\begin{enumerate}
  \item $T\le |C|/2$,\label[property]{pro:cardinality}
  \item $\v{\pi}(T)\ge\t\v{\pi}(C)$, 
  and\label[property]{pro:vol} 
  \item $\capc(T,C\setminus T)\le \phi/10\cdot\v{\pi}(T)$.\label[property]{pro:sparse}
\end{enumerate}
\end{invariant}
\begin{proof}

\textbf{Case 1:} First suppose the set $T$ is due to the assignment in \cref{case:assignTb}. As $R$ was returned by \algSC{}, it holds that $\bm{\pi}(R) \le \bm{\pi}(C \setminus R)$. Thus, we
have $\v{\pi}(C\setminus R)\ge\v{\pi}(R)\ge\t\cdot\v{\pi}(C)$, which
shows \cref{pro:vol} for both
$R$ and $C\setminus R$, and, thus, for $T$. Since the cut $R$ is
$\phi/20$-sparse we have $\v{\pi}(C\setminus R)\ge\v{\pi}(R)\ge20/\phi \cdot \capc(R,C\setminus R)$, which gives \cref{pro:sparse} for both $R$ and $C\setminus R$. As $T$
is chosen as the set of smaller cardinality out of $R$ and $C\setminus R$ we get
\cref{pro:cardinality} for $T$.

\textbf{Case 2:} Now suppose that $T$ is due to the assignment in \cref{case:assignTa}, i.e., it
is the set $A$ that results from \algTWT. \cref{pro:cardinality} directly
follows from the choice of $T$ in that case. We have
\begin{equation*} \textstyle
  \v{\pi}(A)=\v{\pi}(C)-\v{\pi}(B\cup U)\ge
\v{\pi}(C)-\frac{11}{\delta}\v{\pi}(R)\ge
\v{\pi}(C)-\v{\pi}(C)/2\ge\v{\pi}(C)/2 \enspace ,
\end{equation*}
where we used \cref{prop:twt:Avol} of \algTWT{} for the first inequality and $\delta = 1/(20\q)$ by~\cref{claim:condition}. In the second inequality, since $\v{\pi}(R) \leq \t \v{\pi}(C) \leq 1/(440\q) \v{\pi}(C)$, we have that 
$(11/\delta)\v{\pi}(R)\le 220\q\t\v{\pi}(C)\le 
\frac{220\q}{440\q}\v{\pi}(C)
% 220\q/(440\q)\v{\pi}(C)
= \v{\pi}(C)/2
$.

\cref{prop:twt:Acap} of \algTWT{} and the sparsity of $R$ guaranteed by \algSC{} called with $\phi/20$ give us $\capc(A,C\setminus A)\le2\capc(R,C\setminus R)\le\phi/10 \cdot \v{\pi}(R)\le\phi/10 \cdot \v{\pi}(A)$, where the last step uses $\v{\pi}(R)\le\v{\pi}(C)/2\le\v{\pi}(A)$. This establishes \cref{pro:sparse} and concludes the proof.
\end{proof}

\begin{invariant}\label{claim:all-x-small}
  In any iteration of \magicpart{}, all $X \in \X$ satisfy $|X| \leq
  \max\{z,|C|/2\}$, where $z$ is the maximum size of a set in the initial
  partition. 
\end{invariant}
\begin{proof}
For the initial iteration by the definition of $z$.
The algorithm only continues to another iteration in 
\cref{case:repeat} if $|A| < |C|/2$. Then we obtain the new partition $\X'$ 
by setting $\X' = (\X - T) \cup \{T\}$ for some set $T$ that
satisfies the properties of~\cref{claim:T-props}. In particular, we have
$|T| \leq |C|/2$. The inductive step thus follows since no set $X \in \X$ can
increase in cardinality by the fuse operation and $|T| \leq |C|/2$.
\end{proof}

\begin{lemma}\label{lemma:imbalanced-child}
  If \magicpart{} returns from~\cref{case:assignTa}, then it returns a correct (imbalanced) bad child, i.e., \cref{prop:partition:border-routable} and \cref{prop:partition:expansion} of \magicpart{} are fulfilled.
\end{lemma}
\begin{proof}
\cref{claim:condition} shows that the conditions for \algTWT{} are fulfilled
with $\delta = 1/(20\q)$. In this case, the call to \algTWT{} resulted in a
three-partition $(A,B,U)$ such that $|A| \geq |C|/2$. The algorithm returns $U$
alongside the partition $\Y$. We thus have $|U| \leq |C|/2$, $|B| \le |C|/2$
and $U \in \Y$ (if $U \neq \emptyset$) by design. From~\cref{claim:all-x-small}
we further get that all $X \in \X$ have size $|X| \leq z$ at the beginning of
this iteration of the algorithm. Removing the nodes of $B\cup U$ from every set
in $\X$ can only make existing sets in the partition $\X$ smaller and the two
new sets of the resulting partition, $B$ and $U$, have size at most
$|C|/2$ as shown above. Hence, $|Y| \leq\max\{z,|C|/2\}$ holds for all $Y\in\Y$.

\cref{prop:partition:border-routable} of \magicpart{} follows directly,
as the cut $E(U,C\setminus U)$ is $1/\phi$-border routable
through $U$ with congestion at most $2$ by \cref{prop:twt:border} of \algTWT{}.

For \cref{prop:partition:expansion}, observe that we called \algSC{} with
sparsity parameter $\phi/20$.
and \cref{claim:condition} shows that
\cref{req:twt:exp} of \algTWT{} is satisfied with $\delta = 1/(20 \q)$. By
\cref{prop:twt:expanding} of \algTWT{} we thus get that
$G[A \cup B] = G[C \setminus U]$ is
$\frac{1}{500}\phi/\q \cdot \deg_{\partial \Y}$-expanding, as desired,
  where we use the fact that
  $\deg_{\partial \Y}|_{A \cup B} \le \deg_{E(B, V\setminus B)}|_{A \cup B} +  \pi|_A$.

\end{proof}

\begin{lemma}\label{lemma:balanced-child}
If the \magicpart{} algorithm returns from~\cref{case:balanced}, then it
returns a correct (balanced) bad child, i.e.,
\cref{prop:partition:border-routable} and \cref{prop:partition:expansion} of
\magicpart{} are fulfilled.
\end{lemma}
\begin{proof}
In this case, the algorithm outputs $U=T'$ alongside the partition $\Y$. We
have $T' \subseteq T$ and by~\cref{claim:T-props}, $|T| \leq |C|/2$, so
$|U| \leq |C|/2$. Also, we have $U \in \Y$ (if $U \neq \emptyset$) by
construction of $\Y$. Analogous to the argument in the proof
of~\cref{lemma:imbalanced-child}, we have that all $Y \in \Y$ have $|Y| \leq \max\{z,|C|/2\}$
since $|T'| \leq |C|/2$.

Recall that $T' = T \setminus X$, where $X$ is a $2$-fair $(\v{s},\v{t})$-cut in $G[T]$ for
$\v{s} = \deg_{E(T, C\setminus T)}$ and
$\v{t} =  \frac{1}{2} \phi  \deg_{\partial C}|_T$.
Thus, \cref{prop:fc-flows} of \cref{claim:fair-cut-ineq} shows that
there exists a $( \deg_{E(T', X)}|_{T'} + \deg_{E(T,C\setminus T)}|_{T'} , \mb{t'})$-flow $g$ through $G[T']$ 
for some non-negative $\mb{t'} \le 2 \mb{t} = \phi \deg_{E(T, V \setminus C)}|_{T'}$. 
Note that $\deg_{E(T,C\setminus T)}|_{T'} + \deg_{E(T', X)}|_{T'} = \deg_{E(T', C \setminus T')}|_{T'}$, i.e., $g$ is a $(\deg_{E(T', C \setminus T')}|_{T'}, \v{t'})$ flow through $G[T']$.
Thus, $E(T',C \setminus T')$ is $1/\phi$-border routable through $T'$ with congestion 2, i.e.,
\cref{prop:partition:border-routable} holds for $U = T'$.

We prove \cref{prop:partition:expansion} by showing the ``or''-case, namely we prove that $\deg_{\partial \Y}(U) \geq \t/20\cdot \deg_{\partial \Y}(C)$ and $\deg_{\partial\Y}(C)\le \deg_{\partial\X}(C)+2\capc(U,C\setminus U)$. We begin with the first statement.

  By construction, $T' \in \Y$, so we have
  \begin{equation*}
  \deg_{\partial \Y}(U) = \deg_{\partial \Y}(T') \geq \capc(T', V \setminus C) = \capc(T, V\setminus C) - \capc(T\setminus T', V\setminus C)\enspace.
  \end{equation*}
  Since $X = T \setminus T'$ is a 2-fair ($\v{s}, \v{t}$)-cut in $G[T]$ 
   we get from \cref{claim:fair-cut-ineq} that
   \begin{equation*} \textstyle  \capc(T \setminus T', V\setminus C) = \capc(X, V\setminus C) =
   \frac{2}{\phi}\v{t}(X)\le \frac{4}{\phi}\v{s}(X) \leq \frac{4}{\phi}\v{s}(T) =
   \frac{4}{\phi}\capc(T, C \setminus T)\enspace.
   \end{equation*}   

  In addition, $T$ satisfies \cref{pro:sparse} of \cref{claim:T-props}, which gives $\frac{4}{\phi}\cdot\capc(T, C \setminus T)\le\frac{2}{5}\cdot\v{\pi}(T)$. Thus, $\capc(T \setminus T', V\setminus C) \le \frac{2}{5}\cdot\v{\pi}(T)$.
  
  Note that the border edges of $T$ must have a large capacity, namely must fulfill
  $\capc(T, V \setminus C) = \deg_{E(C, V \setminus C)}(T) > \v{\pi}(T)/2$
  because the algorithm reached \cref{case:balanced}.
  With the help of these
  inequalitites we get
  \begin{equation*}
  \begin{split}
  \deg_{\partial \Y}(T')
  &\geq \capc(T, V\setminus C) - \capc(X, V\setminus C)\\
  &\ge\tfrac{1}{2}\v{\pi}(T)-\tfrac{2}{5}\v{\pi}(T) \ge \t/10 \cdot \v{\pi}(C)\enspace,
  \end{split}
  \end{equation*}
  where the last inequality follows from \cref{pro:vol} of
  \cref{claim:T-props}. 
  Next, observe that 
  \begin{equation*}
\begin{split}
\deg_{\partial \Y}(C)
    &\le \v{\pi}(C)-\v{\pi}(T')+2\capc(T', C \setminus T')+\capc(T',V\setminus C)\\
    &\leq \v{\pi}(C) + 2\capc(T', C \setminus T') \\
    & \leq \v{\pi}(C) + 2\phi/5 \cdot \v{\pi}(T) \leq 2
    \v{\pi}(C),\enspace
\end{split}
  \end{equation*}
  where the first inequality holds because $\Y$ is obtained by fusing $T'$ in the
  current partition $\X$ and $\v{\pi}=\deg_{\partial \X}$. This fuse operation can be viewed
  as first removing all $\pi$-weight from $T'$ and then adding the weight of
  border edges, where edges in $E(T',C\setminus T')$ are counted twice.
  The second step follows from \cref{claim:fair-cut-ineq} for
  the $2$-fair cut $T \setminus T'$ and \cref{pro:sparse} of \cref{claim:T-props}, which  gives
  $\capc(T', C\setminus T') \leq 2 \v{s}(T \setminus T') \le 2\v{s}(T)=2\capc(T, C \setminus T) \leq \phi/5 \cdot \v{\pi}(T)$. In total, we thus
  have $\deg_{\partial \Y}(T') \geq \t/20 \cdot \deg_{\partial \Y}(C)$,
  as desired.

  To finish the proof of \cref{prop:partition:expansion} we need to show  the second part of the \enquote{either}-case of \cref{prop:partition:expansion}, i.e.,  that $\deg_{\partial\Y}(C)\le \deg_{\partial\X}(C)+2\capc(U,C\setminus U)$, where
  $U$ is the returned bad child. Recall that $\X$ denotes the partition that was initially supplied to \magicpart{}. \cref{cla:decrease} guarantees that
  $\deg_{\partial\X}(C)$ does not increase during \magicpart{}. The only increase that
  may happen is due to the last step when we fuse the cluster $T'$ in
  the  partition $\X'$ of the current iteration and return the resulting partition $\Y$. By \Cref{claim:fuse}
  \begin{equation*}
  \begin{split}
  \deg_{\partial\Y}(C)
    &\le \deg_{\X'}(C)-\deg_{\X'}(T')+2\capc(T',C\setminus T')+\capc(T',V\setminus C)\\
    &\le \deg_{\X'}(C)+2\capc(T',C\setminus T')
    \le \deg_{\partial\X}(C)+2\capc(U,C\setminus U)\enspace,
  \end{split}
  \end{equation*}
  where $\deg_{\partial\X'}(C) \le \deg_{\partial\X}(C)$ follows from~\cref{cla:decrease}
  and $U=T'$. This finishes the proof of the lemma.
\end{proof}

Observe that the algorithm terminates if it reaches~\cref{case:unbalanced}
or~\cref{case:balanced}. Thus, an iteration of \magicpart{} is
\emph{non-terminating} if it reaches \cref{case:repeat}. We first show that the
$\deg_{\partial \X}(C)$-weight decreases significantly in a non-terminating
iteration.
  \begin{claim}\label{cla:decrease}
  A non-terminating iteration of Algorithm \magicpart{} reduces the weight
  function $\deg_{\partial \X}(C)$ by a factor
  of $(1-\t/4)$.
  \end{claim}
  \begin{proof}
  Let $\X_1$ be the partition at the beginning of the non-terminating iteration and let $\X_2$ be the partition at the end of that iteration. In such an iteration $\X_1$ is only modified by the fusing operation in  \cref{case:repeat}, i.e., $\X_2 = (\X_1 - T) \cup \{T\}$. Note that this  is the only non-terminating case and it happens only if $\deg_{\partial C}(T)\le\deg_{\partial X_1}(T)/2$.
 
  The new partition $\X_2$ has
      \begin{equation*}
      \begin{split}
      \deg_{\partial\X_2}(C)
      &\le\deg_{\partial\X_1}(C)-\deg_{\partial\X_1}(T)+\deg_{\partial \X_1}(T)+2\capc(T,C\setminus T)\\
      &\le\deg_{\partial\X_1}(C)-\deg_{\partial\X_1}(T)+\deg_{\partial \X_1}(T)/2+\phi/5 \cdot \deg_{\partial\X_1}(T)\\
      &=\deg_{\partial\X_1}(C) - (1/2 - \phi/5) \deg_{\partial\X_1}(T)\\
      &\le\deg_{\partial\X_1}(C) -(1/4) \cdot \deg_{\partial\X_1}(T)\\
      &\le(1-\t/4)\deg_{\partial\X_1}(C)\enspace, 
    \end{split}
    \end{equation*}

 The first inequality follows because we can view the fuse operation as first removing all weight from $T$, then adding the weight of edges in $\partial C$ that are incident to $T$, and finally adding the weight of edges in $E(T,C\setminus T)$. 
 Note that the weight of the latter edges has to be added twice for the two endpoints that share an edge.
 The second inequality follows due to the condition $\deg_{\partial C}(T)\le\deg_{\partial X_1}(T)/2$ and \cref{cla:Tproperties}, \cref{pro:sparse}. The third inequality uses $1/2 - \phi/5 \geq 1/4 \ge \t/4$ as $\t \le \bal \le 1/2$ and the last uses \cref{cla:Tproperties}, \cref{pro:vol}.
  \end{proof}

\begin{lemma}\label{lemma:partition-runtime}
Given a cluster $C$ with volume $m_C=\deg(C)$, and algorithms for \algSC{},
\algTWT{}, and \algFC{}, that run in time $\Tsc$, $\Ttwt$ and $\Tfc$,
respectively, the algorithm for \magicpart{} runs in time
$O(\log(m_C)/\t \cdot (\Tsc(m_C) + \Ttwt(m_C) ) +\Tfc(m_C,2))$.
\end{lemma}
\begin{proof}
In every non-terminating iteration through the while loop $\deg_{\partial \X}(C)$ reduces 
by a factor of $(1-\t/4)$. Hence, there can be at most
$\log_{1/(1-\t/4)}m_C=O(\log(m_C)/\t)$ such iterations. In addition
to these iterations there is at most one call
to \algFC{}. The remaining bookkeeping cost is at most linear in $m_C$ and thus
dominated by $\Tsc(m_C)$.
\end{proof}

\begin{corollary}
With the implementation of our building blocks 
the running time of \magicpart{} is
=
$O(\log^7(m_C) \cdot \Tfc(m_C,2))
=\tO(m_c)$ for a cluster $C$ with volume $m_C=\deg(C)$.
\end{corollary}

%%%%%%%%%%%%%%%%%%%%%%%%%%%%%%%%%%%%%%%%%%%%%%%%%%%%%%%%
%%%%%%%%%%%%%% Implementing the Blocks %%%%%%%%%%%%%%%%%
%%%%%%%%%%%%%%%%%%%%%%%%%%%%%%%%%%%%%%%%%%%%%%%%%%%%%%%%
\section{Implementing the Building Blocks}

\begin{lemma}[Fair Cut Properties]\label{lemma:fairCutFlows}
Let $\alpha \ge 1$.
  If $U$ is an $\alpha$-fair $(\v{s},\v{t})$-cut in $G=(V,E, \capc)$, then 
    \begin{itemize}
    \item there exists non-negative vertex weight vectors $\mb{x}$ and $\mb{y}$ such that there exist a $(\mb{x}, \mb{y})$ flow  with congestion at most $\alpha$ in $G[U]$
    and this flow routes the demand $\deg_{E(U,\vu)}-\alpha |\mb{s}-\mb{t}| \le \mb{x}-\mb{y} \le \deg_{E(U,\vu)}+\alpha |\mb{s}-\mb{t}|$,  more precisely, for $u\in U$ if $\v{s}(u)<\v{t}(u)$, then the net-flow out of $u$ is exactly $\deg_{E(U,\vu)}(u) - \v{s}(u)+\v{t}(u)$ and else it is at least $\deg_{E(U,\vu)}(u) -\alpha( \v{s}(u)-\v{t}(u))$ and at most $\deg_{E(U,\vu)}(u)$,  
    and 
     \item there exists non-negative vertex weight vectors $\mb{x}$ and $\mb{y}$ such that there exist a $(\v{x}, \v{y})$ flow with congestion at most $\alpha$ in $G[V\setminus U]$ which routes the demand $\deg_{E(U,\vu)}-\alpha |\mb{s}-\mb{t}| \le \mb{x}-\mb{y} \le \deg_{E(U,\vu)}+\alpha |\mb{s}-\mb{t}|$, for $u\in \vu$ if $\v{s}(u)>\v{t}(u)$, then the net-flow out of $u$ is exactly $\deg_{E(U,\vu)}(u) + \v{s}(u)-\v{t}(u)$ and else it is at most $\deg_{E(U,\vu)}(u)$ and at least $ \deg_{E(U,\vu)} +\alpha( \v{s}(u)-\v{t}(u))$.
    \end{itemize}    
\end{lemma}

\begin{proof}
Let $f$ be a flow such that $(U,f)$ is an $\alpha$-fair
($\v{s}$,$\v{t}$)-cut/flow pair.

Consider the scaled flow $f' := \alpha \cdot f$.
Note that by~\cref{def:fair-cut}, (1) for every vertex $v \in  U$ if $\v{s}(v)-\v{t}(v)\ge 0$ it holds that $\alpha (\v{s}(v)-\v{t}(v)) \ge f'(v) \ge 0$ and otherwise
it holds that $\v{s}(v)-\v{t}(v)\ge  f'(v) \ge \alpha(\v{s}(v)-\v{t}(v))$, and (2) for every vertex $v \in V\setminus U$ 
if $\v{s}(v)-\v{t}(v) \ge 0$ it holds that $\alpha(\v{s}(v)-\v{t}(v)) \ge f'(v) \ge \v{s}(v)-\v{t}(v)$ and otherwise   $0 \ge f'(v) \ge \alpha(\v{s}(v)-\v{t}(v)) $. Also note that $|f'| \le \alpha \deg_{E}$.

\newcommand{\F}[0]{\mathcal{F}}

In the flow $f'$, 
by \cref{prop:fair-cut:cross-cut} of \cref{def:fair-cut}, each edge $(u,v)$ in $E(U, V \setminus U)$ sends at least $\capc$ unit of flow from $U$ to $V \setminus U$. Also, it sends no flow in the reverse direction.
Now consider the graph $G^*$ which consists of $G$ augmented by two artificial nodes $s^*$ and $t^*$ such that for every $v\in V$ there is an edge $(s^*, v)$ with capacity $\v{s}(v)-\v{t}(v)$ if $\v{s}(v) > \v{t}(v)$ and there is an edge $(v, t^*)$ with capacity $\v{t}(v)-\v{s}(v)$ if $\v{t}(v)>\v{s}(v)$. Note that $f'$ can be trivially extended to an $s^*$-$t^*$ flow in $G^*$.
Consider a classic path decomposition of that flow in $G^*$. When restricted to $G$ this gives a 
path decomposition $\F$ of $f'$ in $G$ such that each  path flow starts at a vertex with positive net-flow, ends at a vertex with negative net-flow and has non-negative flow value, called {\em weight}, such that for each edge $e$ the sum of the weights of the path flows is exactly the flow value of $e$. The number of such paths is at most $m+n$.
As the flow only sends flow from $U$ to $V \setminus U$, each path flow in $\F$ contains at most one edge of the cut $E(U, V\setminus U)$.
Furthermore if a path flow contains an edge of the cut $E(U,\vu)$ then its starting point is in $U$ and its ending point is in $\vu$.
In the following we remove path flows from $\F$ (and, thus, from $f'$) until the resulting flow has certain desired properties.
% by \cref{prop:fair-cut:cross-cut} of \cref{def:fair-cut},
% each edge $(u,v)$ in $E(U, V \setminus U)$ sends at least $\capc$ unit of flow
% from $U$ to $V \setminus U$. Also, it sends no flow in the reverse direction. Consider a path
% decomposition $\F$ of $f'$ such that each flow path starts at a vertex with
% positive net-flow, ends at a vertex with negative net-flow and sends exactly
% one unit. As the flow only sends flow from $U$ to $V \setminus U$, each path in
% $\F$ contains at most one edge of the cut $E(U, V\setminus U)$. Furthermore if
% a path contains an edge of the cut $E(U,\vu)$ then its starting point is in $U$
% and its ending point is in $\vu$. In the following we remove paths from $\F$
% (and, thus, from $f'$) until the resulting flow has certain desired properties.

\paragraph{Part 1.}
(a) First, we reduce the weight  of the path flows of $\F$ containing an edge of $E(U,\vu)$ by a positive value  until
each cut edge sends exactly $\capc$ units of flow.
This does not reduce the net-flow into vertices $v \in U$ with negative net-flow as the end point of the path flows whose weight is reduced must be in $\vu$.
Thus, every $v \in U$ with $\v{s}(v) - \v{t}(v) <0$ still has at most $\v{s}(v)-\v{t}(v)$ net-flow out of $v$, i.e., at least net-flow $\v{t}(v)-\v{s}(v)$ into $v$. It follows that the sum of the weights of the path flows ending at $v$ is $\v{t}(v)-\v{s}(v) >0$. Note that no such path contains an edge from $E(U,\vu)$ as there is no flow from $\vu$ to $U$ in $f'$. We reduce the weight of such flow paths by a positive value until the net-flow at $v$ is exactly $\v{t}(v)-\v{s}(v)$. 
By our observation this does not reduce the flow on the edges in $E(U,\vu)$. 
The resulting flow  (ie the sum of the weights of the path flows) is called $f''$. Note that for every $v \in U$, if $\v{s}(v)<\v{t}(v)$ then $f''(v) = \v{s}(v)-\v{t}(v)$.
If $\v{s}(v)=\v{t}(v)$ then $f(v)=f'(v) = 0$ implying that no path flow in $\F$ started or ended at $v$. Thus, $f''(v)=0$.
Furthermore, by \cref{def:fair-cut} and the fact that the weight of path flows in $\F$ is reduced to no smaller than 0, if $\v{s}(v) >\v{t}(v)$ then $\alpha(\v{s}(v)-\v{t}(v)) \ge f'(v) \ge f''(v) \ge 0$. 

% This does not reduce the net-flow into vertices $v \in U$ with negative net-flow as the end point of the path must be in $\vu$.
% Thus, every $v \in U$ with $\v{s}(v) - \v{t}(v) <0$ still has at least $\v{t}(v)-\v{s}(v)$ negative net-flow out of $v$, i.e., positive net-flow $\v{s}(v)-\v{t}(v)$ into $v$. It follows that there are at least $\v{s}(v)-\v{t}(v)$ flow paths ending at $v$. Note that no such path contains an edge from $E(U,\vu)$ as there is no flow from $\vu$ to $U$ in $f'$. We remove such flow paths until there are exactly $\v{s}(v)-\v{t}(v)$ flow paths ending at $v$. By our observation this does not reduce
% the flow on the edges in $E(U,\vu)$. 
% The resulting flow is called $f''$. Note that for every $v \in U$, if 
% $\v{s}(v)<\v{t}(v)$ then $f''(v) = \v{s}(v)-\v{t}(v)$.
% If $\v{s}(v)=\v{t}(v)$ then $f(v)=f'(v) = 0$ implying that no flow path in $\F$ started or ended at $v$. Thus, $f''(v)=0$.
% Furthermore, by \cref{def:fair-cut} and the fact that only existing flow is removed, if $\v{s}(v) >\v{t}(v)$ then $\alpha(\v{s}(v)-\v{t}(v)) \ge f'(v) \ge f''(v) \ge 0$. 

Thus, as $\alpha \ge 1$, $|\mb{f''} |\le \alpha|\mb{s}-\mb{t}|$
{\em The crucial observation is that $f''$ induces a $(\mb{x},\mb{y})$-flow in $G[U]$ for some non-negative vertex weights $\mb{x}$ and $\mb{y}$: }
Consider a new flow $f_U$ defined on $G[U]$ that routes the flow from the cut
edge $E(U, V\setminus U)$ and from vertices $u \in U$ with $f''(u)<0$ with
exactly the same amount on each edge as in $f''$, but in the reverse direction.
More specifically, for every $u \in U$ (a) if $f''(u) <0$, set
$x(u) := \deg_{E(U, V\setminus U)}(u) - f''(u) = \deg_{E(U, V\setminus U)}(u)
-\v{s}(u) + \v{t}(u)$ and $y(u) := 0$ and (b) otherwise (ie $f''(u)\ge 0$), set
$x(u):=\deg_{E(U, V\setminus U)}(u)$ and
$y(u) := f''(u) \le \alpha(\v{s}(u)-\v{t}(u)) $. Note that both $\mb{x}$ and $\mb{y}$
are non-negative vertex weights and
\begin{itemize}
\item
$\sum_{u \in U} x(u) -y(u) = \Big(\sum_{u \in U, f''(u) <0 } -f''(u) + \sum_{u
  \in U} \deg_{E(U, V\setminus U)}(u)\Big) -
\sum_{u \in U, f''(u) > 0} f''(u) = 0$ (as the  flow \enquote{absorbed in} or \enquote{leaving} $G[U]$ equals the flow \enquote{generated in} $G[U]$) and
\item $f_U$ is a $(\mb{x},\mb{y})$-flow which routes exactly the demand
$\mb{x}-\mb{y} =  \deg_{E(U, V\setminus U)}-f''$.
\end{itemize}

Recall that $|\mb{f''}| \le \alpha |\mb{s}-\mb{t}|$. Thus, $\deg_{E(U,\vu)}-\alpha |\mb{s}-\mb{t}| \le \mb{x} - \mb{y} \le \deg_{E(U,\vu)}+\alpha |\mb{s}-\mb{t}|$.

\paragraph{Part 2.}
Now consider the flow $f_{V \setminus U}$ in $G[\vu]$ defined as follows.
First, we reduce the weight  of the path flows of $\F$ containing an edge of $E(U,\vu)$ by a non-negative value  until 
each cut edge sends exactly $\capc$ units of flow. Note that each such path flow starts in $U$ and ends in $\vu$ as there is no flow from $\vu$ to $U$.
Thus no $v \in \vu$ with $f'(v)>0$ can be a starting node of a  flow path whose weight was reduced, i.e., the net-flow for such nodes is unchanged.
Thus, every $v \in \vu$ with $\v{s}(v) - \v{t}(v) >0$ still has at least $\v{s}(v)-\v{t}(v)$ net-flow out of $v$. Thus, the weight of the path flows in $\F$ that start at $v$ is at least $\v{s}(v)-\v{t}(v)$. Note that no such path contains an edge from $E(U,\vu)$ as there is no flow from $\vu$ to $U$ in $f'$. Next, we reduce the weight of the path flows starting at nodes $v$ with $\v{s}(v)-\v{t}(v)>0$ until the weight of the path flows starting at $v$ is exactly $\v{s}(v)-\v{t}(v)$. By our observation this does not reduce the flow on the edges in $E(U,\vu)$.

% First, we remove enough paths from the path decomposition of $f'$ (and, thus, from $f'$) until
% each cut edge sends exactly $\capc$ units of flow. Note that each such path starts in $U$ and ends in $\vu$ as there is no flow from $\vu$ to $U$.
% Thus no $v \in \vu$ with $f'(v)>0$ can be a starting node of a removed flow path, i.e., the net-flow for such nodes is unchanged.
% Thus, every $v \in \vu$ with $\v{s}(v) - \v{t}(v) >0$ still has at least $\v{s}(v)-\v{t}(v)$ positive net-flow out of $v$. Thus, there are at least $\v{s}(v)-\v{t}(v)$ flow paths in the flow decomposition starting at $v$. Note that no such path contains an edge from $E(U,\vu)$ as there is no flow from $\vu$ to $U$ in $f'$. Next, we remove flow paths starting at such nodes $v$ until there are exactly $\v{s}(v)-\v{t}(v)$ flow paths starting at $v$. By our observation this does not reduce the flow on the edges in $E(U,\vu)$.

The resulting flow is called $f'''$. Note that for every $v \in V$, if 
$\v{s}(v)>\v{t}(v)$ then $f'''(v) = \v{s}(v)-\v{t}(v)$.
If $\v{s}(v)=\v{t}(v)$ then $f(v)=f'(v) = 0$ implying that no path flow in the flow decomposition started or ended at $v$. Thus, $f'''(v)=0$.
Furthermore, by \cref{def:fair-cut} and the fact that we only reduce the weight of path flow in $\F$ to at least 0, if $\v{s}(v) <\v{t}(v)$ then $0  \ge f'''(v) \ge f'(v) \ge \alpha(\v{s}(v)-\v{t}(v))$. 
% If $\v{s}(v)=\v{t}(v)$ then $f(v)=f'(v) = 0$ implying that no flow path in the flow decomposition started or ended at $v$. Thus, $f'''(v)=0$.
% Furthermore, by \cref{def:fair-cut} and the fact that we only removed existing flow from $f'$, if $\v{s}(v) <\v{t}(v)$ then $0  \ge f'''(v) \ge f'(v) \ge \alpha(\v{s}(v)-\v{t}(v))$. 
Thus, $|\mb{f'''}| \le \alpha|\mb{s}-\mb{t}|.$
{\em The crucial observation is that $f'''$ induces a $(\mb{x},\mb{y})$-flow in $G[\vu]$ for some non-negative vertex weights $\mb{x}$ and $\mb{y}$: }

% {\em The crucial observation is that $f'''$ when  restricted to $G[\vu]$ induces a $(\mb{x},\mb{y})$-flow in $G[\vu]$ for some non-negative vertex weights $\mb{x}$ and $\mb{y}$: }

In $G[\vu]$ the flow $f'''$ sends from every edge of the cut $E(U,\vu)$ and
from each vertex $v \in \vu$ with positive net-flow $f'''(v)$ to each vertex in
$\vu$ by sending on each edge exactly the same amount as in $f'''$. More
formally, for every $v \in \vu$ (a) if $f'''(v)>0$, set
$\v{x}(v) := \deg_{E(U,\vu)}(v) + f'''(v) = \deg_{E(U,\vu)}(v) + \v{s}(v)-\v{t}(v)$ and
$\v{y}(v) :=0$ and (b) otherwise set $\v{x}(v):= \deg_{E(U,\vu)}(v)$ and
$\v{y}(v) := -f'''(v) \le -\alpha(\v{s}(v)-\v{t}(v))$. Note that both $\mb{x}$ and $\mb{y}$
are non-negative. Now (1)
$\sum_{v \in \vu} \v{x}(v) -\v{y}(v)= \Big(\sum_{v \in \vu, f'''(v)>0}f'''(v) + \sum_{v
  \in \vu, \deg_{E(U,\vu)}(v)>0}\deg_{E(U,\vu)}(v) \Big) - \sum_{v \in \vu,
  f'''(v)<0} (-f'''(v)) = 0$ and (2) $f_{\vu}$ is a $(\mb{x}, \mb{y})$-flow
which routes exactly the demand
$\mb{x}-\mb{y} = \deg_{E(U, V\setminus U)}+f'''$.

Recall $|\mb{f'''}| \le \alpha|\mb{s}-\mb{t}|.$ Thus, $\deg_{E(U,\vu)}-\alpha |\mb{s}-\mb{t}| \le \mb{x} - \mb{y} \le \deg_{E(U,\vu)}+\alpha |\mb{s}-\mb{t}|$.

Note that since $f$ is feasible, both resulting flows have congestion at most $\alpha$.
\end{proof}

Let $\mb{d}$ be a demand vector. By definition, a flow $f$ routes $\mb{d}$ in a graph $G=(V,E)$ if every $v \in V$ has net flow exactly $f(v) = \mb{d}(v)$. For suitable $\mb{s}, \mb{t} \geq 0$ we may split the demand  $\mb{d} =  \mb{s} - \mb{t}$ to allow for the interpretation that each vertex $v \in V$ sends out exactly $\mb{s}(v)$ units of flow and receives exactly $\mb{t}(v)$ units of flow. Note that there may be overlap and a vertex can have nonnegative values for both $\mb{s}(v)$ and $\mb{t}(v)$. In total, the net flow out of each vertex is however $f(v) = \mb{s}(v) - \mb{t}(v) = \mb{d}(v)$, as desired.

\begin{claim}\label{claim:fair-cut-ineq}\label{prop:fair-cut:t}
If $U$ is an $\alpha$-fair $(\v{s},\v{t})$-cut in $G=(V,E, \capc)$, then 

\begin{enumerate}
\item for all $A \subseteq U$ it holds
$\capc(A, V \setminus U) + \v{t}(A) \leq \alpha \cdot \big( \v{s}(A) + \capc(A,
U \setminus A) \big)$. In particular,
$\capc(U, V\setminus U) + \v{t}(U) \leq \alpha \cdot \v{s}(U)$,
and \label[property]{prop:fc-cap}

  \item for all $A \subseteq V \setminus U$ it holds $\capc(A, U) + \v{s}(A) \leq \alpha \cdot \big( \v{t}(A) + \capc(A, V \setminus U \setminus A) \big)$. In particular, $\capc(U, V\setminus U) + \v{s}(V \setminus U) \leq \alpha \cdot \v{t}(V \setminus U)$. And \label[property]{prop:fc-cap-t}

  \item For $\mb{x} = \deg_{E(U, V \setminus U)}$ there exist \label[property]{prop:fc-flows}
  \begin{itemize}
    \item a $(\mb{x} + \mb{t}|_U, \, \mb{s'})$ flow in $G[U]$ with congestion $\alpha$ for some  $\mb{s'} $ with $\mb{0} \le \mb{s'}\leq \alpha \cdot \mb{s}|_U$, and
    \item a $(\mb{x} + \mb{s}|_\vu, \, \mb{t'})$ flow in $G[V \setminus U]$ with congestion $\alpha$ for some  $\mb{t'}$ with $\mb{0} \le \mb{t'} \leq \alpha \cdot \mb{t}|_\vu$.
  \end{itemize}
\end{enumerate}

\end{claim}
\begin{proof}
Let $f$ be a flow such that $(U,f)$ is an $\alpha$-fair cut/flow pair.
  We first establish the properties for the $G[U]$-side, then we show that the results for the $G[V\setminus U]$-side follow with symmetrical arguments.

By~\cref{def:fair-cut} for all nodes in $U$ it holds that $f(v) \le \v{s}(v)-\v{t}(v)/\alpha $. When summed over all nodes in $A$ it follows that $f(A) \le \v{s}(A) - \v{t}(A)/\alpha.$ At most $\capc(U \setminus A, A)$ flow is sent from $U\setminus A$ into $A$. As no flow is sent from $V\setminus U$ to $U$, it also follows that the total net-flow sent {\em out} of the nodes in $A$ plus the flow sent into $A$ (which is at most $\capc(U\setminus A, A)$) must equal the flow out of $A$ which is at least $\capc(A,\vu)/\alpha$, by \cref{prop:fair-cut:cross-cut} of \cref{def:fair-cut}. Thus, $\capc(A,\vu)/\alpha \le f(A) + \capc(U \setminus A, A)$.
Combining the two gives $\capc(A,\vu) + \v{t}(A) \le \alpha (\v{s}(A) + \capc(U \setminus A, A))$.

Consider the split of the vertex set $U$ into `designated sources' and `designated targets' by defining $S = \{u \in U \,|\, \mb{s}(u) \geq \mb{t}(u)\}$ and $T = \{u \in U \,|\, \mb{s}(u) < \mb{t}(u)\}$. Let $g$ be the flow from~\cref{lemma:fairCutFlows} in $G[U]$ and call the demand it routes $\mb{d}$. By the properties of this flow, we have
  \begin{equation*}
    \begin{aligned}
      \mb{d} & \geq \deg_{E(U,\vu)} + (\mb{t} - \mb{s})|_T - \alpha(\mb{s} - \mb{t})|_S 
        = \deg_{E(U,\vu)} +\mb{t}|_T + \alpha \mb{t}|_S - \mb{s}|_T - \alpha\mb{s}|_S \\
        & \geq \deg_{E(U,\vu)} +\mb{t}|_U - \alpha\mb{s}|_U,
    \end{aligned}
  \end{equation*}
  since $\mb{s}$ and $\mb{t}$ are nonnegative and $\alpha \geq 1$.
  \cref{lemma:fairCutFlows} also shows that  $\mb{d}|_S \le \deg_{E(U,\vu)}$, thus it follows that 
  \begin{equation*}
    \begin{aligned}
      \mb{d} & \leq \deg_{E(U,\vu)} + (\mb{t} - \mb{s})|_T 
        \le\deg_{E(U,\vu)} +\mb{t}|_U
    \end{aligned}
  \end{equation*}
   Recall that by definition $\mb{s}$ is non-negative.
  Thus, there is a vertex weight $\mb{s'}$ with  $\mb{0} \le \mb{s'} \leq \alpha \cdot \mb{s}|_U$,  such that the flow $g$ routes exactly the demand $\mb{d} = \deg_{E(U,\vu)} +\mb{t}|_U - \mb{s'}$. This implies that there exists a
  $(\deg_{E(U,\vu)} +\mb{t}|_U, \mb{s'})$-flow for some non-negative vertex weights $\mb{s'}$ with  $\mb{0} \le \mb{s'} \leq \alpha \cdot \mb{s}|_U$.

  \smallskip We now show that the results for the $G[V\setminus U]$-side follow
  in a completely symmetrical manner. By the definition~\ref{def:fair-cut}
    for all nodes in $\vu$ it holds that $f(v) \ge \v{s}(v)/\alpha -\v{t}(v)$. When
    summed over all nodes in $A$ it follows that $-f(A) \le \v{t}(A)-\v{s}(A)/\alpha.$
    The total flow sent from $A$ to $\vu\setminus A$ is at most
    $\capc(A, \vu \setminus A)$. As no flow is sent from $V\setminus U$ to $U$,
    it follows that the total flow sent into $A$ is at most the total net-flow
    into the nodes of $A$ plus the flow sent out of $A$. Thus, it follows that
    $\capc(A,\vu)/\alpha \le -f(A) + \capc(A, \vu \setminus
    A)$. Combining the two gives $\capc(U,\vu) + \v{s}(A) \le \alpha (\v{t}(A)+
    \capc(A, \vu \setminus A))$. 

For \cref{prop:fc-flows}, observe that
we can again consider the split of the vertex set $V \setminus U$ into \enquote{designated sources} $S$ and \enquote{designated targets} $T$, similar to above but for $\vu$. Let $g$ be the flow in $G[V \setminus U]$ from \cref{lemma:fairCutFlows} and assume it routes the demand $\v{d}$. Then,
  \begin{align*}
      \deg_{E(U,\vu)} + \mb{s}|_\vu \ge \deg_{E(U,\vu)} +(\mb{s} - \mb{t})|_S \ge \mb{d} \textrm{ and }\\
      \mb{d} \geq \deg_{E(U,\vu)} + (\mb{s} - \mb{t})|_S - \alpha(\mb{t} - \mb{s})|_T \geq \deg_{E(U,\vu)} + \mb{s}|_\vu - \alpha \cdot \mb{t}|_\vu.
  \end{align*}
  Thus there exists a non-negative vertex weight $\mb{t'}$ with $\mb{t'} \le \alpha \cdot \mb{t}$ such that $\mb{d} = \deg_{E(U,\vu)} + \mb{s}|_\vu - \mb{t'}$ and $g$ routes $\mb{d}$.
  Hence, $g$ is the flow that routes the demand required by the second statement of \cref{prop:fc-flows}, concluding the proof of \cref{claim:fair-cut-ineq}.
\end{proof}

\subsection{\texorpdfstring{Algorithm \algFC{}}{Algorithm FairCut}}\label{sec:fair-cut}

Here we construct the desired an algoritm for \algFC{}
using a reduction from $\alpha$-fair $(s, t)$-cut where $s$ and $t$ are
\emph{vertices}. Note the difference to \cref{def:fair-cut}, where
$\v{s}, \v{t}$ are functions from $V$.

\begin{definition}
    Let $G=(V,E,\capc)$ be a weighted graph and $s,t$ be two vertices in $V$. For any parameter $\alpha \ge 1$ a cut $(U, V\setminus U)$ is called an $\alpha$-fair $(s,t)$-cut if there exists an $(s,t)$-flow with congestion 1 such that $f(u,v)\ge \capc(u,v)/\alpha$ for every edge $(u,v)$ with $u \in S$ and $v \in V\setminus S$.
\end{definition}

An $\alpha$-fair $(s,t)$-cut for \emph{vertices} $s$ and $t$ can be computed in time $\tO(m/(\alpha-1))$ in a graph with volume $m$ using the recent result by Li and Li~\cite{LL25}. Note that this method is randomized and correct with high probability.

\begin{proof}[Proof of \cref{lemma:fair-cut}]
To avoid confusing we denote $G=(V_G,E_G)$ in this proof.
Construct a graph $H=(V_G\cup\{s,t\}, E_H)$ by adding a super-source $s$ and
super-target $t$ to $G$. For all $v\in V_G$, insert edges $\{s,v\}$ with capacity
$\v{s}(v) - \v{t}(v)$ if $\v{s}(v) > \v{t}(v)$ and edges $\{v,t\}$ with capacity $\v{t}(v) - \v{s}(v)$ if $\v{s}(v) < \v{t}(v)$. The remaining edges have the same capacity in $H$ as in $G$.
In $H$, we can compute an $\alpha$-fair $(s, t)$-cut/flow pair $(U,f)$ in time $\tO(|E_H|/(\alpha-1))$ using Theorem 1 in \cite{LL25}. Since $|E_H| \leq |E| + |V|$ and $G$ is connected, we have a running time of $\tO(|E|/(\alpha-1))$ for this computation.

We set $B := U \setminus \{s\} \subseteq V_G$. Together with the truncated flow
$f' := f|_{E_G}$ , the pair $(B, f')$ then satisfies all properties required by
\cref{def:fair-cut}. Observe that by removing edges incident to $s$, and $t$, each $v \in V_G$ has a net flow $f'(v) = f(s,v) - f(v,t)$.

  Let $v \in V_G$ be a vertex with $\v{s}(v) - \v{t}(v) \geq 0$. By construction, there is no edge $\{v, t\}$ in $H$, hence $0 \leq f'(v) = f(s,v) \leq \v{s}(v) - \v{t}(v)$, since $f$ is feasible in $H$. This gives \cref{prop:fc:net-s}.
  By a completely analogous argument, vertices $v \in V_G$ with $\v{s}(v) - \v{t}(v) \geq 0$ have no edge $\{s, v\}$ in $H$, giving $0 \geq f'(v) = -f(t,v) \geq \v{s}(v) - \v{t}(v)$ and establishing \cref{prop:fc:net-t}.

  Now consider all edges $\{u, v\} \in E_H(B, V_H \setminus B)$, where $u \in B$ and $v \in V_H \setminus B$. Since $f$ is a $\alpha$-fair cut, these edges must be saturated up to a factor $1/\alpha$, giving $f(u,v) \geq 1/\alpha \cdot \capc(u, v)$.
  We can group these edges into three categories: (1) edges leaving $s$, (2) edges entering $t$, and (3) edges in $E_G$.

  Let $e = \{s, v\}$ be an edge from category (1). By construction, it must hold that $v \in V_G \setminus B$ and $\v{s}(v) - \v{t}(v) > 0$ since $v$ has an incoming edge from $s$. It consequently has no edge going to $t$.
  As $e$ it is almost saturated, this gives $f'(v) = f(s,v) \geq 1/\alpha \cdot (\v{s}(v) - \v{t}(v))$, and thus \cref{prop:fair-cut:generate}.
  Analogously, edges from category (2) are of the form $\{v,t\}$ for a vertex $v \in B$ with $\v{s}(v) - \v{t}(v) < 0$. The vertex $v$ has an edge to $t$ but no edge to $s$ and hence $f'(v) = - f(v, t) \leq 1/\alpha \cdot (\v{s}(v) - \v{t}(v))$, which gives \cref{prop:fair-cut:absorb}.
  Lastly, edges in category (3) are also edges in $G$. As they are all only sending flow in the direction from $B$ to $V \setminus B$, and are saturated to at least a factor $1/\alpha$, this gives \cref{prop:fair-cut:cross-cut}.

\end{proof}

\subsection{\texorpdfstring{Algorithm \algTWT{}}{Algorithm TwoWayTrim}}\label{subsec:twt}

In this section we show our algorithm for \algTWT{} and prove its correctness.
Recall that the algorithm has to fulfill the following requirement:
\RoutineTWT
In particular, we show the following result.

\begin{lemma}\label{claim:twt}
    \algTWT{} can be implemented using two executions of \algFC{} with $O(m)$ additional work.
\end{lemma}

To obtain our Algorithm for \algTWT{}, we execute the following steps.

\bigskip\noindent
Algorithm \algTWT$(G, C, R, \v{\pi}, \phi)$:

\begin{enumerate}
  \item Define a flow problem in $G[C\setminus R]$: The source function is $\v{s}_1 = \deg_{E(R, C\setminus R)}$, the target function is $\v{t}_1 = 1/5\cdot\delta \phi\v{\pi}$ 
  and run \algFC$(G[C \setminus R], \v{s}_1, \v{t}_1, \alpha_1)$ with $\alpha_1=2$. This results in a $2$-fair $(\v{s}_1, \v{t}_1)$-cut $X_0$ of $C \setminus R$. 
  Let $X_1 = X_0 \cup R$.
  We set $A := C\setminus X_1$ and thus obtain the partition $(X_1, A)$ of $C$.

  \item Define a flow problem in $G[X_1]$: The source function is $\v{s}_2 = \deg_{E(X_1, A)}$, the target function is $\v{t}_2 = \frac{1}{2}\phi \deg_{E(C, V \setminus C)}$. Then run \algFC$(G[X_1], \v{s}_2, \v{t}_2, \alpha_2)$ with $\alpha_2=2$. This results in a $2$-fair cut $X_2$ and we set $U := X_1\setminus X_2$ and $B := X_2$. Notice that $(A,B,U)$ now forms a three-partition of $C$, return it.
\end{enumerate}

We establish correctness of the algorithm by verifying that the output $(A,B,U)$ fulfills the claimed properties. The first three properties are shown to be satisfied in the following claim.

\begin{claim}
  The output of $(A, B, U)$ satisfies \cref{prop:twt:Acap,prop:twt:Avol,prop:twt:border}.
\end{claim}
\begin{proof}
  We have $A = C \setminus X_1 \subseteq C \setminus R$ by design. Then, since $X_0$ is a $2$-fair $(\v{s}_1, \v{t}_1)$-cut, by \cref{claim:fair-cut-ineq}
  we have
  \begin{equation*}\textstyle
  \begin{aligned}
      \capc(A, C \setminus A) &= \capc(A, X_0) + \capc(A, R) \leq \alpha_1 \v{s}_1(X_0) + \capc(A,R) \\ &=  2  \capc(X_0, R) + \capc(A,R) \le 2\capc(R, C \setminus R).
  \end{aligned}
  \end{equation*}
This establishes \cref{prop:twt:Acap}. For \cref{prop:twt:Avol}, we can again make use of \cref{claim:fair-cut-ineq} applied to the fair cut $X_0$ to obtain 
  \begin{equation*}
    \textstyle 
      \frac{1}{5}\delta\phi\cdot\v{\pi}(X_0) = \v{t}_1(X_0) \leq 2 s_1(X_0) \le 2 \v{s}_1(X_0) = 2\capc(R, C \setminus R).
  \end{equation*}
  Recall that $B \cup U = X_0 \cup R$, so we get
  \begin{equation*}\textstyle
    \v{\pi}(B \cup U) = \v{\pi}(X_0) + \v{\pi}(R) \leq \frac{10}{\delta\phi} \capc(R, C \setminus R) + \v{\pi}(R) \leq (1 + \frac{10}{\delta}) \v{\pi}(R) \le \frac{11}{\delta} \v{\pi}(R),
  \end{equation*}
  where the previous to last inequality follows since $\capc(R, C \setminus R) \leq \phi \v{\pi}(R)$ by \cref{req:twt:Rsparse} and the last one follows because $\delta \le 1$.
  
  Finally, \cref{claim:fair-cut-ineq}, \cref{prop:fc-flows} applied to the fair cut $X_2$ gives a ($\v{x} + \v{s}_2, \, \v{t}'$)-flow $g$ with $\v{x} = \deg_{E(X_1 \setminus X_2, x_2)}$ in $G[X_1 \setminus X_2] = G[U]$ \ with congestion $\alpha_2 = 2$ and some non-negative $\v{t}' \leq \alpha_2 \v{t}_2 = \phi\deg_{E(C, V \setminus C)}$.
  Note that 
\begin{equation*}  
\begin{split}
\v{x} +\v{s}_2 
&= \deg_{E(U, X_1 \setminus U)}|_U + \deg_{E(X_1, C \setminus X_1)}|_U 
= \deg_{E(U, X_1 \setminus U)}|_U + \deg_{E(U, C \setminus X_1)}|_U \\
&= \deg_{E(U, C \setminus U)}|_U\enspace,
\end{split}
\end{equation*}
  where the equality only holds on the weight vectors restricted to $U$.
  As $\v{t'}$ is non-negative, it holds that $|\v{t'}| \le  \phi\deg_{E(C, V \setminus C)}$.

  This flow proves that $E(U, C\setminus U)$ is $1/\phi$-border-routable through $U$ with congestion $2$, as desired for \cref{prop:twt:border}.
\end{proof}

Before we can show that \cref{prop:twt:expanding} is met, we first establish the auxiliary property that $G[A]$ is sufficently expanding.

\begin{claim}
  $G[A]$ is $1/5 \cdot \delta\phi\v{\pi}|_A$-expanding.
\end{claim}
\begin{proof} 
  For this proof, fix $\v{\psi} = \delta\phi\v{\pi}$ for better readability. We need to show that $G[A]$ is $1/5\cdot\v{\psi}$-expanding.
  Recall that $(X_1,A)$ is a partition of $C$, $R \subseteq X_1$, and $A \subseteq C\setminus R$. Let $S \subseteq A$ be some cut with $\v{\psi}(S) \leq \v{\psi}(A \setminus S)$ and assume by contradiction that $\capc(S, A \setminus S) < \frac{1}{5}\v{\psi}(S)$. Then, since $G$ is $\v{\psi}|_{C \setminus R}$ expanding by \cref{req:twt:exp}, we have $\capc(S, C \setminus S) \geq \v{\psi}(S)$ and
  \[
    \capc(S, X_1) = \capc(S, C \setminus S) - \capc(S, A \setminus S) >
    \textstyle \frac{4}{5}\v{\psi}(S).
  \]
  As $X_1$ is a $2$-fair cut, by \cref{prop:fair-cut:cross-cut} of
  \cref{def:fair-cut}, there is a flow $f_1$ such that each edge in $E(X_1,A)$
  carries at least $1/2$ of its capacity units of $f_1$-flow. As
  $S \subseteq A$, some of this flow also enters $S$, namely at least
  \[
    f_1(X_1, S) \geq \textstyle\frac{1}{2} \cdot \capc(S, X_1)
    > \textstyle \frac{1}{2} \cdot \frac{4}{5} \v{\psi}(S) = \textstyle
    \frac{2}{5}\v{\psi}(S).
  \]
  The total flow absorbed within $S$ is however at most $\v{t}_1(S) = 1/5 \cdot \v{\psi}(S)$. The remaining flow of more than $1/5 \cdot \v{\psi}(S)$ has to leave $S$ using the edges with capacity $\capc(S,A\setminus S)$ in order to be absorbed. It follows that 
  \[ \textstyle 
    \capc(S,A\setminus S) \geq f_1(X_1, S) -
    \v{t}_1(S) > \frac{1}{5}\v{\psi}(S),
  \]
  which is a contradiction to our initial assumption.
\end{proof}

Next, we can show that \cref{prop:twt:expanding} is fulfilled. We introduce some simplifying notation. Let $Y = A \cup B$ and 
$\bm{\sigma} = \deg_{E(B,V\setminus B)}|_{Y} + \v{\pi}|_A$. Thus, the goal for \cref{prop:twt:expanding} is to show that $G[Y]$ is $\frac{1}{25} \delta \phi \cdot \bm{\sigma}$ expanding.
Unraveling our notation, this is equivalent to showing that 
 $\capc(S, Y \setminus S) \geq \frac{1}{25} \delta\phi \cdot \bm{\sigma}(S)$ 
 for all sets $S \subseteq Y$ with $\bm{\sigma}(S) \le \bm{\sigma}(Y \setminus S)$.

\begin{claim}
  $G[A\cup B]$ is $\frac{1}{25}\cdot \delta \phi \, \big(\deg_{E(B,V\setminus B)}|_{A \cup B} + \v{\pi}|_A \big)$-expanding.
\end{claim}
\begin{proof}
  Let $S \subseteq A \cup B$ be a subset with $\bm{\sigma}(S) \le \bm{\sigma}(Y \setminus S)$.
  We can assume without loss of generality that the condition $\v{\pi}(S \cap A) \leq \v{\pi}(A \setminus S)$ holds for $S$. If this assumption does not hold for the chosen $S$, apply the following arguments to the complement $Y \setminus S$ instead, for which the condition then has to hold. This will give the bound $\capc(S, Y \setminus S) \geq \frac{1}{25} \delta\phi \cdot \bm{\sigma}(Y \setminus S)$, which also implies the desired bound for $S$, since $\bm{\sigma}(S) \le \bm{\sigma}(Y \setminus S)$.

  The proof proceeds in two parts: First we argue that we can use the flows of the fair cuts to attribute the capacity of the cut edges in $E(B, V \setminus B)$ to the units of $\v{\pi}(S \cap A)$ with little overhead. This is the main part of the proof. Then, in a second step, we use the expansion of $G[A]$ from the claim above to establish the desired result.
  
  \textbf{Part 1.} Observe that we can categorize the edges leaving $S \cap B$ to obtain
  \begin{equation}
    \label{eq:deg-bvb}
    \deg_{E(B, V \setminus B)}(S) \leq \capc(S \cap B, U) + \capc(S \cap B, V \setminus C) + 2 \capc(S \cap B, A).
  \end{equation}
  We can bound the first two summands in terms of the third.
  As $X_2$ is a $2$-fair $(\v{s}_2, \v{t}_2)$-cut in $G[X_1]$, by \cref{prop:fc-cap} of \cref{claim:fair-cut-ineq}  we have for $S \cap B \subseteq B = X_2$ that

  \begin{equation*}
    \capc(S \cap B, U) + \v{t}_2(S \cap B) \leq 2 \big( \v{s}_2(S \cap B) + \capc(S \cap B, B \setminus S) \big),
  \end{equation*}
  which, using the definitions of $\v{t}_2$ and $\v{s}_2$, gives
  \begin{equation}\label{eq:twt-wd}\textstyle
    \capc(S \cap B, U) + \frac{1}{2}\phi\capc(S \cap B, V \setminus C) \leq 2 \capc(S \cap B, A) + 2 \capc(S \cap B, B \setminus S).
  \end{equation}
  Note that $\capc(S \cap B, A) = \capc(S \cap B, S \cap A) + \capc(S \cap B, A \setminus S)$, and $\capc(S \cap B, A \setminus S)$ can be upper-bounded together with $\capc(S \cap B, B \setminus S)$ by  $\capc(S \cap B, Y \setminus S)$.
  Thus,
  \begin{equation}\label{eq:bound-together}
    \capc(S \cap B,A) + \capc(S \cap B, B \setminus S) \le \capc(S \cap B, S \cap A) + \capc(S \cap B, Y \setminus S).
  \end{equation}
  Scaling \cref{eq:twt-wd} by a factor $2/\phi \geq 1$ and combining it with \cref{eq:deg-bvb} thus gives
  \begin{equation*}\label{eq:twt-bvb-2} 
    \begin{aligned}
    \deg_{E(B, V \setminus B)}(S) &\leq \textstyle (2 + \frac{4}{\phi}) \capc(S \cap B, A) + \frac{4}{\phi} \capc(S \cap B, B \setminus S)\\
    & \leq \textstyle (2 + \frac{4}{\phi}) \capc(S \cap B, S \cap A) + (2 + \frac{4}{\phi}) \capc(S \cap B, Y \setminus S) \\
    & \leq 5/\phi \cdot \big( \capc(S \cap B, S \cap A) + \capc(S \cap B, Y \setminus S) \big),
    \end{aligned}
  \end{equation*}
  using \cref{eq:bound-together} in the second inequality and for the last inequality we use that $\phi \leq 1/2$, i.e, $2 \le 1/\phi$, so $2 + \frac{4}{\phi} \leq 5/\phi$.

  Next, we attribute $\capc(S \cap B, S \cap A)$ to the units of $\bm{\pi}(S \cap A)$ following a similar argument as above. 
  Since $X_0$ is a 2-fair ($\v{s}_1, \v{t}_1$)-cut in $G[C \setminus R]$ with $\v{t}_1 = 1/5\cdot\delta \phi\bm{\pi}$ and $A= (C \setminus R) \setminus X_0$, by \cref{claim:fair-cut-ineq}, \cref{prop:fc-cap-t} we have for $S \cap A$ that
  \begin{equation*}\textstyle
    \begin{aligned}
        \capc(S \cap A, S \cap B) &\le \capc(S \cap A, X_0) +  \capc(S \cap A, R) \\
          &= \capc(S \cap A,X_0) + \v{s_1}(S \cap A) \\
          & \leq \textstyle  2(\frac{1}{5} \delta\phi \cdot \bm{\pi}(S \cap A) + \capc(S \cap A, A \setminus S)),
    \end{aligned}
  \end{equation*} 
  where we use that $(C \setminus R) \setminus X_0 = A$ and, thus, $(C \setminus R) \setminus X_0  \setminus (A \cap S) = A \setminus S.$
  Now we can again upper-bound disjoint sets of edges leaving $S$ together, by observing that $2\capc(S \cap A, A \setminus S) + \capc(S \cap B, Y \setminus S) \leq 2\capc(S, Y \setminus S)$.
  Note that $(2/5)\delta\phi (5/\phi) = 2\delta$ and $\delta \leq 1$, so $1 + 2\delta \leq 3$ and
  \begin{equation*}
    \begin{aligned}
      \deg_{E(B, V \setminus B)}(S) + \bm{\pi}(S \cap A) & \leq  \textstyle 5/\phi \cdot \big( \frac{2}{5} \delta\phi \cdot \bm{\pi}(S \cap A) + 2\capc(S, Y \setminus S) \big) + \bm{\pi}(S \cap A)  \\
      &\leq \textstyle 3 \cdot \bm{\pi}(S \cap A) + \frac{10}{\phi} \capc(S, Y \setminus S).
    \end{aligned}
  \end{equation*}

  \textbf{Part 2.} Finally, for the second step, since $G[A]$ is $1/5 \cdot \delta\phi\v{\pi}|_A$-expanding by the previous claim, we have $\v{\pi}(S \cap A) \leq \frac{5}{\delta\phi}\capc(S\cap A, A \setminus S)$, since $S \cap A$ is the smaller side in terms of $\v{\pi}$ volume by assumption. Therefore, we can conclude that
  \begin{equation}
    \begin{aligned}
      \deg_{E(B, V \setminus B)}(S) + \v{\pi}(S \cap A) & \leq \textstyle 3 \cdot \frac{5}{\delta\phi} \cdot \capc(S\cap A, A \setminus S) + \frac{10}{\phi} \capc(S, Y \setminus S) \\
      & \leq \textstyle 25\frac{1}{\delta\phi} \capc(S, Y \setminus S) ,
    \end{aligned}
  \end{equation}
  where the last step follows since $\delta \leq 1$.
\end{proof}

With the above claims showing the correctness of the algorithm, we conclude the proof of \cref{claim:twt} by noting that the procedure can clearly be implemented by two executions of \algFC{} with an overhead of $O(m)$ for setting up the flow instances.

%%%%%%%%%%%%%%%%%%%%%%%%%%%%%%%%%%%%%%%%%%%%%%%%%%%%%%%%
%%%%%%%%%%%%% General Cut Matching Game %%%%%%%%%%%%%%%%
%%%%%%%%%%%%%%%%%%%%%%%%%%%%%%%%%%%%%%%%%%%%%%%%%%%%%%%%
\section{A General Cut Matching Game}\label{sec:cmgame}

In this section we describe a general Sparsest Cut Oracle and provide a proof
for \cref{lemma:spx} via a so-called \emph{cut matching game}
(described below) with an extension of the cut player by Orecchia et
al.~\cite{OSVV08} for a so-called non-stop cut matching game. 
Crucially, even if the sparse cut is very imbalanced, we can guarantee expansion of the larger side of the cut.
The first step is to prove the ability for a general cut player to produce a well
expanding graph against any matching player. Then we show that with a suitable
matching player for a given input graph with non-negative, integral vertex weights, we either find a sparse cut or ensure its expansion relative to those weights.

The cut matching game is a technique introduced by Khandekar, Rao and Vazirani~\cite{KRV06} to compute an approximately sparsest cut using a polylogarithmic number of maximum flows. Given some input graph $G=(V,E)$, the rough idea is to repeatedly find a bisection of the graph (cut player) and then match node pairs on opposing sides by computing a maximum flow (matching player). 
Conceptually, the algorithm maintains an (initially empty) graph $H$ over the same vertex set as $G$. For each matched pair there is an edge added to $H$ and, crucially, all edges of $H$ can be embedded in $G$ with small congestion.
Khandekar et al. show that $O(\log^2 n)$ rounds suffice to ensure that either (1) one of the bisections exposes a sparse cut in $G$, or (2) the graph $H$ consisting of the union of all matching edges constitutes a $\Theta(1)$-expander.

Orecchia, Schulman, Vazirani and Vishnoi~\cite{OSVV08} later gave an improved cut player algorithm, which ensures that $H$ is a $\Omega(\log n)$-expander after $O(\log^2 n)$ rounds for any matching player. Our cut player (\cref{sec:cut-player}) builds on this algorithm.

\paragraph{Handling Deletions and Vertex Weights.}
For the task of designing an improved Sparsest Cut Oracle (cf. \cref{sec:impr-sc}) it is crucial that no very \enquote{unbalanced} cut is returned if a more \enquote{balanced} one exists, as this is necessary to keep recursion depths small. This is addressed by so-called \emph{non-stop} 
versions of the cut-matching game (see e.g. \cite{RST14,SW19,ADK22}), by maintaining a subset of \emph{active vertices} $A$.
If the matching player ever fails to match all pairs across a given bisection with low congestion, this implies the existence of a sparse cut $S$ in $G$. In this case, the matching player can decide to either terminate the game early (for example if the sparse cut $S$ is considered sufficiently balanced) or it may delete the vertices in $S$ by removing them from the set of active vertices $A$ and keep playing on the remaining vertices. Handling these deletions \emph{without restarting the game} is challenging but essential for ensuring a near-linear running time. 

While it is originally stated to operate on the nodes of a given graph, our observation for handling vertex weights is that the cut-matching game can be understood to run directly on the internal graph $H$. In particular, the cut player can operate completely independently from $G$ and only the matching player is given access to $G$. Given an integral, nonnegative vertex weight function $\bm{\pi}$ alongside the graph $G$, our main insight is that we can consider each unit of $\bm{\pi}$-volume as participating in the cut-matching game separately. For each vertex $v$, we let $\bm{\pi}(v)$ (so-called) \emph{units} participate in the game instead of $v$ itself and keep a fixed mapping, associating each unit with its vertex in $G$. 
Via this mapping, we can ensure that any expansion achieved by the cut-matching game will be with respect to the vertex weights $\bm{\pi}$.
Note that this requires $\bm{\pi}$ to be integral and nonnegative, but does allow for vertices to have a $\bm{\pi}$-value of $0$.

To make this work we introduce a new matching player that maintains two key invariants. First, they ensure that the units corresponding to a vertex in $G$ are either all active in the cut-matching game, or all inactive.
Consequently, there are no vertices in the graph that have both active and inactive units at the same time. Hence, at all times there is a subset $R$ of \enquote{inactive} vertices from the graph $G$, which the matching player maintains and can output whenever queried for it.
Secondly, it maintains an embedding of all matching edges into $G$ with the crucial property that \emph{if $H$ is $\1_A$-expanding with quality $r$ and the embedding of the matching edges causes congestion $1/\phi$ in $G$, then $G$ is $r\phi\cdot \bm{\pi}|_{V \setminus R}$-expanding}, where $A$ is the set of active units and $V \setminus R$ is the set of vertices in $G$ with active units.
Upon termination of the game, the cut player ensures that $r$ is sufficiently large, giving the desired $\bm{\pi}$-expansion of $G$.

By adjusting the congestion allowed for embedding a single round of matchings, one can trade off between approximating the sparsity of the returned cut and the final expansion quality. A feature of our matching player definition is that we make this trade-off explicit by parametrizing the algorithm with a value $c \geq 1$. To implement \algSC{}, we require the returned cut to have a sparsity of at most $\phi$ and accept the slightly worse expansion guarantee of $\phi/q$ with quality parameter $q \geq 1$.
For example, when computing a $\phi$-expander decomposition, one usually prefers to ensure an expansion quality of $\phi$ and accepts a cut of sparsity $q\phi$.
With the parametrization, our matching player can be seemlessly reused in this setting.

Recall that the cut player can be understood to work independently from the input graph $G$. We introduce a new cut player that (1) operates on $k$ units for any integer $k$, (2) allows for the deletion of units during the game and (3) achieves an expansion of $\Omega(\log{k})$ on the active units against any matching player.
For this, we generalize the result of \cite{ADK22}, which also builds on the cut player by~\cite{OSVV08}, in two ways. First, we specify the cut player independent of the input graph $G$ and use the more versatile concept of units instead. Secondly, we strengthen the achieved expansion to our notion of $\1|_A$-expansion instead of the property that $A$ is a \enquote{near-expander} in $G$ (as used in \cite{SW19, ADK22}).

\paragraph{General Definition.}
Fix any positive integer $k$ and let $U := [k]$ be the set of units participating in the cut-matching game.
For two subsets $X_1, X_2 \subseteq U$, we define a \emph{matching between} $X_1$ and $X_2$ as a set of pairs from $X_1 \times X_2$ where each element from $X_1$ is contained
in exactly one pair and each element of $X_2$ is contained in at most one pair. Note that $X_1$ and $X_2$ may intersect, so self-loops are
allowed. A family of matchings can be interpreted as an edge set of the graph with node set $U$, where there is an edge between any pair in the matching.

With this, we can give the following very general definition of a cut matching
game that allows for unit-deletions. It gives a clean interface for the cut and matching player, and states a set of minimal requirements on these two algorithms for producing an expanding graph $H$. 
In order for this expansion of $H$ to be useful for finding sparse cuts or verifying expansion of some given input graph, the matching player is usually extended with additional capabilities. In particular, it usually maintains an embedding of $H$ into the input graph (see \cref{sec:matching-player}). 

\begin{itemize}
\item A \emph{matching player} $\mathcal{M}$ is an algorithm that, given a set
$A \subseteq U$ as well as a pair of subsets $A^{\ell}, A^r \subseteq A$ outputs a
set $S \subseteq A$ and a matching $M$ between $A^{\ell} \setminus S$ and
$A^r \setminus S$. We use $\mathcal{M}(A, A^{\ell},A^r)$ to denote this output pair.

\item A \emph{cut player} $\mathcal{C}$ is an algorithm that, given a subset
$A \subseteq U$ as well as a family of matchings $Y$ outputs a pair of subsets
$A^{\ell}, A^r \subseteq A$. We use $\mathcal{C}(A, Y)$ to denote this output pair.

\item The \emph{cut-matching-game} $\mathcal{G}(\mathcal{C}, \mathcal{M}, T)$
is an algorithm that alternates between a cut player and a matching player for $T$
rounds. More precisely, it initializes a subset $A_0 \gets U$ and an empty
ordered set of matchings $Y_0 \gets \emptyset$. Then, it proceeds for $T$
rounds: In round $t+1$, it computes
$(S_{t+1}, M_{t+1}) \gets \mathcal{M}\big(A_t, \, \mathcal{C}(A_t, Y_t)\big)$,
then it updates $A_{t+1} \gets A_t \setminus S_{t+1}$ and
$Y_{t+1} \gets Y_t \cup \{M_{t+1}\}$. Finally, it returns the pair
$(A_T, Y_T)$.
\end{itemize}

In this view of the cut-matching framework, a game always takes exactly $T$
rounds. Note also, that this game is defined independently from any input
graph. The goal is to design cut players that can ensure that when the game is completed against any
matching player, its output $A$ is well-expanding in the graph $H=(U, Y)$.

\paragraph{Our Cut and Matching Players.}

With this definition, we offer a fresh and versatile perspective on the cut-matching game by cleanly separating the cut and the matching player into independent, modular blocks. 
We prove the following theorem about the cut player $\mathcal{C}^X$, which is
an extension of the cut player of~\cite{OSVV08} with the additional ability to handle the deletion of units during the game. 

\begin{lemma}[Cut Player]\label{theorem:new-cut-player}
For any integer $k$, there is a cut player $\mathcal{C}^X$ such that for some $T = O(\log^2{k})$ the cut-matching-game $\mathcal{G}(\mathcal{C}^X, \mathcal{M}, T)$ outputs a pair $(A, Y)$ such that if $|A| \geq (1 - \frac{1}{2\log{k}}) \cdot k$, then the graph $(U, Y)$ is $\1|_A$-expanding with quality $\Omega(\log k)$ with high probability against any matching player $\mathcal{M}$.
\end{lemma}

In \cref{sec:matching-player} we present and analyze a new matching player algorithm. Given an input graph $G$ with integral vertex weights $\bm{\pi}$, it allows the execution of the cut-matching game on the individual units of the $\bm{\pi}$ volume, while simultaneously maintaining an embedding of the matching edges into $G$. This is crucial in our implementation of \algSC{}, which we present in \cref{sec:spx}.

In \cref{sec:spx-prl} we further show that the cut matching game using our new cut and matching players can be parallelized efficiently.

\subsection{\texorpdfstring{Algorithm \algSC{}}{Algorithm SparsestCutApx}}
\label{sec:spx}

Equipped with the new cut and matching players from \cref{sec:cut-player,sec:matching-player}, respectively, we are ready to give our algorithm for \algSC{} for integral, nonnegative weights $\bm{\pi}$.
It is a simple extension of the basic cut-matching game on $\bm{\pi}(V)$ units using these two players with an additional check whether too many units (and thus too much $\bm{\pi}$-volume) have been deleted. In that case we can terminate the game early and return a balanced sparse cut without guaranteeing expansion on either side.

\bigskip

Algorithm \algSC$(G=(V,E),\v{\pi},\phi)$:
\begin{enumerate}
  \item Fix the parameter $c = \lceil 10 / \phi \rceil$ and let $k = \bm{\pi}(V)$ be the number of units for the cut-matching game. Initialize $\mathcal{M}_c^X$ using $G$ and $\bm{\pi}$. Choose $T = O(\log^2{k})$ according to \cref{claim:num-iterations}. 

  \item Start the cut matching game $\mathcal{G}(\mathcal{C}^X, \mathcal{M}_c^X, T)$ with $k$ units using the new cut player $\mathcal{C}^X$ and matching player $\mathcal{M}_c^X$. For this, initialize $t = 0$, $A_0 = \{1, \ldots, k\}$ and $Y_0 = \emptyset$. Then execute the following steps while $t \leq T$: \label[step]{step:mp:cmgame}
  \begin{enumerate}
    \item Compute an update as $(D_{t+1}, M_{t+1}) = \mathcal{M}_c^X(A_t, \, \mathcal{C}^X(A_t, Y_t))$ and then set $A_{t+1} = A_t \setminus D_{t+1}$ and $Y_{t+1} = Y_t \cup \{ M_{t+1} \}$.\label[step]{step:spx:cmgame}
    \item \textbf{If} $|A_{t+1}| < (1 - \frac{1}{2\log{k}}) \cdot k$, stop the cut matching game and go to \cref{step:spx:return}. \\
    \textbf{Otherwise} continue with the next iteration $t+1$.
    \label[step]{step:spx:stop}
  \end{enumerate}

  \item Upon termination of the cut matching game, query $\mathcal{M}_c^X$ for the set of inactive vertices $R \subseteq V$. If $\bm{\pi}(R) \leq \bm{\pi}(V \setminus R)$, return $R$, otherwise return $V \setminus R$.
  
  \label[step]{step:spx:return}
\end{enumerate}

\noindent
This algorithm gives our implementation of \algSC{} with the parameters $\q = O(\log{\bm{\pi}(V)})$ and $\bal =  1/(2 \log{\bm{\pi}(V)})$. In \cref{sec:spx-analysis}, we prove these guarantees and establish the correctness of \cref{lemma:spx} by building on the following analysis of the cut matching game using our new cut and matching players.

\subsection{The Cut Player}\label{sec:cut-player}
The goal of the cut player is to assign each node $i$ a value $u_i$ based on
all previous matchings such that well-connected vertices have similar values.
After sorting these values, a sweep cut over the $u_i$ values exposes cuts in the graph across which the expansion is not yet good enough. By outputting this cut in the
cut-matching game, the matching player is forced to increase the expansion.

Similar to \cite{KRV06,OSVV08, ADK22}, the assignment of the $u_i$ values is
based on a random walk that distributes the initial random charges through the
graph using the information from the previous matchings. We use the following standard lemma to perform the sweep cut, once the $u_i$ values are found. Finding good $u_i$
values is the main challenge in this approach.

\begin{lemma}[Lemma 3.3 in~\cite{RST14}]\label{lemma:RST14}
Given is a set $A \subseteq U$ and values $\textbf{u} \in \mathbb{R}^A$ such that $\sum_i \v{u}_i = 0$. In time $O(|A| \log{|A|})$, we can compute two sets $A^{\ell}, A^r \subset A$ and a separation value $\eta$ such that:
\begin{enumerate}
  \item $\eta$ separates $A^{\ell}$ from $A^r$, i.e., either we have that $\operatorname{max}_{i \in A^{\ell}} \textbf{u}_i \leq \eta \leq \operatorname{min}_{i \in A^r} \textbf{u}_j$, or $\operatorname{max}_{i \in A^{\ell}} \textbf{u}_i \geq \eta \geq \operatorname{min}_{j \in A^r}\textbf{u}_j$ \label[property]{prop:rst14:separate}
  \item $|A^r| \geq |A|/2$ and $|A^{\ell}| \leq |A|/8$ \label[property]{prop:rst14:size}
  \item for every source node $i \in A^{\ell}$: $|\textbf{u}_i - \eta|^2 \geq \frac{1}{9} |\textbf{u}_i|^2$ \label[property]{prop:rst14:source}
  \item $\sum_{i \in A^{\ell}} |\textbf{u}_i|^2 \geq \frac{1}{80} \sum_{i \in A} |\textbf{u}_i|^2$ \label[property]{prop:rst14:sumsource}
\end{enumerate}
\end{lemma}

The algorithm \algSC{}, which uses the new cut player, maintains the following invariant to ensure that at any point in the cut matching game, not too many units are deleted. This is important for the analysis.

\begin{invariant}\label{assumption:few-deletions}
  In any round $t$, $|U \setminus A_t | \leq \frac{1}{2\log{k}} \cdot k$.
\end{invariant}

\subsubsection{The Algorithm}\label{sec:cp-alg}

We define and analyze the cut player $\mathcal{C}^X$ in the context of the
corresponding cut-matching game. Fix some $k$ and let $\mathcal{M}$ be an
arbitrary matching player. Let $t < T$ be some round of
$\mathcal{G}(\mathcal{C}^X, \mathcal{M}, T)$ where $T = O(\log^2{k})$ and $\delta = \Theta(\log{k})$ are parameters that are fixed later. Note that $\delta$ is chosen as a power of $2$.

\paragraph{\mathversion{bold}Set $A_t$, vectors $\1_t$ and $d_t$, and matrix $I_t$.\mathversion{normal}} 
Let $A_t$
denote the set of active vertices \emph{at the start of} round $t+1$. Note that $A_0 = U$ and $A_{i+1} \subseteq A_i$ for all $i$. The vector
$\1_t := \1|_{A_t}$ is the ($k$ dimensional) indicator vector for $A_t$, where
the $i$-th entry is $1$ if $i \in A_t$ and $0$ otherwise. When normalized to
unit length, we write $d_t:=1 / \sqrt{|A_t|} \cdot \1_t$. The corresponding
restricted identity matrix is $I_t:=\diag(\1_t)$, which is the identity on
$A_t$ and $0$ elsewhere.

In the following we define two types of matrices and introduce the following
notation: (a) Matrices where all rows and columns $i$ with $i \not\in A_t$ are
all zero. We call them \emph{\zero} matrices. (b) Matrices where all rows and
columns $i$ with $i \not\in A_t$ are all zero except for the diagonal entries
which are 1. Thus the part of the matrix formed by rows $i$ and columns $j$
with $i,j \not\in A_t$ forms the identity matrix. We call them \emph{\id}
matrices. (c) We call the matrix spanned by the rows $i$ and columns $j$ with
$i,j \in A_t$ the \emph{$A_t \times A_t$ block} of the matrix.

Intuitively, multiplication of a vector $\v u$ with a \zero matrix
\enquote{deletes} the contribution to $\v u$ by the vertices in
$U \setminus A_t$. Multiplication of a \id matrix keeps the $\v u$ value
of vertices not in $A_t$ unchanged. If a \zero or \id matrix is additionally
doubly stochastic on the block $A_t \times A_t$, then the contribution to
$\v{u}$ by the vertices in $A_t$ is ``spread'' (in some way) over the
vertices in $A_t$, without changing the total contribution to $\v{u}$ of
the vertices in $A_t$.

Also recall that for every doubly stochastic matrix, the largest eigenvalue
is 1 and corresponds to the eigenvector $\1_V$.

\paragraph{\mathversion{bold}Matrix $M_t$.\mathversion{normal}} We consider a matching $M_t$ as a matrix, where
$(M_t)_{uv} = 1$ iff $u$ is matched to $v$ in this round. Note that self-loops
are allowed. Also, some nodes may be unmatched in this round, but we add
self-loops to all unmatched nodes. Note that $M_t$ is symmetric,
doubly-stochastic on the block $A_t \times A_t$ (i.e.,
$\1_t M_t = M_t \1_t = \1_t$), and a \zero matrix.

\paragraph{\mathversion{bold}Matrix $N_t$.\mathversion{normal}} We are now ready to define the relevant matrices for our
random walk. The \emph{slowed down} matching matrix is
$N_t := I - \frac{1}{ \delta}(I_t - M_t).$ Note that $N_t$
is an \id matrix and on the block $A_t \times A_t$ it is equal to
$I_t - \frac{1}{\delta}(I_t - M_t)$.
It is doubly stochastic and symmetric, and also doubly-stochastic on the block
$A_t \times A_t$.

\paragraph{\mathversion{bold}The flow matrix $F_t$.\mathversion{normal}}  The \emph{flow matrix} $F_t$ is defined recursively as
\begin{equation*}
  F_t := N_t F_{t-1} N_t \quad \text{with} \quad F_0 = I.
\end{equation*}
By induction on $t$ it follows that, as the product of doubly stochastic
matrices, $F_t$ is also doubly stochastic. It follows that all entries in $F_t$
are non-negative. Note that it is neither a \zero matrix nor an \id matrix.

With the eigenvalue properties of doubly stochastic matrices the following claim follows.
\begin{claim} For all $t \in \mathbb{N}$ it holds that
    $\1_V^T F_t = F_t \1_V  = \1_V$.
\end{claim}

Under our \cref{assumption:few-deletions}, that in any round not too
many nodes have been deleted, we can derive the following claim establishing
that the remaining flow volume is still very large in every round.
  \begin{claim}\label{claim:volume-preserved}
    $\1_t^T F_t \1_t \geq (1-1/\log{k}) k$ .
  \end{claim}
  \begin{proof}
    Let $Z := U \setminus A_t$, then
    \begin{equation*}
      \begin{aligned}
        \1_t^T F \1_t & = \1^T F \1 - \1^T F \1_{Z} - \1_Z^T F \1 + \1_{Z}^T F \1_{Z} \\
                      & \geq \1^T \1 - \1^T \1_{Z} - \1^T \1_{Z} \\
                      & = k - 2 |Z| \geq (1 - 1/\log{k})k .
      \end{aligned}
    \end{equation*}
    The first two steps used that $F$ is symmetric and doubly stochastic and
    for the last inequality we crucially rely on
    \cref{assumption:few-deletions}.
  \end{proof}

  \paragraph{\mathversion{bold}Matrix $Q_t$.\mathversion{normal}} We further
  define $Q_t := d_t d_t^T$ as an averaging matrix, where every entry
  ($Q_t)_{i,j}$ is $1/|A_t|$ if $i \in A_t$ and $j \in A_t$ and every other
  entry is $0$. Thus $Q_t$ is a \zero matrix and doubly stochastic on the block
  $A_t \times A_t$. Note that $Q_t$ is a projection matrix, i.e.~$Q_t^2 = Q_t$.

  \paragraph{\mathversion{bold}Matrix $P_t$.\mathversion{normal}} We define the matrix $P_t := I_t - Q_t$, which is also symmetric and a \zero matrix. Further, in $P_t$ all rows and columns sum to $0$ and it is a projection matrix,
  as we will show in \cref{claim:properties}\labelcref{claim:props:p2}, i.e.~$P_t^2 = P_t$.

\paragraph{\mathversion{bold}The walk matrix $W_t$.\mathversion{normal}} Finally, we define the \emph{walk matrix}
\begin{equation*}
  W_t := (P_t F_t P_t)^\delta.
\end{equation*}
As it is the result of a multiplication with a \zero matrix, it is a \zero matrix.
To summarize, all the above matrices, i.e., $I$, $I_t$, $M_t$, $N_t$, $F_t$,
$Q_t$, $P_t$ and $W_t$, are symmetric.

\paragraph{\mathversion{bold}Definition of the cut player $\mathcal{C}^X$.\mathversion{normal}}
In round $t+1$, upon receiving an input of $A_t \subseteq U$ and
$Y_t =(M_1, M_2, \ldots, M_t)$, the cut player $\mathcal{C}^X$ executes the
following steps:

\begin{enumerate}
  \item Define $W_t$ as above using the given matchings $M_1, M_2, \ldots, M_t$.
  \item Pick a random unit vector $\v{r}$ and compute $\v{u} := W_t \v{r}$.
  \item Use \cref{lemma:RST14} to compute and output $A^{\ell}, A^r \subseteq A$ from $\v{u}$.\label[step]{step:cp:sweep}
\end{enumerate} 
Regarding the application of \cref{lemma:RST14} in \cref{step:cp:sweep}, we need to ensure that $\sum_i \bm{u}_i = 0$, i.e., $\bm{u}^T \1 = 0$. For this, observe that $\bm{u}^T \1 = (W_t \v{r})^T \1 = \v{r}^T(W_t \1) = \v{r}^T (P_t F_t P_t)^\delta \1 = 0$, where the last step follows since $P_t \1 = 0$, which implies $(P_t F_tP_t)^\delta \1 = 0$.

\paragraph{The Potential Function.}
% We track the progress of the cut matching game with a standard potential
% function that measures the difference of the column vectors of $W_t$ to to
% their average. Let $\v{w}_i^{(t)}$ be the $i$-th column vector of $W_t$
% and define $\mu_t := \frac{1}{|A_t|} \sum_{i \in A_t} \v{w}_i^{(t)}$.
% Then we define the potential

In order to prove correctness of the cut-player algorithm, we introduce a potential function that tracks
the converge of the random walk. Ideally, one would like to minimize the
potential function that is the sum of all eigenvalues except the largest, as is
done in~\cite{OSVV08}. The challenge is finding a proper
generalization of the potential function that can be shown to decrease at every
step even when vertices are removed from $A_t$. We define

\begin{equation}\label{eq:potential}
  \varphi(t) := \Tr[W_t^2] = \Tr[(P_t F_t P_t)^{2\delta}] \enspace .
\end{equation}

Note that
$\varphi(t) = \sum_{i \in A_t} (W_t^2)_{i,i} = \sum_{i\in A_t} \lVert \v{w}_i
\rVert^2$, where $\v{w}_i$ is the $i$-th column of $W_t$. While at first glance
this potential function may seem very different from the potential used by
Orecchia et al.~\cite{OSVV08} it actually simplifies to 
$\varphi_{\text{no-del}}(t) = \Tr[F_t^{2\delta}] - 1$ if there are no
deletions. The latter is exactly the
function used in~\cite{OSVV08}.

We first analyze how this potential is useful in proving expansion of the flow
matrix $F_t$, then we show that it actually decreases significantly in every
round.

\subsubsection{Expansion}
In this section, we establish the crucial relation between the potential
$\varphi(t)$ and the edge expansion of the flow matrix $F_t$ in any round $t$.
This relation holds in any round, but the bound only becomes meaningful if the
potential is very small.

Let $t$ be some fixed round. We may omit the subscript $t$, as all matrices in
this section are understood to be from this round. We consider the graph $H_F$ whose
weighted adjacency matrix is the current flow-matrix $F$. We write $\volF$ and
$\capc_F$ to denote the volume and capacity in this graph, respectively. For
any set $Z \subseteq U$, we denote by $\1_Z$ the indicator vector for the set
$Z$.

The goal of this section is to prove the following Lemma, similar to Lemma 5.9
in~\cite{ADK22}. Using a slightly refined analysis, we can show expansion of all subsets of $U$,
not just those within $A$.

\begin{lemma}\label{lemma:expansion}
  Let $S \subseteq U$ and denote $s := |S \cap A|$ and $a := |A|$. Then
\begin{equation*}
\textstyle
\capc_{F}(S, U \setminus S) \geq \big(1 - \frac{3}{2}s/a - \varphi(t)^{\frac{1}{2\delta}}\big) \cdot s \enspace .
\end{equation*}
\end{lemma}

This lemma directly gives a bound on the $\1_A$-expansion of $H_F$. This can be
thought of a variant of edge expansion, where only nodes in $A$ contribute to
the volume.
We get the following lower bound, which improves as the potential decreases.

\begin{corollary}\label{cor:H-exp}
  The graph $H_F$ is $\1_A$-expanding with quality $1/4 - \varphi(t)^\frac{1}{2\delta}$.
\end{corollary}
\begin{proof}
Let $S \subseteq U$, such that $|S \cap A| \leq |(U\setminus S) \cap A|$, i.e.,
$S$ is the smaller side in terms of the $\1_A$ volume. Note that this implies
$|S \cap A|/|A| \leq 1/2$. By \cref{lemma:expansion}, it follows that
$\capc_F(S, U\setminus S)\, / \, |S \cap A| \ge 1 - 3/4 -
\varphi(t)^\frac{1}{2\delta}$, as desired.
\end{proof}

The bound is not meaningful when the potential is large, but becomes relevant
when the potential is close to 0. In the next section, we show that after
$O(\log^2{k})$ rounds, the cut player can ensure that the potential is smaller
than $1/k$. By choosing an appropriate $\delta = \Theta(\log{k})$ such that
$(1/k)^{\frac{1}{2\delta}} \leq 1/20$, we can ensure that $H_F$ is $\1_A$
expanding with a constant quality of $1/5$.

\begin{proof}[Proof of \cref{lemma:expansion}]

  Consider the matrix $X := PFP$ at the core of the potential function. Expanding the terms, we get
  \begin{equation*}
    X = F - FQ - QF + QFQ.
  \end{equation*}
  Thus, we can use $X$ to obtain an expression for the current flow matrix
  \begin{equation}\label{eq:f-x-q}
    F = X + FQ + QF - QFQ \enspace .
  \end{equation}

  Recall that, by design, we have $F\1_V = \1_V$ and for any two sets
  $Y,Z \subseteq U$, we have $\capc_F(Y,Z) = \1_Y^T F \1_Z$. Thus, for every
  set $Z \subseteq U$,
  \begin{equation}\label{eq:vol-s}
    \textstyle
    \volF(Z) = \capc_F(Z, U) = \1_Z^T F \1_V = \1_Z^T \1_V = |Z| \enspace.
  \end{equation}

  Fix any set $S \subseteq U$.
  Regarding the weight of edges crossing the cut $(S, U\setminus S)$, we have
  \begin{equation}\label{eq:cap-S-VS}
      \capc_F(S, U\setminus S) = \volF(S) - \capc_F(S, S) = |S| - \capc_F(S,S) ,
  \end{equation}
  using \cref{eq:vol-s} for the last equality. The main part of the proof
  is the derivation of an upper bound for $\capc_F(S,S)$, which is the total
  capacity of edges with both endpoints inside $S$. With the above equation,
  this will directly give a lower bound for $\capc_F(S, U\setminus S)$. For
  this, we make use of \cref{eq:f-x-q} and obtain
  \begin{equation*}\label{eq:four-summands}
  \capc_F(S, S) = \1_S ^T F \1_S = \underbrace{\1_S ^T X \1_S}_{=: x_1} + \underbrace{\1_S ^T F Q \1_S}_{=: x_2} + \underbrace{\1_S ^T Q F \1_S}_{=: x_3} - \underbrace{\1_S ^T Q F Q \1_S}_{=: x_4} \enspace ,
  \end{equation*}
  where we bound each summand separately. Recall that $s := |S \cap A|$ and
  observe that $\1_A ^T \1_S = s$ implies that
  $Q\1_S = d d^T \1_S = 1/a \cdot \1_A \1_A^T \1_S = s/a \cdot \1_A$, which
  also holds for the transpose,
  $\1_S ^T Q =1/a \cdot \1_S^T \1_A \1_A^T =s/a \cdot \1_A^T$. By transposition
  and since $Q$ and $F$ are symmetric, we get that $x_2 = x_3$ and further
  \begin{equation*}
    x_2 = x_3 = \1_S ^T F Q \1_S = s/a \cdot \1_S^T F \1_A \leq s/a \cdot s \enspace .
  \end{equation*}
  For the last inequality, note that since $F$ has entries between $0$ and $1$,
  it follows that $\1_S^T F \1_A \leq \1_S^T \1_A = s$. For $x_4$, we obtain
  \begin{equation*}
  x_4 = \1_S ^T Q F Q \1_S = s^2/a^2 \cdot \1_A^T F \1_A \geq s^2/a^2 \cdot (1-1/\log{k}) k \geq \textstyle 
  % \frac{1}{2}sn/a^2 \cdot s \geq 
  \frac{1}{2}s/a \cdot s ,
  \end{equation*}
  using \cref{claim:volume-preserved} for the lower bound on $\1_A^T F \1_A$, and $k \geq 4$.
  Lastly, we bound $x_1$. Note that $X$ is a zero-$A$ matrix and hence
  $\1_S^T X \1_S = \1_{S \cap A}^T X \1_{S \cap A}$. We derive the bound using
  the largest eigenvalue of $X$ from the following fact.

  \begin{fact}
    Let $X \in \mathbb{R}^{k \times k}$ be a symmetric matrix with largest eigenvalue $\lambda_k[X]$. Then  
    $
      \lambda_k[X] = \operatorname{max}_{v} \frac{v^T X v}{v^T v}
    $ .
  \end{fact}

  In particular, the above fact and the observation that $\1_{S \cap A}^T \1_{S \cap A}= s$ gives 
  \begin{equation*}
  x_1 = \1_S^T X \1_S = \1_{S \cap A}^T X \1_{S \cap A} \leq \lambda_n[X] \cdot s \leq \varphi(t)^{\frac{1}{2\delta}} \cdot s \enspace .
  \end{equation*}
  For the last inequality, observe that the eigenvalues of $X$ are directly
  related to the current potential, since
  $\varphi(t) = \Tr[X^{2\delta}] = \sum_{i=1}^k \big( \lambda_i [X]
  \big)^{2\delta} \geq \big( \lambda_k[X]\big)^{2\delta}$.

  Combining the bounds for all summands $x_i$, we thus have
  \begin{equation*}
    \textstyle
    \capc_F(S, S) = x_1 + x_2 + x_3 - x_4 
    \leq \big(\frac{3}{2} s/a  + \varphi(t)^{\frac{1}{2\delta}} \big) \cdot s.
  \end{equation*}
  In combination with~\cref{eq:cap-S-VS} this gives the desired result, since $|S| \geq s$, and
  \begin{equation*}\textstyle
      \capc_F(S, U\setminus S) = |S| - \capc_F(S,S) \geq (1 - \frac{3}{2} s/a - \varphi(t)^{\frac{1}{2\delta}}) \cdot s .
  \end{equation*}

\end{proof}

\subsubsection{Convergence}

We now show that after $O(\log^2{k})$ rounds, the potential has dropped below $1/k$.

\begin{lemma}\label{lemma:pot-reduction}
  For any round $t$, 
  \begin{equation*}
  \varphi(t) - \varphi(t+1) \geq \tfrac{1}{4} \sum_{\{i, j\} \in M_{t+1}} \lVert \v{w}_i - \v{w}_j \rVert^2 + \sum_{i \in A_t \setminus A_{t+1}} \lVert \v{w}_i \rVert^2
  \end{equation*}
  where $\v{w}_i$ is the $i$-th column vector of $W_t$.
\end{lemma}

\begin{proof}
The proof of this lemma relies on the following properties of the relevant
matrices and on some technical properties of the trace. The following claim
follows easily from the structural properties of $Q_t$, the fact that $Q_t$,
$N_t$, and $F_t$ are doubly stochastic, and that $P_t = I_t - Q_t$.
\begin{claim}\label{claim:properties}
  For any $t$, the following properties hold
  \begin{enumerate}[label=\roman*)]
      \item $P_t^2 = P_t$ and $P_t$ is symmetric \label[property]{claim:props:p2}
      
    \item $P_{t+1}P_t = P_t P_{t+1} = P_{t+1}$ \label[property]{claim:props:pp}

    \item $Q_t N_t = Q_t = N_t Q_t$, $P_t N_t = N_t P_t$ and $P_t N_t$ is symmetric.\label[property]{claim:props:np-nq} 
    \item  $P_t F_t P_t$ is symmetric.\label[property]{claim:props:PFP}
  \end{enumerate}
\end{claim}
\begin{claim}[Trace Properties]\label{claim:trace-props}
\hspace*{0pt}
  \begin{enumerate}[label=\roman*)]
  \item Let $X,Y \in \mathbb{R}^{n \times n}$ be symmetric matrices. Then
  $\Tr[(XYX)^{2k}] \leq \Tr[X^{2k} Y^{2k} X^{2k} ]$ for any integer $k$.\label{claim:trace:exp}
  \item Let $X \in \mathbb{R}^{n \times n}$ be a symmetric matrix with column vectors $x_i$. Then, for any $t$: \\
  $\Tr[I_t X^2] = \sum_{i \in A_t} \lVert x_i \rVert^2$ and
  $\Tr[\mathcal{L}(M_t) X^2] = \sum_{\{i, j\} \in M_t} \lVert x_i - x_j
  \rVert^2$, where $\mathcal{L}(M_t) := I_{t} - M_{t}$. \label{claim:trace:sum}
  \end{enumerate}
\end{claim}

We prove these claims in \cref{sec:omitted-proofs} and now return to the proof of
\cref{lemma:pot-reduction}. Ideally, we would like to express $W_{t+1}$ in
terms of $W_t$. This is not quite possible for the matrices, yet we can get the
following equality:
\begin{align*}
  P_{t+1} F_{t+1} P_{t+1} &= P_{t+1} \, (N_{t+1} F_t N_{t+1}) \, P_{t+1} \\
                          &= N_{t+1} P_{t+1} \, F_t \, P_{t+1} N_{t+1} && \text{by \cref{claim:properties}\labelcref{claim:props:np-nq}}\\
                          &= N_{t+1} \, P_{t+1} \,\, P_t F_t P_t \,\, P_{t+1} \, N_{t+1} && \text{by \cref{claim:properties}\labelcref{claim:props:pp}.}
\end{align*}
The central part already looks close to what we would need for $W_t$, but it is
missing the exponent. We can however use this to obtain a bound on
$\varphi(t+1)$ using \cref{claim:trace-props}. Note that  \cref{claim:trace-props} can be applied as
\cref{claim:properties}\labelcref{claim:props:np-nq}
and \cref{claim:properties}\labelcref{claim:props:PFP} show that $N_{t+1}P_{t+1}$,
$P_tF_tP_t$ and $P_{t+1}N_{t+1}$ all are symmetric.
\begin{align*}
  \varphi(t+1) &= \Tr[(N_{t+1} \, P_{t+1} \,\, P_t F_t P_t \,\, P_{t+1} \, N_{t+1}) ^{2\delta}] \\
               &\leq \Tr[(N_{t+1} P_{t+1})^{2\delta} \, \underbrace{(P_t F_t
                 P_t)^{2\delta}}_{=W_t^2} \, (P_{t+1} N_{t+1})^{2\delta}] &&
     \text{by \cref{claim:trace-props}\labelcref{claim:trace:exp}.}
\end{align*}

Now, since $N_{t+1}$ and $P_{t+1}$ commute by
\cref{claim:properties}\labelcref{claim:props:np-nq}, and $P_{t+1}$ is a
projection matrix by \cref{claim:properties}\labelcref{claim:props:p2}, it
follows that $(N_{t+1} P_{t+1})^{2\delta} = N_{t+1}^{2\delta} P_{t+1}$ and
$(P_{t+1} N_{t+1})^{2\delta} = P_{t+1} N_{t+1}^{2\delta}$.

Using this and the cyclic shift property of the trace, we obtain
\begin{align*}
  \varphi(t+1) &\leq \Tr[ P_{t+1} N_{t+1}^{4\delta} W_t^2] = \Tr[I_{t+1}
                 N_{t+1}^{4\delta}W_t^2] - 
\underbrace{\Tr[ Q_{t+1} N_{t+1}^{4\delta} W_t^2]}_{\geq 0} \leq \Tr[I_{t+1} N_{t+1}^{4\delta}W_t^2] \enspace ,
\end{align*}
For the last inequality, observe that $Q_{t+1} N_{t+1}^{4\delta} = Q_{t+1} = d_{t+1}d_{t+1}^T$ by \cref{claim:trace-props}\labelcref{claim:props:np-nq}. Then we have that
$\Tr[Q_{t+1}W_t^2] = \Tr[d_{t+1}^T W_t^2 d_{t+1}] = d_{t+1}^T W_t^2 d_{t+1} = \lVert W_t d_{t+1}\rVert_2^2 \geq 0$ using the cyclic shift property of the trace for the first step. This results in a scalar, which is nonnegative since it is the length of the vector $ W_t d_{t+1}$. 

From now on we continue similar to the analysis
by~\cite{OSVV08}, making use of the following claim to
rewrite $N_{t+1}^{4\delta}$ in terms of the Laplacian matrix,
$\mathcal{L}(M_{t+1}) := I_{t+1} - M_{t+1}$ of the graph whose edges are
exactly the matching edges.
Recall that in $M_t$ the rows and columns corresponding to vertices not in $A_t$ are all zero and that 
$N=I-\frac{1}{\delta}(I_t-M_t)$. We use the following claim, which is adapted from~\cite{OSVV08} and proved in \cref{sec:omitted-proofs}.

\begin{claim}\label{claim:N-osvv}
$N_t^{4\delta}= I - \lambda(I_t-M_t)$ for $\lambda := \frac{1}{2} - \frac{1}{2}(1 - \frac{2}{\delta})^{4\delta}\ge1/4$.
\end{claim}

Note that $I_{t+1}\mathcal{L}(M_{t+1}) = \mathcal{L}(M_{t+1})$. With this, we can finally express the new potential in
terms of the old potential minus the gain from deletions minus the gain from
the matching steps
\begin{align*}
  \varphi(t+1) &\leq \Tr[I_{t+1} N_{t+1}^{4\delta}W_t^2] = \Tr[I_{t+1} (I - \lambda \mathcal{L}(M_{t+1})) W_t^2]\\
               &= \Tr[I_{t+1} W_t^2] - \lambda \Tr[I_{t+1}\mathcal{L}(M_{t+1}) W_t^2] \\
               &= \sum_{i \in A_{t+1}} \lVert \v{w}_i \rVert^2 - \lambda  \Tr[\mathcal{L}(M_{t+1}) W_t^2] \\
              &= \underbrace{\sum_{i \in A_t} \lVert \v{w}_i \rVert^2}_{=\varphi(t)} - \sum_{i \in A_t \setminus A_{t+1}} \lVert \v{w}_i \rVert^2 - \, \lambda \sum_{\{i, j\} \in M_{t+1}} \lVert \v{w}_i - \v{w}_j \rVert^2 \enspace .
\end{align*}
\noindent
Here, the last equality follows from
\cref{claim:trace-props}\labelcref{claim:trace:sum} using the fact that the
columns of $W_t$ are $0$ outside of $A_t$. Thus, since $\lambda \geq 1/4$,
 \begin{equation*}
  \begin{aligned}
    \varphi(t) - \varphi(t+1) 
    &\geq \tfrac{1}{4} \sum_{\{i, j\} \in M_{t+1}} \lVert \v{w}_i - \v{w}_j \rVert^2 + \sum_{i \in A_t \setminus A_{t+1}} \lVert \v{w}_i \rVert^2.
  \end{aligned}
  \end{equation*}
This concludes the proof of \cref{lemma:pot-reduction}. 
\end{proof}

Next we argue similar to Lemma 5.9 in \cite{ADK22} that the expected gain of every round is actually a $\Omega(1/\log{k})$ fraction of the current potential $\varphi(t)$.

\begin{lemma}[Projection Lemma, Lemma E.3 in \cite{ADK22}]\label{lem:projection}
  Let $\{v_i\}_{i=1}^\ell$ be a set of $\ell \leq k+1$ vectors in $\mathbb{R}^k$. For $i \in [\ell]$ let $u_i = v_i^T \v{r}$ be the projection of $v_i$ onto a random unit vector $\v{r} \in \mathbb{S}^{k-1}$. Then
  \begin{enumerate}
    \item $\E_{\v{r}}[u_i^2] = \frac{1}{k}\lVert v_i \rVert_2^2$ for all $i$, and $\E[(u_i - u_j)^2] = \frac{1}{k}\lVert v_i - v_j \rVert_2^2$ for all pairs $(i,j)$.
    \item For all indices $i$ and pairs $(i, j)$ with probability of at least $1-k^{-\alpha/8}$, for every $\alpha \geq 16$ and large enough $k$, it holds that
    \begin{equation*}
      \begin{aligned}
        u_i^2         ~\leq~ & \frac{\alpha \log{k}}{k} \lVert v_i \rVert_2^2\\
        (u_i - u_j)^2 ~\leq~ & \frac{\alpha \log{k}}{k} \lVert v_i - v_j\rVert_2^2\\
      \end{aligned}
    \end{equation*}
   \label[property]{prop:proj-stretch}
  \end{enumerate}
\end{lemma}

We say that a round is \emph{good}, if \cref{prop:proj-stretch} of \cref{lem:projection}
holds. Thus, a round is good with high probability. The following claim shows that a good round ensures that
there is sufficient progress, i.e.,~the potential decreases by a $\Omega(1/\log{k})$
fraction. Note that even if a round $t$ is not good, by
\cref{lemma:pot-reduction}, we still have $\varphi(t+1) \leq \varphi(t)$.

\begin{lemma}\label{lemma:pot-good-round}
  In every good round $t$,
  \begin{equation*}
  \mathbb{E}_{\v{r}}\Bigg[ \tfrac{1}{4} \sum_{\{i, j\} \in M_{t+1}} \lVert \v{w}_i - \v{w}_j \rVert^2 + \sum_{i \in A_t \setminus A_{t+1}} \lVert \v{w}_i \rVert^2 \Bigg] \geq \frac{1}{2880\alpha \log{k}} \cdot \varphi(t)
  \end{equation*}
  for $\alpha \geq 16$, where $\v{w}_i$ is the $i$-th column vector of $W_t$.
\end{lemma}
\begin{proof}
Recall that we have $u_i = \v{w}^T_i \v{r}$ for all $i \in A_t$ and since $t$ is a good round, \cref{lem:projection}, \cref{prop:proj-stretch} holds with probability $1$.
For the gain from the matchings, we have 
\begin{equation*}
  \begin{aligned}
    \tfrac{1}{4} \sum_{\{i, j\} \in M_{t+1}} \lVert \v{w}_i - \v{w}_j \rVert^2 
    &\geq \frac{k}{4\alpha \log{k}} \sum_{\{i, j\} \in M_{t+1}} (u_i - u_j)^2 &&& \text{by~\cref{lem:projection}.\labelcref{prop:proj-stretch})}\\
    &= \frac{k}{4\alpha \log{k}} \sum_{i\in A^{\ell} \setminus S_t} (u_i - \eta)^2 &&& \text{by~\cref{lemma:RST14}.\labelcref{prop:rst14:separate})}\\
    & \geq \frac{k}{36\alpha \log{k}} \sum_{i\in A^{\ell} \setminus S_t} u_i^2 &&& \text{by~\cref{lemma:RST14}.\labelcref{prop:rst14:source})}
  \end{aligned}
\end{equation*}
where for the second inequality we also used that each node in $A^{\ell} \setminus S_t$ is matched exactly once in $M_{t+1}$. Similarly, we have for the gain from deletions that 
\begin{equation*}
     \sum_{i \in A_t \setminus A_{t+1}} \lVert \v{w}_i \rVert^2
    \geq \frac{k}{\alpha \log{k}} \sum_{i \in A_t \setminus A_{t+1}} u_i^2
    \geq \frac{k}{\alpha \log{k}} \sum_{i \in A^{\ell} \cap S_t} u_i^2
\end{equation*}
using~\cref{lem:projection}.\labelcref{prop:proj-stretch}). Note that by definition, $A_t \setminus A_{t+1} = S_t \supseteq A^{\ell} \cap S_t$.
So combined, we have
\begin{equation*}
  \begin{aligned}
    \tfrac{1}{4} \sum_{\{i, j\} \in M_{t+1}} \lVert \v{w}_i - \v{w}_j \rVert^2 + \sum_{i \in A_t \setminus A_{t+1}} \lVert \v{w}_i \rVert^2
    \geq \frac{k}{36\alpha \log{k}} \sum_{i \in A^{\ell}} u_i^2
    \geq \frac{k}{80\cdot36\alpha \log{k}} \sum_{i \in A_t} u_i^2,
  \end{aligned}
\end{equation*}
where the last inequality is due to~\cref{prop:rst14:sumsource} of \cref{lemma:RST14}.

Finally, recall that the $u_i$ values depend on the randomly chosen $\v{r}$, where we have $\E_{\v{r}}[u_i^2] = \frac{1}{k}\lVert \v{w}_i \rVert_2^2$ for all $i$ by \cref{lem:projection}. Also, we have $\sum_{i \in A_t} \lVert \v{w}_i \rVert^2 = \varphi(t)$ as noted above. We thus conclude

\begin{equation*}
  \begin{aligned}
    \mathbb{E}_{\v{r}}\Bigg[ \tfrac{1}{4} \sum_{\{i, j\} \in M_{t+1}} \lVert \v{w}_i - \v{w}_j \rVert^2 + \sum_{i \in A_t \setminus A_{t+1}} \lVert \v{w}_i \rVert^2 \Bigg] 
    & \geq \frac{k}{80 \cdot 36\alpha \log{k}} \cdot\sum_{i \in A_t} \mathbb{E}_{\v{r}} [u_i^2] \\
    & = \frac{1}{2880\alpha \log{k}} \cdot\sum_{i \in A_t} \lVert \v{w}_i \rVert^2  \\
    & = \frac{1}{2880\alpha \log{k}} \cdot\varphi(t).
  \end{aligned}
\end{equation*}
\end{proof}

\begin{claim}\label{claim:num-iterations}
  There is a $T = O(\log^2{k})$ such that with high probability over the choices of $\v{r}$, we have $\varphi(T) \leq 1/k^3$. 
\end{claim}
\begin{proof}
  The initial potential is $\varphi(0) = k-1$. After $T$ rounds, the potential is 
  \begin{equation*} \textstyle
    \varphi(T) = \prod_{i=1}^T(1-X_i)\cdot\varphi(0) \leq \exp(-\sum_{i=1}^T X_i) \cdot \varphi(0)
  \end{equation*}
  where $X_i$ is the factor by which the potential drops in round $i \leq T$. Note that $X_i$ is a random variable depending on the random choice of $\bm{r}$ in round $i$. By \cref{lemma:pot-reduction} we have $X_i \geq 0$ in any round $i$ and by \cref{lemma:pot-good-round} we further have that $\E_{\bm{r}}[X_i] \geq \frac{1}{c\log{k}}$ for some constant $c > 0$ if round $i$ is good.

  Assume all $T$ rounds were good, i.e., condition on the event that all $T$ rounds are good. Since we have a sum of $T$ independent random variables with values in $[0, 1]$ and expectation at least $\frac{1}{c\log{k}}$, we can use a Chernoff bound to see that $\sum_{i=1}^T X_i \geq \frac{T}{2c\log{k}}$ with probability at least $1 - \exp(-\frac{T}{8c\log{k}})$. By choosing $T := 8cd \log^2{k}$, we thus get that $\sum_{i=1}^T X_i \geq 4d\log{k}$ with probability $1 - 1/k^d$ for any $d\geq1$. So with high probability
  \begin{equation*} \textstyle
    \varphi(T) \leq \exp(-4d\log{k}) \cdot(k-1) = \frac{k-1}{k^{4d}} \leq 1/k^3.
  \end{equation*}

  Finally, recall that a round is good with probability at least $1-k^{-\alpha/8}$ for $\alpha \geq 16$. So a sequence of $T$ rounds consists only of good rounds with probability at least $(1-k^{-\alpha/8})^T \geq 1 - T/k^{\alpha/8} \geq \Omega(1-1/k)$. Thus, the probability that all $T$ rounds are good and $\phi(T) \le 1/k^3$ happens with probability $ \Omega(1-1/k)$, i.e.,~the claim follows.
\end{proof}

\subsubsection{\texorpdfstring{Proof of~\cref{theorem:new-cut-player}}{Proof of Lemma 10}}

\begin{proof}[Proof of~\cref{theorem:new-cut-player}]
  Let $\mathcal{M}$ be any matching player.
  We choose the value $T = O(\log^2{k})$ according to~\cref{claim:num-iterations} and a value $\delta = \Theta(\log{k})$ such that it is a power of $2$ and $(1/k)^{\frac{3}{2\delta}} \leq 1/20$.
  Then we start the cut matching game $\mathcal{G}(\mathcal{C}^X, \mathcal{M}, T)$ to obtain a pair $(A,Y)$. Assume that $|A| \geq (1-\frac{1}{2\log{k}}) \cdot k$. Upon termination, after $T$ rounds, the internal potential as defined in \cref{eq:potential} has been reduced to below $\varphi(T) \leq 1/k^3$ with high probability by \cref{claim:num-iterations}. Consider the flow matrix $F_T$ of the cut player and let $H_F$ be the graph whose weighted adjacency matrix is $F_T$. By \cref{cor:H-exp}, we get that $H_F$ is $\1_A$-expanding with quality $1/4 - \varphi(T)^{\frac{1}{2\delta}} \geq 1/4 - (1/k)^{\frac{3}{2\delta}} \geq 1/5$.

  % Finally, $H_F$ can be embedded in the graph $(U,Y)$ with congestion $2/\delta = \Omega(1/\log{k})$, giving that $(U,Y)$ is $\1_A$-expanding with quality $\Omega(\log{k})$.

Finally, recall that $F_T$ is the result of a slowed down random walk using the walk matrices $N_t = I - \frac{1}{\delta}(I_t - M_t)$, $t \leq T$. Since $Y$ contains all the (unweighted) matching edges from all $M_t$, $t\leq T$, it follows that $H_F$ can be embedded in the graph $(U,Y)$ with congestion $2/\delta = O(1/\log{k})$. Since $H_F$ is $\1_A$-expanding with quality at least $1/5$, this gives that the graph $(U,Y)$ is $\1_A$-expanding with quality at least $\delta/2 \cdot 1/5 = \Omega(\log{k})$, as claimed.
\end{proof}

\subsubsection{Efficient Implementation}\label{sec:cm-time}

We conclude the presentation of the cut player algorithm by proving that it can be implemented efficiently. Crucially, none of the matrices defined in~\cref{sec:cp-alg} have to be given explicitly. Instead, each of the relevant matrices corresponds to an operation that can be implemented in time linear in the number of active elements.

\begin{claim}\label{lemma:cp-runtime}
  For any $A \subseteq [k]$ and ordered set of matchings $Y$, the output of the cut player $\mathcal{C}^X(A, Y)$ can be computed in time $O(|A| \, (t+1) \log{k})$, where $t = |Y|$. 
\end{claim}
\begin{proof}
  Let $t = |Y|$ be the number of matchings in $Y$, i.e., we are executing round $t+1$ in the cut matching game.
  We show that $W_t \bm{r}$ can be computed in time $O(|A| \cdot t \log{k})$ if $t \geq 1$. For $t=0$, we simply have $\bm{u} = W_0 \bm{r} = \bm{r}$.
  The computation of the output $A^{\ell}$ and $A^r$ using \cref{lemma:RST14} takes $O(|A| \log{(|A|)})$ time. Thus, the total time for executing step $t+1$ is $O(|A| \, (t+1) \log{k})$.
  
  Recall that $W_t = (P_t F_t P_t)^\delta  = (P_t N_t N_{t-1} \ldots N_2 N_1 N_1 N_2 \ldots N_{t-1} N_t P_t)^\delta$, i.e., it is an application of either a matrix $P_t$ or $N_\ell$, $\ell \leq t$. 
  Let $a_\ell = |A_\ell|$ be the size of the active set in step $\ell \leq t$.
  For any vector $\bm{v}$, computing the product $P_t \bm{v} = I_t \bm{v} - 1/{a_t} \cdot (\v{v}^T \1_t) \1_t$ requires subtracting the average over all elements in $A_t$ from each individual element in $A_t$ and setting the other elements to $0$. This can be done in time $O(a_t)$. For any $\ell \leq t$, computing the product $N_\ell \v{v}$ is equivalent to multiplying the value at each index $i$ by $(1 - 1/\delta)$ and adding $(1 - 1/\delta) \cdot \v{v}_i$ to the index $j$ corresponding to the matching partner $j$ of $i$ in $M_\ell$. In particular, the $\v{v}_i$ value of unmatched elements $i$ remains unchanged, which includes all elements in $U \setminus A_\ell$.
  Clearly, this can be implemented in time linear in the number of matched pairs in $M_\ell$, which is at most $a_\ell \leq |A|$.

  As each matrix operation corresponds to an operation that can be implemented in time $O(|A|)$ and there are $\delta \cdot (2t + 2)$ matrices, we conclude that $W_t \bm{r}$ can be computed in time $O(\delta \cdot t \cdot |A|) = O(|A| \cdot t \log{k})$.
\end{proof}

\subsection{The Matching Player}\label{sec:matching-player}

In this section we define and analyze a novel matching player that allows the execution of the cut-matching game on a graph with non-negative, integral weights on the vertices. The expansion achieved by such a run of the cut-matching game will then be with respect to the given weights. Conceptually, we achieve this by playing the game on the individual weight units instead of the vertices.

Assume we are given a graph $G=(V,E, \capc)$ and an \emph{integral} vertex weight function $\v{\pi} : V \rightarrow \{0, \ldots, W\}$. Since $\v{\pi}$ is integral, we can split the $\v{\pi}$-volume of each vertex $v$ into $\v{\pi}(v)$ individual units. Fix $U = \{1, \ldots, \bm{\pi}(V)\}$ as the set of all individual units of $\v{\pi}$-volume.
Then we can define a \emph{mapping} $\theta: U \rightarrow V$ as a function that associates each unit of $\v{\pi}$-volume with its corresponding vertex in $V$. We define the natural inverse $\theta^{-1}(v) = \{u \in U \, | \, \theta(u) = v \}$ which gives the set of all units of $\bm{\pi}$-volume that map to a vertex $v \in V$. By design, we thus have $|\theta^{-1}(v)| = \bm{\pi}(v)$. We use the shorthands $\theta(X) = \{\theta(x) \, | \, x \in X \}$ for $X \subseteq U$ and $\theta^{-1}(S) = \bigcup_{v \in S} \theta^{-1}(v)$ for $S \subseteq V$.

Recall that our definition of the cut matching game is independent of the input graph, and only requires a number of units. We consider an execution of the cut matching game using $|U|$ units.
Our matching player is parametrized by a value $c \geq 1$, which allows to trade off between approximating the cut size and the expansion quality.
In addition to responding to outputs of the cut player with a proper matching, our matching player has two tasks. The first task is to ensure that the returned matching can be embedded in $G$ with congestion $4c$, deleting vertices and corresponding units if necessary. For deleting vertices, however, the algorithm is restricted to only removing $1/c$ sparse cuts. The second task is to maintain a subset of \enquote{inactive} graph-vertices $R \subseteq V$, which is returned at the end of the cut-matching game or whenever queried for it.
Altogether, we thus ensure the following invariants hold through the rounds of the cut matching game.

\begin{enumerate}
  \item $\theta^{-1}(R)$ is the set units that have so far been deleted.
  \item $R$ is a $1/c$ sparse cut in $G$ w.r.t. $\bm{\pi}$.
  \item In round $t$, the congestion of embedding all matchings in $G$ is at most $4ct$. 
\end{enumerate}

\paragraph{\mathversion{bold}Definition of the Matching Player $\mathcal{M}_c^X$.\mathversion{normal}}
We assume that $\mathcal{M}_c^X$ has access to an implementation for \algFC{}. For a fixed value $c$, the algorithm $\mathcal{M}_c^X$ is given as follows.
With the given graph $G$ and weight function $\bm{\pi}$, we initialize the matching player $\mathcal{M}_c^X$ before the cut matching game begins by fixing a mapping $\theta$. Also fix the value $\alpha = 3/2$. 
Then, in each round, $\mathcal{M}_c^X$ is given a set $A \subseteq U$ and two subsets $A^{\ell}, A^r \subseteq A$ and executes the following steps. In the following we use the variables $u$ and $v$ to denote vertices in $V$ and $x$ and $y$ to denote units, i.e., elements of $U$.
\begin{enumerate}
  \item Define $V' = \theta(A)$ and fix a source function $\v{s} \in \mathbb{N}_0^{V'}$ in $G[V']$ where $\v{s}(v) = |\theta^{-1}(v) \cap A^{\ell}|$ for all $v \in V'$. Similarly, let $\v{t} \in \mathbb{Q}_{\geq 0}^{V'}$ be a target function in $G[V']$ with $\v{t}(v) = 1/\alpha \cdot |\theta^{-1}(v) \cap A^r|$ for all $v \in V'$.
  \label[step]{step:mp:def}

  \item Obtain the graph $G'$ from $G[V']$ by scaling the capacities by the integral factor $\lceil c \alpha \rceil$.
  \label[step]{step:mp:graph}

  \item Call \algFC$(G', \v{s}, \v{t}, \alpha)$ to obtain a $\alpha$-fair $(\mathbf{s}, \mathbf{t})$-cut/flow pair $(S, g)$. Update the set $R \gets R \cup S$ and define the set of to-be-deleted units as $D = \theta^{-1}(S)$. \label[step]{step:mp:fc} 

  \item Let $g' = \alpha g$ be a scaled flow. Discard all fractional flow in $g'$ to obtain a flow $f$ that routes an integral demand.
  \label[step]{step:mp:flow}

  \item Let $P$ be a path decomposition of $f$.
  \label[step]{step:mp:decomp}

  \item Initialize $M$ as an empty matching and mark all $x \in A^r \setminus D$ as
  \emph{unmatched}. Then, for each $y \in A^{\ell} \setminus D$: \textbf{If} there
  is an unmatched $x \in A^r \setminus D$ with $\theta(x) = \theta(y)$, add the pair $y, x$
  to $M$ and mark $x$ as \emph{matched}. \textbf{Otherwise} 
  execute these steps
  \label[step]{step:mp:match}
  \begin{enumerate}
    \item Find a path $P_y$ in $P$ that starts at $\theta(y)$ and remove it from $P$. \label[step]{step:mp:findpath}
    \item Let $v \in V' \setminus S$ be the vertex where $P_y$ ends. \label[step]{step:mp:findvertex}
    \item Select an arbitrary, unmatched element $x$ from $\theta^{-1}(v) \cap A^r$, add the pair $y, x$ to $M$ and mark $x$ as \emph{matched}. \label[step]{step:mp:findpartner}
  \end{enumerate}

  \item Return the set $D$ and the matching $M$.
\end{enumerate}

\subsubsection{Analysis}

We first verify that the algorithm is indeed correct and produces matchings that can be used within the cut matching game.

\begin{claim}\label{claim:mp:correct}
  When given $A$, $A^{\ell}$ and $A^r$ as input, the set $M$ returned by $\mathcal{M}_c^X$ is a matching, i.e.~matches each unit of $A^{\ell} \setminus D$ to a distinct partner in $A^r \setminus D$.
\end{claim}
\begin{proof}
  We argue the correctness of the algorithm and show that each of the steps in the algorithm is well-defined. This is obvious for \cref{step:mp:def,step:mp:graph,step:mp:fc,step:mp:flow,step:mp:decomp}. In \cref{step:mp:match} we have to show that we always find the required paths and vertices.

  First, we show that for all $v \in V \setminus S$ with $\mathbf{s}(v) - \mathbf{t}(v) \geq 0$, there are at least $\mathbf{s}(v) - \alpha\mathbf{t}(v)$ paths starting at $v$ in $P$. To see this, note that as $g$ is a $\alpha$-fair cut in $V'$, by \cref{def:fair-cut}, \cref{prop:fair-cut:generate}
   we have 
  $g(v) \ge  1/\alpha \cdot (\mathbf{s}(v) - \mathbf{t}(v))$ and, thus, $g'(v) = \alpha g(v) \geq \mathbf{s}(v) - \mathbf{t}(v)$. 
  As only the fractional flow is discarded to obtain $f$ from $g'$ it follows that
  $f(v)\ge  \lfloor  \v{s}(v) -\v{t}(v) \rfloor = \v{s}(v) - \lceil  \v{t}(v) \rceil \ge \v{s}(v) - \alpha \v{t}(v)$ and $\v{s}(v) - \alpha \v{t}(v)$ is an integer.
  This gives the number of paths starting in $v$ in $P$ is at least $\mathbf{s}(v) - \alpha  \mathbf{t}(v)$.

  Assume we reach \cref{step:mp:findpath} in order to match a unit $y \in A^{\ell} \setminus D$ and let $v = \theta(y)$. Note that $v \in V' \setminus S$ as all units belonging to vertices in $S$ belong to $D$. By definition, there are $\mathbf{s}(v)$ units at $v$ that need to be matched and $\alpha \mathbf{t}(v)$ of these can be paired up directly at $v$ in \cref{step:mp:match}.
  Consequently, as  $v$  still has units that need to be matched when we reach \cref{step:mp:findpath} there are $\mathbf{s}(v) - \alpha \mathbf{t}(v) \geq 1$ units that need to be matched to a unit that is sitting at some vertex other than $v$.
  In particular, $v$ thus satisfies $\mathbf{s}(v) - \mathbf{t}(v) \geq 0$, so from above we get that $\mathbf{s}(v) - \alpha \mathbf{t}(v)$ paths are available, which is sufficient to match all the units that need to leave $v$. Thus, when we reach \cref{step:mp:findpath}, sufficiently many paths $P_y$ starting at $v$ exist.

  In \cref{step:mp:findvertex}, recall that the path $P_y$ must start in $V' \setminus S$ by the definition of $D$. Thus, the end vertex $v$ of $P_y$ cannot be in $S$ since $P_y$ is a flow path that starts in $V' \setminus S$,  $g'$ is a fair flow and by \cref{def:fair-cut}, there is no flow sent from $V' \setminus S$ into $S$. Thus, it follows that $v \in V' \setminus S$. We hence have $\theta^{-1}(v) \subseteq U \setminus D$ and consequently the element $x$ matched to $y$ fulfills $x \in A \setminus D$.

  It remains to show that there is always an unmatched element in $\theta^{-1}(v) \cap A^r$ when we reach \cref{step:mp:findpartner}. For this, observe that $v$ must be a net target of $f$ as it is an endpoint of a path decomposition of $f$. Then, by \cref{def:fair-cut}, \cref{prop:fc:net-t} we have $|f(v)| \le |g'(v)| \leq \alpha \v{t}(v)= |\theta^{-1}(v) \cap A^r|.$ It follows that at most $|f(v)|$ paths can end at $v$ in $P$, and hence, for each such path there is a unit available in $\theta^{-1}(v) \cap A^r$.
\end{proof}

The set of deleted vertices $R$, that is maintained by the matching player, is in fact at all times a $1/c$ sparse cut in $G$ with respect to the weight function $\bm{\pi}$. This justifies its use in our algorithm for \algSC{}.

\begin{invariant}\label{claim:mp:sparse}
  At any point, the set of deleted vertices $R \subseteq V$ is a $1/c$ sparse cut in $G$ w.r.t. $\bm{\pi}$, i.e., it holds $\capc(R, V \setminus R) \leq 1/c \cdot \bm{\pi}(R)$.
\end{invariant}
\begin{proof}
  First, observe that each $S$ that is computed is a $1/c$ sparse cut in $G[A]$. For this, recall that $G'$ is obtained from $G[A]$ by scaling the capacities $\capc$ by the factor $\lceil c\alpha \rceil$. Since $S$ is a $\alpha $-fair $(\v{s}, \v{t})$-cut in $G'$, we thus get from~\cref{claim:fair-cut-ineq} that 
  \begin{equation*}
    c \alpha \cdot \capc(S, V' \setminus S) \leq \alpha  \cdot \v{s}(S) = \alpha  \cdot \textstyle \sum_{v \in S} |\theta^{-1}(v) \cap A^{\ell}| \leq \alpha  \cdot \bm{\pi}(S) \enspace ,
  \end{equation*}
  since we have $|\theta^{-1}(v)| = \bm{\pi}(v)$ by design. At any point, $R$ is a union of disjoint $1/c$ sparse cuts, so it is sparse itself. Formally, let $S_1, \ldots, S_t$ be the sequence such that $R = \bigcup_i S_i$ and let $V'_1, \ldots, V'_t$ be the sequence of subsets in which each $S_i$ was computed. Note that $V \setminus R \subseteq V'_i \setminus S_i$ for all $i$, therefore
  \begin{equation*}
    \capc(R, V \setminus R) = \textstyle \sum_i \capc(S_i, V \setminus R) \leq \sum_i \capc(S_i, V'_i \setminus S_i) \leq 1/c \cdot\sum_i \bm{\pi}(S_i) = 1/c \cdot \bm{\pi}(R).
  \end{equation*}
\end{proof}

The following invariant is clearly satisfied by the algorithm throughout all calls, since it always deletes all units mapped to vertex in the set $S$ (by definition of $D = \theta^{-1}(S)$) and this set $S$ is added to the set $R$ in \cref{step:mp:fc}.

\begin{invariant}\label{inv:mp}
  Let $Z \subseteq U$ be the set of units that $\mathcal{M}_c^X$ has deleted so far. Then, $\theta^{-1}(R) = Z$.
\end{invariant}

We now show that the matching $M$ returned by $\mathcal{M}_c^X$ can be embedded in $G$ with low congestion.

\begin{claim}\label{claim:mp:embed}
  For all $X \subseteq V$, it holds
  $|\{ (i,j) \in M \,| \, i \in \theta^{-1}(X), \, j \in \theta^{-1}(V \setminus X) \}| \leq 4c \cdot \capc(X, V \setminus X)$.
\end{claim}
\begin{proof}
  Fix some $X \subseteq V$ and define $M_X = \{ (i,j) \in M \,| \, i \in \theta(X), \, j \in \theta(V \setminus X) \}$ as the set of relevant pairs from $M$.
  Let $X' = X \cap V'$, hence $X'$ and $V'$ are the respective restrictions of $X$ and $V$ to active vertices.
  Since $M$ only contains pairs of active units, we know for each $(i,j) \in M_X$, that $i \in \theta^{-1}(X')$ and $j \in \theta^{-1}(V' \setminus X')$. 

  As $X'$ and $V' \setminus X'$ are disjoint, so are $\theta^{-1}(X')$ and $\theta^{-1}(V' \setminus X')$ and for each unit $u \in V'$ we know $\theta(u)$ is either contained in $X'$ or in $V' \setminus X'$. Therefore, for each $(i, j) \in M_X$ we have $\theta(i) \in X'$ and $\theta(j) \in V' \setminus X'$. In particular, it follows $\theta(i) \neq \theta(j)$ and the pair must have been matched in \cref{step:mp:findpartner}. Thus, it corresponds to a path from $\theta(i)$ to $\theta(j)$ in $G'$, which notably crosses the cut $(X', V' \setminus X')$.
  
  Since each path of $P$ is used at most once (in \cref{step:mp:findpath}), it follows that $|M_X|$ is at most the number of paths in $P$ crossing the cut $(X', V' \setminus X')$. We conclude the proof by arguing that there are at most $4c \cdot \capc(X', V' \setminus X')$ such paths.
  
  Recall that $P$ is a path decomposition of the flow $f$, which has congestion at most $\alpha = 3/2$ in the graph $G'$ as it is obtained by scaling the feasible flow $g$ by $\alpha$. In $G'$, the capacities are scaled by a factor $\lceil c \alpha \rceil$ compared to $G$, giving a total congestion of $f$ in $G$ of $\alpha \cdot \lceil c \alpha \rceil = 3/2 \cdot \lceil c \cdot 3/2 \rceil \leq 4c$. It follows that $|M_X|\leq 4c \cdot \capc(X', V' \setminus X') \leq 4c \cdot\capc(X, V \setminus X)$, as desired.
\end{proof}

Next, we show that the matching player does not delete too many units in a single iteration. This will be relevant for the \algSC{} algorithm.

\begin{claim}\label{claim:mp:few-deletions}
  Let $\epsilon, \delta \geq 0$ be values such that $|A^{\ell}| \leq \epsilon \cdot |A|$ and $|A^r| \geq \delta \cdot |A|$. Then, the output $D$ of the matching player satisfies $|A \setminus D| \geq (\delta - \epsilon \alpha^2) \cdot |A|$.
\end{claim}
\begin{proof}
  Consider a set $S$ computed in \cref{step:mp:fc} by the matching player. We show that there must be many unsaturated target capacity in $V' \setminus S$. This implies the existence of a sufficient number of units in $V' \setminus S$: Since $S$ is a $\alpha$-fair cut in $G'$, it must saturate the targets in $S$ up to a factor $1/\alpha$. So, by \cref{claim:fair-cut-ineq} and using the definitions, we get 
  \begin{equation*}
    \v{t}(S) \leq \alpha \v{s}(S) = \alpha |\theta^{-1}(S) \cap A^{\ell}| \leq \alpha |A^{\ell}| \leq \epsilon \alpha |A| \enspace .
  \end{equation*}
  Recall that $A = \theta^{-1}(V')$,  $D = \theta^{-1}(S)$, and that $|\theta^{-1}(V' \setminus S) \cap A^r| = \alpha\v{t}(V' \setminus S)$ by the definition of $\v{t}$. We thus get for the number of remaining units that 
  \begin{equation*}
    \begin{aligned}
      |A \setminus D| & = |\theta^{-1}(V' \setminus S)| \geq  |\theta^{-1}(V' \setminus S) \cap A^r| = \alpha\v{t}(V' \setminus S) = \alpha \v{t}(V') -  \alpha\v{t}(S) \\
      & = |A^r| -  \alpha\v{t}(S) \geq \delta |A| - \epsilon \alpha^2 |A| \enspace .
    \end{aligned}
  \end{equation*}
\end{proof}

We conclude the presentation of the matching player by analyzing the running time required for a single execution. Note that this running time is independent of the value of $c$.

\begin{claim}[Running Time]\label{claim:mp:runtime}
  Given a graph $G$ with $m$ edges, $n$ nodes and an algorithm for \algFC{} that runs in time $\Tfc$, the algorithm $\mathcal{M}_c^X$ can be implemented in time $O(\Tfc(m,3/2) + m \log{n})$.
\end{claim}
\begin{proof}
  \cref{step:mp:def,step:mp:graph} can be implemented in time $O(m)$. In \cref{step:mp:fc}, we run \algFC{} on a graph with $m$ edges, giving a running time of $\Tfc(m,3/2) + O(m)$ for that step. Using dynamic trees~\cite{ST81} the path decomposition in \cref{step:mp:decomp} can be computed in time $O(m \log{n})$. By iterating these paths, we can implement~\cref{step:mp:match} in time $O(m \log{n})$. Thus, the total time spent is $O(\Tfc(m,3/2) + m \log{n}) = \tO(m)$.
\end{proof}

\subsection{\texorpdfstring{Analysis of \algSC{}}{Analysis of SparsestCutApx}}
\label{sec:spx-analysis}

Before we show the correctness of our algorithm \algSC{} we first analyze the running time of the cut matching game using our new cut and matching players. This dominates the running time of \algSC{}.

\begin{claim}[Running Time]\label{lemma:cmgame-runtime}   
  Given is a graph $G=(V,E)$ with $m$ edges, $n$ nodes, integral vertex weights $\bm{\pi}$ and an algorithm for \algFC{} that runs in time $\Tfc$.
  The output of the cut-matching game $\mathcal{G}(\mathcal{C}^X, \mathcal{M}^X, T)$ using our new cut and matching players can be computed in time $O\big(T^2 \cdot (\Tfc(m) + \bm{\pi}(V) \log{(\bm{\pi}(V))} + m \log{n})\big)$ for any $T \geq 1$.
\end{claim}
\begin{proof}
  The running time is dominated by the execution of the cut and the matching player. In round $t \leq T$, we first compute the output of the cut player $\mathcal{C}^X(A_t, Y_t)$, which takes $O((t+1) k \log{k})$ time by \cref{lemma:cmgame-runtime}. Then, we call the matching player, which, by \cref{claim:mp:runtime}, takes $O(\Tfc(m) + m \log{n})$ time, where $m$ and $n$ is the number of edges and nodes of $G$, respectively. As $t$ ranges to at most $T$, when summing up the times for all iterations, we get a total running time $O(T^2 \cdot(\Tfc(m) + \bm{\pi}(V) \log{(\bm{\pi}(V))} + m \log{n}))$.
\end{proof}

Finally, we can conclude the analysis of the \algSC{} algorithm by proving the correctness of the procedure from \cref{sec:spx} and establishing \cref{lemma:spx}.

\begin{proof}[Proof of \cref{lemma:spx}]
  We distinguish two cases, based on whether the cut matching game is terminated normally, or stopped early in the execution of the above algorithm for \algSC{}. Note that in either case, we have $\bm{\pi}(R) \leq \bm{\pi}(V \setminus R)$ by design. We show how to establish \cref{prop:SC:sparse,prop:SC:exp} in each case and conclude by showing how to obtain the claimed running time.

  \textbf{Case 1 [Unbalanced Sparse Cut].} First, assume that the algorithm never reaches the ``If''-branch in \cref{step:spx:stop} and executes all $T$ iterations of the cut matching game without stopping early.
  We thus have $|A_T| \geq (1 - 1/(2\log{k})) \cdot k$. Recall that $k = \bm{\pi}(V) = |U|$. Let $ U \setminus A_T$ be the set of all deleted units. With \cref{inv:mp}, we get $\bm{\pi}(R) = |\theta^{-1}(R)| = |U \setminus A_T| \leq 1/(2\log{k} )\cdot k$. And in particular, $\bm{\pi}(R) \leq \bm{\pi}(V \setminus R)$, which means $R$ is the set returned by the algorithm. From \cref{claim:mp:sparse} we get that $R$ is a $1/c$ sparse cut in $G$ w.r.t $\bm{\pi}$, which directly gives \cref{prop:SC:sparse}, since $1/c \leq \phi/10 \leq \phi$.

  We establish \cref{prop:SC:exp} by showing expansion of $G$ with high probability. Since the cut matching game terminated normally and $A_T$ is sufficiently large, we get from \cref{theorem:new-cut-player} that the graph $H=(U, Y_T)$ is $\1_{A_T}$-expanding with quality $r = \Omega(\log{k})$ with high probability.
  Observe that $H$ is an unweighted multi-graph and hence $\capc_H(\partial F) = |F|$ for any $F \subseteq Y_T$.
  Via \Cref{claim:mp:embed} which bounds the congestion caused by the embedding in $G$ maintained by the matching player, the expansion of $H$ implies the expansion of $G$ as follows.

  Let $C \subseteq V$ be a subset such that $\bm{\pi}(C \setminus R) \leq \bm{\pi}((V \setminus R) \setminus C)$, i.e., $C$ is the smaller side w.r.t the $\bm{\pi}|_{V \setminus R}$-volume.
  The goal is to show that $\capc(C, V \setminus C) \geq \q \cdot \bm{\pi}(C \setminus R)$. Define $Z = \theta^{-1}(C \setminus R)$ as the set of units mapped to $C \setminus R$.
  Observe that by \cref{inv:mp}, all the units in $Z$ are still active and we have $Z \subseteq A_T$, as these are the units mapped to $C$ which have not been deleted.
  From the expansion of $H$, since $|Z| \leq |A_T \setminus Z|$ by assumption, we get
  $\capc_H(Z, U \setminus Z) \geq r \cdot |Z| = r \cdot \bm{\pi}(C \setminus R)$.
  % As the matching player maintains an embedding of all matched pairs into $G$, we also obtain
  Since the matching player guarantees that each returned matching $M_t$ for $t \leq T$ can be embedded in $G$ with congestion $4c$, we also obtain

  \begin{equation}\label{eq:embed-H}
    \begin{aligned}
      \capc_H(Z, U \setminus Z) & = \big| \{ (i,j) \in Y_T \,| \, i \in Z, \, j \in U \setminus Z \} \big| \\
      & = \textstyle \sum_{t=1}^T \big| \{ (i,j) \in M_t \,| \, i \in Z, \, j \in U \setminus Z \} \big| \\
      & \leq \textstyle T \cdot 4c \cdot \capc(C, V \setminus C) \enspace .
    \end{aligned}
  \end{equation}
  Here the last inequality follows from \cref{claim:mp:embed}, using the definition $Z = \theta^{-1}(C \setminus R)$ and the fact that $\capc(C \setminus R, \, \, V \setminus (C \setminus R)) \leq \capc(C, V \setminus C)$.
  Together, this implies the desired $\bm{\pi}|_{V \setminus R}$ expansion of $G$ with a quality of
  \begin{equation*}\textstyle
    r/(4cT) = \phi \cdot \frac{\Omega(\log{\bm{\pi}(V)})}{O(\log^2{(\bm{\pi}(V))})} = \phi / \q, \quad \text{where} \quad \q = O\big(\log{\bm{\pi}(V)}\big) \enspace ,
  \end{equation*}
  with high probability, thus proving \cref{prop:SC:exp} and concluding the first case.

  \smallskip
  \textbf{Case 2 [Balanced Sparse Cut].} Now suppose the algorithm reaches the ``If''-branch in \cref{step:spx:stop} and terminates the cut matching game early, after completing $t < T$ iterations. We first analyze the $\bm{\pi}$-volume in $R$ and $V \setminus R$.

  The iteration $t$ must have begun with $|A_t| \geq (1 - 1/(2\log{k})) \cdot k$, as otherwise, the cut matching game would have been stopped in the iteration preceding $t$ (and clearly it also holds for $t=0$). In iteration $t$ itself, not too many units are deleted. Formally, recall that by \cref{prop:rst14:size} of \cref{lemma:RST14}, we have $|A^{\ell}| \leq 1/8 \cdot |A_t|$ and $|A^r| \geq 1/2 \cdot |A_t|$. 
  Thus, by \cref{claim:mp:few-deletions}, since $1/2 - 1/8 \cdot \alpha^2 = 1/2 - 1/8 \cdot (3/2)^2 \geq 1/5$, we obtain 
  \begin{equation}\label{eq:pivr-k}
    \bm{\pi}(V \setminus R) = |A_{t+1}| = |A_t \setminus D_{t+1}| \geq 1/5 \cdot |A_t| \geq 1/5 \cdot  (1 - 1/(2\log{k})) \cdot k \enspace .
  \end{equation}
    For $\bm{\pi}(V) \geq 8$ we thus have $\bm{\pi}(V \setminus R) \geq k / (2 \log k)$. Recall that $\bm{\pi}(R) = |U \setminus A_{t+1}| \geq k / (2 \log{k})$, since the stopping condition of \cref{step:spx:stop} was true. Hence, for all $\bm{\pi}(V) \geq 8$, we get 
  \begin{equation*}
    \min \{ \bm{\pi}(R), \, \bm{\pi}(V \setminus R) \} \geq \bal \cdot \bm{\pi}(V) \quad \text{where} \quad \bal = \textstyle \frac{1}{2 \log{\bm{\pi}(V)}} \enspace .
  \end{equation*}
  This proves \cref{prop:SC:exp} by showing the balance property for both $R$ and $ V \setminus R$ (instead of the expansion). Finally, we prove \cref{prop:SC:sparse}. From \cref{claim:mp:sparse}, we get that $\capc(R, V \setminus R) \leq 1/c \cdot \bm{\pi}(R) \leq \phi/10 \cdot \bm{\pi}(R)$, which shows \cref{prop:SC:sparse} in case that $R$ is returned. If instead $V \setminus R$ is returned, note that \cref{eq:pivr-k} implies $\bm{\pi}(V) \leq 10 \bm{\pi}(V \setminus R)$ for $\bm{\pi}(V) \geq 10$. We then get $\capc(R, V \setminus R) \leq (\phi/10 ) \cdot \bm{\pi}(V) \leq \phi\bm{\pi}(V \setminus R)$, which concludes the proof of the second case.

  \smallskip
  \textbf{Running Time.} The running time is dominated by the execution of the cut-matching game.
  As $t$ ranges to at most $T = O(\log^2{k})$, we get from \cref{lemma:cmgame-runtime} a total running time for \algSC{} of $O(T^2 \cdot(k \log{k} + \Tfc(m) + m \log{n}))$, as claimed.
\end{proof}

\subsection{Omitted Proofs}\label{sec:omitted-proofs}

We now provide the omitted proofs from before.

\begin{proof}[Proof of \cref{claim:properties}]\hfill\\
\cref{claim:props:p2}.~~
Note that
$P_t^2 = (I_t - Q_t)^2 = I_t - I_t Q_t - Q_t I_t + Q_t^2= I_t - 2Q_t + Q_t =
P_t$ which uses the fact that $Q_t$ is a projection matrix. As both $I_t$ and
$Q_t$ are symmetric, it follows that $P_t$ is symmetric.

\medskip\noindent
\cref{claim:props:pp}.~~
First, observe that $\1_t^T \1_{t+1} = |A_{t+1}|$
and $ \1_t^T I_{t+1} = \1_{t+1}^T$. Hence
    \[
    Q_t Q_{t+1} = \frac{1}{|A_t|} \cdot \frac{1}{|A_{t+1}|} \1_t \, (\1_t^T
    \1_{t+1}) \, \1_{t+1}^T = \frac{1}{|A_t|} \1_t \1_{t+1}^T = Q_t I_{t+1}
    \enspace.
    \]
    With this, we get the following equality. The other case can be shown
    symmetrically.
    \[
    P_t P_{t+1} = (I_t - Q_t)(I_{t+1} - Q_{t+1}) = I_{t+1} - Q_{t+1}
    \underbrace{- Q_t I_{t+1} + Q_t Q_{t+1}}_{=0} = P_{t+1} \enspace .
    \]
     
\medskip\noindent
\cref{claim:props:np-nq}.~~ Let $B$ be a doubly stochastic, \id matrix.
    Then $Q_t B = Q_t = B Q_t$. It follows that
    $P_t B = (I_t - Q_t)B = I_t B - Q_tB = B I_t - B Q_t = B(I_t-Q_t) = BP_t$.
    If $B$ is also symmetric, then it follows that
    $(P_t B)^T = (B P_t)^T = P_t^T B^T = P_t B$, which shows that $P_t B$ is
    symmetric.
    
    As $N_t$ is a symmetric, doubly stochastic \id matrix, this shows \cref{claim:props:np-nq}

    \medskip\noindent
    \cref{claim:props:PFP}.~~ We use induction on $t$. The claim holds for
    $t=0$ as $F_0 = I$ and $P_t I P_t = P_t^2 = P_t$ is symmetric by
    definition. For $t>0$ note that
    $P_t F_t P_t = P_t N_{t} F_{t-1} N_{t} P_t = N_t P_{t} F_{t-1} P_{t} N_t.$
    Using \cref{claim:props:pp} this is equal to
    $N_t P_t P_{t-1} F_{t-1} P_{t-1} P_t N_t$. We know by induction that
    $ P_{t-1} F_{t-1} P_{t-1}$ is symmetric. Let us call it $Y$ for the rest of
    the proof to simplify the notation, i.e.~we have to show that
    $N_t P_t Y P_t N_t$ is symmetric. By the fact that $N_t$ and $P_t$ are
    symmetric it follows that
    $(N_t P_t Y P_t N_t)^T = ( P_t N_t)^T Y^T (N_t P_t)^T = N_t^T P_t^T Y P_t^T
    N_t^T = N_t P_t Y P_t N_t$, which is what we wanted to show.
\end{proof}

\begin{proof}[Proof of \cref{claim:trace-props}]
  \mbox{}
  \begin{enumerate}[label=\roman*)]
    \item \cite[Theorem A.2]{OSVV08}
    \item The first statement can be found in e.g. \cite[Lemma 5.6, item 6]{ADK22}, the other statement is central in e.g. \cite{OSVV08}.
  \end{enumerate}
\end{proof}

\begin{proof}[Proof of \cref{claim:N-osvv}]
Observe that $I_tM_t=M_tI_t=M_t$ and that $M_t^2=I_t$. This 
implies $(I_t-M_t)^2=I_t^2-2M_t+M_t^2=2(I_t-M_t)$, and also
$(I_t-M_t)^j=2^{j-1}(I_t-M_t)$ for $j\ge 1$.
\begin{equation*}
\begin{split}
N_t^{k}
&=\big(I-\tfrac{1}{\delta}(I_t-M_t)\big)^{k}\\%=\big(P-\tfrac{1}{d}(P-M)\big)^{k}\\
&= I + \sum_{i=1}^k\binom{k}{i}(-1)^i\tfrac{1}{\delta^i}(I_t-M_t)^i\\
&= I + \tfrac{I_t-M_t}{2}\sum_{i=1}^k\binom{k}{i}(-1)^i(\tfrac{2}{\delta})^i\\
&= I + \tfrac{I_t-M_t}{2}\big((1-\tfrac{2}{\delta})^k-1\big)\\
&= I - \tfrac{I_t-M_t}{2}\big(1-(1-\tfrac{2}{\delta})^k\big)\\
&= I - (I_t-M_t)\lambda.\\
\end{split}
\end{equation*}
Note that $\lambda\ge 1/4$ for $k=4\delta$ since $(1-\tfrac{2}{\delta})^{4\delta}
=(1-\tfrac{2}{\delta})^{\frac{\delta}{2}8}\le e^{-8}\le 1/2$.
\end{proof}

%%%%%%%%%%%%%%%%%%%%%%%%%%%%%%%%%%%%%%%%%%%%%%%%%%%%%%%%%%%%
%%%%%%%%%%%%%  Parallel Implementation   %%%%%%%%%%%%%%%%%%%
%%%%%%%%%%%%%%%%%%%%%%%%%%%%%%%%%%%%%%%%%%%%%%%%%%%%%%%%%%%%

\section{Parallel Implementation}

The goal of this section is to prove~\cref{thm:main:parallel}, which gives our result for a parallel congestion approximator. We first analyze the work and span that each of our subroutines requires in the parallel setting. Then we derive the time required for \algCH{}, which builds the congestion approximator.
While most of our routines can be adapted in a rather straight-forward manner, special care has to be taken when parallelizing the cut matching game.
Our algorithm utilizes a known result for computing fair cuts \cite{AKL+24}, which directly gives parallel implementations for the first two subroutines.

\begin{claim}
  Given a graph $G=(V,E)$ with $n$ nodes and $m$ edges, we can implement

  \begin{itemize}
    \item \algFC{} in $O\big(m \operatorname{poly}(\frac{1}{\alpha-1}, \log{n})\big)$ work and $O(\operatorname{poly}(\frac{1}{\alpha-1}, \log{n}))$ span, w.h.p. for any $\alpha > 1$, and \label{it:prl:fc}

    \item \algTWT{} in $\tO(m)$ work and $O(\operatorname{polylog}{n})$ span, w.h.p. \label{it:prl:twt}
  \end{itemize}
\end{claim}
\begin{proof}
  The result for \algFC{} directly follows from using the algorithm from \cite[Theorem 9.3]{AKL+24} in our reduction from \cref{sec:fair-cut}.
  As \algTWT{} consists of two applications of \algFC{} called with constant $\alpha$, its result follows directly since all parameters of \algTWT{} depend only on local information and can be constructed in constant span. 
\end{proof}
  
The main challenge in parallelizing our approach is proving that \algSC{} can be adapted to the parallel setting. In \cref{sec:spx-prl} we show how to implement both the cut and the matching player in parallel and prove the following lemma. Crucially, the quality and balance guarantee remain equal to the sequential setting.

\begin{lemma}\label{lemma:spx-prl}
Given a graph $G=(V,E)$ with $n = |V|$ and $m=\deg(V)$, we can implement \algSC{} with parameters $\q = O(\log{\bm{\pi}(V)})$ and $\bal =  1/(2 \log{\bm{\pi}(V)})$ in $\tO(m)$ work and $O(\operatorname{polylog}{n})$ span.
\end{lemma}

We represent a partition by storing an identifier of its corresponding cluster at each vertex. In particular, edges can detect in constant span to which cluster each of their endpoints belong. A hierarchical decomposition is a sequence of partitions, which we can store accordingly.
For the basic operations and computations within our algorithms, we show the following parallel implementations.

\begin{claim}\label{claim:prl:time}
  Let $\v{w}$ be a vertex weight function, and $C \subseteq V$ be a set of vertices. Then 
  \begin{enumerate}
    \item we can compute $\v{w}(C)$ and $|C|$ in $O(n \log{n})$ work and $O(\log{n})$ span, and
    \item for any partition $\X$ of $C$ and $S \subset C$, we can compute $\deg_{\partial \X}(S)$ in $O(m \log{n})$ work and $O(\log{n})$ span; and we can fuse $S$ in $\X$ in $O(m)$ work and $O(1)$ span.
  \end{enumerate}
\end{claim}
\begin{proof}
  The sum of $|C| \leq n$ values can be computed in $O(n \log{n})$ work and $O(\log{n})$ span with a binary tree reduction. This gives the time for computing $\v{w}(C)$ and the computation of $|C|$ follows for $\v{w} = \1_C$.

  To compute $\deg_{\partial \X}(S)$, observe that each edge can check in $O(1)$ span whether it is contained in $\partial \X$ and $\deg_{\partial \X}(v)$ can be computed in span $O\log{n}$. Then, $\deg_{\partial \X}(S)$ is the sum of at most $n$ values. For the fuse operation, each edge independently checks whether it is in $E(S,S)$, $E(S,C)$ or $E(C,C)$ and updates its value accordingly. This takes $O(m)$ work and $O(1)$ span.
\end{proof}

With the above claim in place to establish the work and span required for some basic operations in our algorithm, we can give the time for \magicpart{}. 

\begin{claim}
Given a cluster $C$ with volume $m_C=\deg(C)$, the algorithm for \magicpart{} can be implemented in $\tO(m_C)$ work and $O(\operatorname{polylog}{n})$ span.
\end{claim}
\begin{proof}
  For \magicpart{}, first observe that the arguments for the call to \algSC{} are constructed with constant span. To discern \cref{case:assignTa} and \cref{case:assignTb}, we can compute both of the values of $\bm{\pi}(R)$ and $\t \cdot \bm{\pi}(C)$ in $O(n \log{n})$ work and $O(\log{n})$ span by \cref{claim:prl:time}. In \cref{case:assignTa}, we can setup the arguments for the call to \algTWT{} in constant span and check whether $|A| \geq |C|/2$ in $O(n \log{n})$ work and $O(\log{n})$ span. If $|A| \geq |C|/2$, then the two fuse operations (on $B$ and $U$) can be performed in $O(m_C)$ work and $O(1)$ span by \cref{claim:prl:time}.
  In \cref{case:assignTb} we can identify the smaller set by computing the cardinality of each in $O(n \log{n})$ work and $O(\log{n})$ span.
  To perform the check and fuse operation in \cref{case:repeat}, we need $O(m_C \log{n})$ work and $O(\log{n})$ span. For the call to \algFC{} in \cref{case:balanced}, we prepare the arguments with constant span, as the source and target can be computed from local information. Hence this step is dominated by the work and span of the fair cut computation.

  Observe that there are at most $O(\log{m_C}/\t) = O(\log^2{m_C})$ iterations and at most one execution of \algFC{} with constant $\alpha$. We thus obtain a total work of $\tO(m_C)$ and span of $O(\operatorname{polylog}{n})$ for \magicpart{}.

\end{proof}

We conclude the proof of \cref{thm:main:parallel} by showing the following claim which establishes the claimed work and span bounds for the algorithm \algCH{}. By the analysis in \cref{sec:hierarchy}, this gives our result for a parallel congestion approximator with approximation guarantees $O(\log^2{n} \log{\log{n}})$ with high probability.

\begin{claim}
  Given a graph $G=(V,E)$ with volume $m=\deg(V)$, the algorithm \algCH{} can be implemented in $\tO(m)$ work and $O(\operatorname{polylog}{n})$ span.
\end{claim}
\begin{proof}
  In \algCH{} we can construct the first two levels by a single call to \magicpart{}.
  To build the remaining levels of the hierarchy observe that the clusters in the While loop of \cref{step:two} are independent and can thus be processed in parallel. By \cref{claim:prl:time}, we can detect in $O(n \log{n})$ work and $O(\log{n})$ span whether the set $U$ returned by \magicpart{} is empty. If it is not empty, a bad child event occurs. Note that for the split of cluster $C$ into $U$ and $C \setminus U$ there is no change necessary to adapt the partition $\Y$ as $U \in \Y$. We only have to detect whether both clusters should be marked unprocessed or just $U$. This can be decided by computing the values of $\deg_{\partial \Y}(U)$, $\t \cdot \deg_{\partial \Y}(C)$, $\deg_{\partial\Y}(C)$, and $\capc(U,C\setminus U)$, each of these values can be computed in $O(m \log{n})$ work and $O(\log{n})$ span by \cref{claim:prl:time}.

  As there are at most $O(\log{m}/\t) = O(\log^2{m})$ iterations of the While loop by \cref{lem:runningtimeCH}, this results in a total work of $\tO(m)$ and total span of $O(\operatorname{polylog}{n})$.

\end{proof}

\subsection{Parallel Cut Matching Game}
\label{sec:spx-prl}

We now discuss the parallellization of our cut matching game from \cref{sec:cmgame} in order to prove \cref{lemma:spx-prl} and establish our parallel implementation of \algSC{}.
While the cut player is parallelized rather easily, the matching player requires more attention. A subtle issue here is that computing the path decomposition (\cref{step:mp:decomp} of the algorithm in \cref{sec:matching-player}) is non-trivial in the parallel setting. We first present a result for the cut player, then turn to the analysis of the matching player.

\begin{claim}\label{claim:cp:prl}
  An execution of the cut player when given a set of $t$ matchings can be performed in $O((t+1) \, n\log{n})$ work and $O(t\log^2{n})$ span.
\end{claim}
\begin{proof}
  As observed in the analysis of the running time of the cut player in \cref{sec:cm-time}, an execution of the cut player with $t$ matchings consists of a sequence of applications of either a $P_t$ operation or a $N_k$ operation, $k \leq t$. For a $P_t$ operation, we subtract the average of the active elements from each active element. This can be done in $O(n)$ work and $O(\log{n})$ span. The damped-averaging of a $N_k$ operation can be done in $O(n)$ work and $O(1)$ span. There are $O(\log{n})$ many $P_t$ operations and $O(t\log{n})$ many $N_k$ operations, which gives the claimed work and span.
\end{proof}

As noted above, the critical part for the matching player is computing a suitable path decomposition, as is needed in \cref{step:mp:decomp} of the matching player algorithm. To compute it, we make use of an existing parallel procedure \cite{AKL+24} for computing a path decomposition, which captures at least a $1-\delta$ fraction of the routed flow, for any $\delta \in (0,1)$. By applying this procedure repeatedly on the remaining flow, we can ensure that each unit is matched by a fraction of at least $1/2$. 

\begin{claim}\label{claim:prl:decomp}
  Given is a graph $G=(V,E)$ with $n$ nodes and $m$ edges, as well as a flow $f$ that routes the demand $\v{s} - \v{t}$ for nonnegative, integral vertex weights $\v{s}, \v{t}$ with $\v{s}(V) = \operatorname{poly}(n)$. We can compute a weighted path decomposition $P=\{(u_i,v_i,c_i)\}_i$, i.e., a collection of paths $P_i$ going from $u_i \in V$ to $v_i \in V$ and associated weights $c_i>0$,
  such that for all $u \in V$ we have $\sum_{x: (u,x,c) \in P} c \geq 1/2 \cdot (\v{s}(u) - \v{t}(u))$ with high probability. This can be implemented in $\tO(m)$ work and $O(\operatorname{polylog}{n})$ span.
\end{claim}
\begin{proof}
  We set $\delta = 1/\log{n}$ and apply the algorithm from Theorem 8.1 of \cite{AKL+24} to the flow $f$. With high probability, this results in a data structure $\mathcal{D}$ that captures at least a $(1-\delta)$ fraction of the flow $f$. In particular, the flow $g$ that is captured in $\mathcal{D}$ routes a demand $\v{s}' - \v{t}'$ such that  $\sum_v |\v{s}'(v) - \v{t}'(v)| \geq (1-\delta) \cdot \sum_v |\v{s}(v) - \v{t}(v)|$.

  This computation takes $\tO(m)$ work and $O(\operatorname{polylog}{n})$ span. Note that $\mathcal{D}$ is a layered data structure of $\ell=O(\log{n})$ layers that, crucially, consists of at most $O(m \log^2{n})$ nodes and edges.
  By \cite[Lemma 8.2]{AKL+24} we can further obtain a fractional matching of all source-target-pairs matched in $\mathcal{D}$ in $O(m \log^2{n})$ work and $O(\log{n})$ span. 

  We now verify if the produced fractional matching satisfies the required
  condition $\sum_{x: (u,x,c) \in P} c \geq 1/2 \cdot (\v{s}(u) - \v{t}(u))$
  and terminate if it holds. Otherwise, we subtract the captured flow $g$ from
  the current flow $f$ and restart the path decomposition procedure on the
  remaining flow $f-g$. Subtracting the flow can be done by propagating the
  changes of removing the flow paths top-down through the layers of
  $\mathcal{D}$. More precisely, starting at layer $l = \ell$, for each node
  $x$ of the form $(s,t,h)$ subtract $h$ from each of the predecessors of $x$.
  Then delete all these nodes and proceed with all the modified nodes of level
  $l-1$. Recurse until level $l = 1$ is reached. Since each edge is used at
  most once and $\ell = O(\log{n})$, this takes work $O(m \log^2{n})$ and span
  $O(\log{n})$.

  After this top-down removal of flow values, the lower-most layer of $\mathcal{D}$ (which contains the original graph edges of $G$), holds exactly the flow $f-g$. To restart the procedure on $f-g$, we can build up the other layers starting from this base and obtain a new path decomposition $\mathcal{D}'$ of $f-g$.

  Observe that with every repetition of the flow decomposition procedure, the amount of remaining flow decreases by a factor of at least $1-\delta = 1 - 1/\log{n}$. Hence, after $O(\log^2(\v{s}(V))) = O(\operatorname{polylog}(n))$ iterations, the remaining flow that was never captured in any $\mathcal{D}$ is less than $1/2$ and the required condition must hold. The total work for a single iteration of this procedure is $\tO(m)$ and the total span is $O(\operatorname{polylog}{n})$, giving the claimed bounds.
\end{proof}

This restriction from an imperfect path decomposition forces a slight adjustment to the analysis of the cut-matching game. In particular, the matching player is now allowed to return a fractional matching of the undeleted sources with the condition that each source must be matched to a factor of at least $1/2$.

First, observe that we can achieve a matching player that satisfies the analysis of \cref{sec:matching-player} in the parallel setting if it is allowed to return such fractional matchings. This follows from using the above \cref{claim:prl:decomp} to compute the path decomposition in \cref{step:mp:decomp}. Crucially, the data structure in the above claim also allows for an efficient parallel iteration of the paths which yield the fractional matching. Given our implementation for \algFC{}, the matching player can thus be implemented in $\tO(m)$ work and $O(\operatorname{polylog}{n})$ span.

It remains to show that the cut-matching game still works with fractional matchings. For this, recall the analysis of the cut player in \cref{sec:cut-player}. The first slight difference occurs in the proof of \cref{lemma:pot-reduction} for the expected decrease in potential from a matching step. If the matching $M_{t+1}$ is no longer integral, we have the relation $\Tr[\mathcal{L}(M_{t+1})W_{t+1}^2] = \sum_{i,j \in A_{t+1}} c_{ij} \cdot \lVert\v{w}_i - \v{w}_j\rVert^2$, where $c_{ij}$ is the weight of the matching edge between $i$ and $j$ (as also seen in e.g. \cite[Lemma D.4]{ADK22}). This can be seen as a slight generalization of the lemma using the matching weights.

The analysis of the drop in potential in a good round in \cref{lemma:pot-good-round} thus changes to

\begin{equation*}
  \begin{aligned}
    \tfrac{1}{4} \sum_{i,j \in A_{t+1}} c_{ij} \lVert \v{w}_i - \v{w}_j \rVert^2 
    &\geq \frac{k}{4\alpha \log{k}} \sum_{i,j \in A_{t+1}} c_{ij}(u_i - u_j)^2 &&& \text{by~\cref{lem:projection}.\labelcref{prop:proj-stretch})}\\
    &\geq \frac{k}{4\alpha \log{k}} \sum_{i,j \in A_{t+1}} c_{ij}(u_i - \eta)^2 &&& \text{by~\cref{lemma:RST14}.\labelcref{prop:rst14:separate})}\\
    & \geq \frac{k}{36\alpha \log{k}} \sum_{i,j \in A_{t+1}} c_{ij} u_i^2 &&& \text{by~\cref{lemma:RST14}.\labelcref{prop:rst14:source})}\\
    & = \frac{k}{36\alpha \log{k}} \sum_{i\in A^{\ell} \setminus S_t} u_i^2 \cdot \sum_{j \in A_t \setminus S} c_{ij} \\
    & \geq \frac{k}{36\alpha \log{k}} \sum_{i\in A^{\ell} \setminus S_t}  u_i^2 \cdot \frac{1}{2}
  \end{aligned}
\end{equation*}
where the last inequality crucially uses that each node in $A^{\ell} \setminus S_t$ is matched to at least a factor of $1/2$. As this additional factor of $1/2$ does not affect the asymptotic guarantees of the cut-matching game, we conclude this analysis of its parallel implementation.

\smallskip Lastly, with the correctness of the cut-matching game intact, for
\cref{lemma:spx-prl} it remains to show that the algorithm \algSC{} from
section \cref{sec:spx} can be implemented in $\tO(m)$ work and
$O(\operatorname{polylog}(n))$. From the considerations above and
\cref{claim:cp:prl} we get that the required work and span for computing the
cut matching game update in \cref{step:spx:cmgame} are $\tO(m)$ and
$O(\operatorname{polylog}{n})$, respectively. As we can compute the value of
$|A_{t+1}|$ in \cref{step:spx:stop} and the values $\bm{\pi}(R)$ and
$\bm{\pi}(V\setminus R)$ in \cref{step:spx:return} in $O(n \log{n})$ work and
$O(\log{n})$ span each, the claimed bounds follow.

\section*{Acknowledgements}

We thank Evangelos Kosinas for helpful discussions on this topic.

This research was funded in whole or in part by the Austrian Science Fund (FWF)
\href{https://www.doi.org/10.55776/I5982}{DOI~10.55776/I5982}. For open access
purposes, the author has applied a CC BY public copyright license to any
author-accepted manuscript version arising from this submission.

This project has received funding from the Deutsche Forschungsgemeinschaft
(DFG, German Research Foundation) – 498605858.

% \ifbiber
% \printbibliography
% \else
\bibliography{references}
% \fi

\appendix
\clearpage

\section{Appendix}

\subsection{A Lower Bound  for Congestion Approximators via Oblivious Routing}
\label{sec:lowerbound}
In this section we show
a lower bound on the approximation guarantee for congestion approximators of $\Omega(\log n)$.
More specifically we prove 
that a single-commodity congestion approximator with
 approximation guarantee $\alpha$ implies an oblivious routing strategy with competitive ratio
$O(\lambda \alpha)$, where $\lambda$ is the flow-cut gap. It is known that the flow-cut gap is constant for series-parallel graphs. We then present a series-parallel graph in which every oblivious routing scheme has competitive ratio $\Omega(\log n)$. Note that a single-commodity congestion approximator with approximation guarantee $o(\log n)$ on this graph would lead to a an oblivious routing strategy with competitive ratio $o(\log n)$, leading to a contradiction.
Thus, there is a lower bound of $\Omega(\log n)$ for the approximation guarantee of a single-commodity congestion approximator.

%which will give a logarithmic lower bound on the approximation
% guarantee of single-commodity congestion approximators.
We first show the following auxiliary lemma about the relationship of the approximation guarantee for single-commodity versus multi-commodity congestion approximators.
\begin{lemma}\label{lemma:single-to-multi}
    A single-commodity congestion approximator with approximation guarantee $\alpha$ is a multi-commodity congestion approximator with approximation guarantee $\lambda\alpha$, where $\lambda$ is the flow-cut gap of the graph.
\end{lemma}
\begin{proof}
Let $\mathcal{C}$ be a single-commodity congestion approximator with
approximation guarantee $\alpha$. Fix some multi-commodity flow demand $d$ for
which $\mathcal{C}$ predicts a congestion of at most $1$, i.e., 
we have $d(C,V\setminus C) \leq \capc(C,V\setminus C)$ for
all $C \in \C $.

    % Suppose that I have a multi-commodity flow where the
    % cut approximator says congestion 1. This means that
    % the flow that goes over a cut in the congestion
    % approximator is at most the capacity of the cut.

    %Now assume for contradiction that there exists a subset of vertices $S$ in the graph ($S$ is not necessarily in $\C$)
    
Assume for contradiction that there exists a cut $S$ in the graph (not necessarily
in $\mathcal{C}$) for which
$d(S, V \setminus S)>\alpha \capc(S,V\setminus S)$,
%$< ,
%i.e.,~$\capc(S,V\setminus S)/d(S, V \setminus S)< 1/\alpha,$ 
which implies that
every multi-commodity flow that routes $d$ must have congestion more than
$\alpha$.
    %the optimum multi-commodity flow routing $d$ has a congestion of more than $\alpha$. 
    We show that then
    there exists a single-commodity flow demand $d'$ for which the congestion
    approximator predicts a congestion of at most $1$, but at the same time
    $|d'(S)|>\alpha\capc(S,V\setminus S)$, which is a contradiction
    to the approximation guarantee of $\mathcal{C}$.

    We construct $d'$ as follows. Initially, every vertex $v$ has $d'(v)=0$.
    For every demand pair $(s,t)$ in the multi-commodity flow demand $d$ where
    exactly one vertex out of $\{s,t\}$ lies in $S$, we increase $d'$ by
    $d(s,t)$ for the vertex that lies in $S$ and decrease $d'$ by $-d(s,t)$ for
    the vertex that lies outside $S$. All other demand pairs of $d$ are
    ignored. By construction (a) $d$ is a single-commodity demand, i.e.,
    $\sum_{v \in V} d(v)=0$, (b) for every $C$, % \in \mathcal{C}$,
    $d'(C)\le d(C,V\setminus C)$,
    and (c) $d'(S) = d(S,V\setminus S)$.
    %, i.e.~$d'$ has the same demand, and, thus, the
    %same ratio $\capc(S,V\setminus S)/d'(S, V \setminus S)$ over the cut $S$ as
    %the multi-commodity demand. 
    Property~(c) gives $d'(S)=d(S,V\setminus S)>\alpha \capc(S,V\setminus S)$.
    However, Propety~(b) gives
    $d'(C)\le d(C,V\setminus C)\le\capc(C,V\setminus C)$ for every
    $C\in \mathcal{C}$. This is a contradiction to the assumption that $\C$ has
    approximation guarantee $\alpha$.
    
    Hence,  for \emph{any} cut in $G$, 
    $d(S, V \setminus S)\le \alpha\capc(S,V\setminus S)$.
    %{\color{red} NEW: 
    This implies
      $\Phi(d) = \max_S d(S,V\setminus S)/ \capc(S, V \setminus S) \le \alpha$.
      Now recall that the flow-cut gap $\lambda$ w.r.t.~the multi-commodity
      flow demand $d$ is defined to be
      $\max_{d'}\operatorname{opt}_G(d')/\Phi(d')$, where
      $\operatorname{opt}_G(d')$ is the optimum congestion for routing $d'$ in
      $G$. But then the optimum flow can route the multi-commodity flow demand $d$
      with congestion at most $\lambda\alpha$. So our approximator
      $\mathcal{C}$ has approximation guarantee $\lambda\alpha$ for
      multi-commodity flows.
\end{proof}

An oblivious routing scheme defines a unit flow $f_{s,t}$ between every
source-target pair $s$, $t$ in a graph. \emph{Routing using an oblivious routing scheme} works as follows: A multi-commodity demand $d$ is
routed by scaling the flow $f_{s,t}$ by the demand $d_{s,t}$ for every pair
$s,t$ to obtain the multi-commodity flow. It has competitive ratio
$\alpha$ if for any demand $d$ the congestion obtained by routing the demand via
the oblivious routing scheme is at most an $\alpha$-factor larger than the
optimum possible congestion for $d$.

%Recall that given a laminar family of cuts that form a hierarchical
%congestion approximator we can associate with it a tree $T$, such that the
%leaves of $T$ correspond to the vertices of $G$ and each internal vertex $v$
%represents a subset $L_v\subseteq V$, namely the subset of leave vertices
%contained in  $T_v$ (the subtree rooted at $v$). The capacity of an edge $(v,p)$ of a
%node $v$ to its parent $p$ is defined as the capacity of the cut
%$(L_v,V\setminus L_v)$ in $G$.
% 
%We first preprocess the tree to obtain a new tree $T'$ in which an edge $(c,p)$ from
%some vertex $c$ to its parent $p$ has at most half the capacity of all edges of
%the form $(c,x)$, $x\neq p$ (i.e., child edges of $c$). Formally,
%$\capc_T(c,p)\le\frac{1}{2}\sum_{x\in\text{children(c)}}\capc(c,x)$. In order
%to construct $T'$ we traverse $T$, from bottom to top; whenever we identify a
%vertex $v$ with parent $p$ for which the condition does not hold we delete $v$ 
%and attach its children as direct children of $p$. 

\begin{lemma}\label{lem:obliviousrouting}
Given a single-commodity hierarchical congestion approximator with  approximation guarantee $\alpha$
for a graph $G$, we can design an oblivious routing scheme with competitive
ratio $8\lambda \alpha$, where $\lambda$ is the flow-cut gap of $G$. 
\end{lemma}
\begin{proof}
Recall that given a laminar family of cuts that form our hierarchical
congestion approximator we can associate with it a tree $T$, such that the
leaves of $T$ correspond to the vertices of $G$ and each internal vertex $v$
represents a subset $L_v\subseteq V$, namely the subset of leave vertices
contained in  $T_v$ (the subtree rooted at $v$). The capacity of an edge $(v,p)$ of a
node $v$ to its parent $p$ is defined as the capacity of the cut
$(L_v,V\setminus L_v)$ in $G$.

We first preprocess the tree to obtain a new $T'$ in which an edge $(c,p)$ from
some vertex $c$ to its parent $p$ has at most half the capacity of all edges of
the form $(c,x)$, $x\neq p$ (i.e., child edges of $c$). Formally,
$\capc_T(c,p)\le\frac{1}{2}\sum_{x\in\text{children(c)}}\capc(c,x)$. In order
to construct $T'$ we traverse $T$, from bottom to top; whenever we identify a
vertex $v$ with parent $p$ for which the condition does not hold we delete $v$ 
and attach its children as direct children of $p$. 

\begin{claim}
$T'$ is a single-commodity congestion approximator with  approximation guarantee $2\alpha$.
\end{claim}
\begin{proof}
First observe that the congestion prediction of $T'$ can only be smaller than
the congestion prediction of $T$ for any demand. This holds because $T'$ is
obtained from $T$ by just deleting cuts. 

Now, we argue that the congestion prediction can at most change by a factor of
$2$. Fix a single-commodity demand $d$, for which $T$ makes a congestion
prediction of $C$, and let $e=(c,p)$ be the edge in $T$ that has 
this congestion, with $p$ being the parent of $c$. We have
$|d(L_c)|/\capc(c,p)=C$, where $L_c$ are the leaf
vertices in the sub-tree $T_c$. We will show below that $T'$ makes a congestion prediction of at least $C/2$. Thus the approximation guarantee that is $\alpha$ for $T$ becomes at most $2\alpha$ for $T'$.

Now, we define a set of vertices $X$ in $T$ as follows. For every $p$-$\ell$ -path,
where $\ell$ is a leaf in $T_c$, we choose the first vertex on this path that is not
deleted and add it to $X$. Observe that in $T'$ all these vertices are connected by an edge to
either $p$ or a parent of $p$ in case $p$ is deleted. Let $p^*$ denote this parent.

\begin{fact}
$\capc_T(c,p)\ge\frac{1}{2}\sum_{x\in X}\capc_{T'}(x,p^*)$.
\end{fact}
\begin{proof}
If vertex $c$ is not deleted the claim is immediate, because then $c$ is the only
vertex in set $X$, and $\capc_{T'}(c,p^*)=\capc_G(L_c,V\setminus L_c)=\capc_{T}(c,p)$.
Otherwise, observe that the nodes in $X$ are exactly the children of $c$ at the
time that we decide to delete $c$. Therefore, by the deletion condition we have
$\sum_{x\in X}\capc_{T'}(x,p^*)=\sum_{x\in X}\capc_G(L_x,V\setminus L_x)\le
2\capc_G(L_c,V\setminus L_c)=2\capc_T(c,p)$.
\end{proof}
\noindent
Now, we claim that one of the edges $(x,p^*)$ in $T'$ will predict congestion
at least $C/2$. Indeed,
\begin{equation*}
C=\frac{|d(L_c)|}{\capc_T(c,p)}
=\frac{|\sum_x d(L_x)|}{\capc_T(c,p)}
\le2\frac{\sum_x |d(L_x)|}{\sum_{x}\capc_{T'}(x,p^*)}
\le2\max_x\Big\{\frac{|d(L_x)|}{\capc_{T'}(x,p^*)}\Big\}
\end{equation*}
where $L_x$ is the set of leaf vertices in $T_x$. This means one of the edges
$(x,p^*)$ will predict a congestion of at least $C/2$. This finishes the proof
of the claim.
\end{proof}

For proving the lemma we now
give a randomized embedding of $T$ into $G$ such that the expected load
of an edge is at most $4$. This is an embedding of a decomposition
tree and gives therefore an oblivious routing scheme with competitive ratio
$O(\lambda \alpha)$ (see e.g.~\cite{Rae02}).

We map each vertex of $T$ to a random leaf node as follows. A vertex $v$
is mapped to a random child by choosing one of its child-edges $(v,c)$, 
$c\in\operatorname{children}(v)$ at random with probability 
$\capc_{T'}(c,x)/\sum_{c'\in\operatorname{children}(v)}\capc_{T'}(c,v)$. 
This is repeated until a leaf vertex is reached.

This mapping embeds the vertices of $T$ into $G$ in a randomized way (recall
that there is a one-to-one correspondence between leaves of $T$ and vertices of
$G$). In order
to embed the edges we set up a multicommodity flow demand. For every tree edge
$(u_T,v_T)\in E_T$ we introduce a demand $\capc(u_T,v_T)$ between 
$\pi_V(u_T)$ and $\pi_V(v_T)$ in $G$, where
$\pi_V$ is the randomized vertex mapping.

\begin{claim}
This demand has expected congestion at most $4$ in $T'$. 
\end{claim}
\begin{proof}
Fix a tree edge $(c,p)$. Consider another tree edge $(x,y)$ and let
$x_\ell$ and $y_\ell$ denote the leaf vertices that $x$ and $y$ are mapped to,
due to the randomized vertex embedding. We analyze the probability that the
tree path between $x_\ell$ and $y_\ell$ goes through $(c,p)$. For this to
happen $(x,y)$ must be an ancestor edge of $c$ in the tree and exactly
one of $x$ and $y$ must be mapped into the subtree $T_c$. 

Order the ancestor edges in increasing distance to $c$. 
So $e_0=(c_0,p_0)$, $e_1=(p_0=c_1,p_1)$,\dots
The expected load induced on $e_0$ by all ancestor edges (including itself) is

\begin{equation*}
\operatorname{load}(e_0)
=\sum_{i\ge0}\capc(e_i)\cdot\Pr[\text{exactly one of $c_i$, $p_i$ mapped to $T_c$}]
\end{equation*}
Let for an edge $e_i$, $\operatorname{sib}(e_i)$ denote the \emph{sibling
  edges} of $e_i$, i.e., the edges $(x,p_i)$ with $x$ being a child of $p_i$
($e_i$ is a sibling of itself). The probability that $c_i$  is mapped into
the sub-tree $T_c$ is
\begin{equation*}
\Pr[\text{$c_i$ mapped to $T_c$}]=\prod_{j=i-1}^{0}\frac{\capc(e_j)}{\capc(\operatorname{sib}(e_j))}
\end{equation*}
With this we can estimate the expected flow on $e_0$ when routing the demand in
$T'$ as follows:
\begin{equation*}
\begin{split}
\operatorname{flow}(e_0)
&\le\sum_i\capc(e_i)\Pr[\text{$c_i$ or $p_i$ mapped to $T_c$}]
\le2\sum_i\capc(e_i)\prod_{j=i-1}^0\frac{\capc(e_j)}{\capc(\operatorname{sib}(e_j))}\\
&=2\sum_i\capc(e_0)\prod_{j=i-1}^0\frac{\capc(e_{j+1})}{\capc(\operatorname{sib}(e_j))}
\le  2\capc(e_0)\sum_i\frac{1}{2^i}\le4\capc(e_0)\qedhere
\end{split}
\end{equation*}
\end{proof}

\begin{claim}
    There is a randomized embedding of $T'$ into $G$ with congestion $8\lambda \alpha$.
\end{claim}
\begin{proof}
    As we showed in \Cref{lemma:single-to-multi} a single-commodity congestion approximator with approximation guarantee $\alpha$ is a
    multi-commodity congestion approximator with approximation guarantee $\lambda\alpha$, where
    $\lambda$ is the flow-cut gap of the graph. As $T'$ has approximation guarantee $2 \alpha$
    and the congestion for routing the demand in $T'$ is $4,$ the claim follows.
\end{proof}
\noindent
This completes the proof of~\cref{lem:obliviousrouting}.
\end{proof}

% the ratio determines the factor between the l-r length and the s-t length 
% this has to decrease for smaller dimensions (deeper recursion levels) as
% otherwise the graphs are plotted on top of each other
% dimension source target name factor params
\def\diamond #1#2#3#4#5#6{%
    \coordinate (#4S) at (#2);
    \coordinate (#4T) at (#3);
    \node at (#2) (n#4S) [shape=circle,fill=black,inner sep=0pt,#6,minimum width=#1*1.5pt] {};
    \node at (#3) (n#4T) [shape=circle,fill=black,inner sep=0pt,#6,minimum width=#1*1.5pt] {};   
    \pgfmathsetmacro\nextdim{int(#1-1)}%
    \pgfmathsetmacro\nextratio{#5*0.47}% ratio is decreased by some factor
    \ifcsstring{nextdim}{-1}{%
       \draw[#6] (n#4S.center) -- (n#4T.center);
    }{%
       \coordinate (#4L) at ($(#2)!0.5!(#3)!#5! 90:(#3)$);
       \coordinate (#4R) at ($(#2)!0.5!(#3)!#5!-90:(#3)$);
       \edef\temp{%
            \noexpand\diamond{\nextdim}{#4S}{#4L}{#4A}{\nextratio}{#6}% 
            \noexpand\diamond{\nextdim}{#4S}{#4R}{#4B}{\nextratio}{#6}% 
            \noexpand\diamond{\nextdim}{#4L}{#4T}{#4C}{\nextratio}{#6}% 
            \noexpand\diamond{\nextdim}{#4R}{#4T}{#4D}{\nextratio}{#6}% 
       }%
       \temp%
    }%
}

\def\dim{4} % dimension of the graph
\pgfmathsetmacro\dimMinusOne{int(\dim-1)}%
\begin{figure}[t]
\centering
\begin{tikzpicture}
\coordinate (s) at (0,0);
\coordinate (t) at (12,0);
% recursively draw diamond graph 
\diamond{\dim}{s}{t}{diam}{0.4}{line width=0.3pt}%

%labels
\node at (ndiamS.west)   [anchor=east]  {$s$};
\node at (ndiamT.east)   [anchor=west]  {$t$}; 
\node at (ndiamAT.north) [anchor=south] {$\smash{\ell}$};
\node at (ndiamBT.south) [anchor=north] {$r$};
\node at ($(s)!0.5!(diamL)$) {$D_\dimMinusOne^{s,\ell}$};
\node at ($(s)!0.5!(diamR)$) {$D_\dimMinusOne^{s,r}$};
\node at ($(t)!0.5!(diamL)$) {$D_\dimMinusOne^{\ell,t}$};
\node at ($(t)!0.5!(diamR)$) {$D_\dimMinusOne^{r,t}$};
\node at ($(s)!0.5!(t)$)     {$D_\dim^{s,t}$};

\end{tikzpicture}
\caption{The recursive diamond graph of order $\dim$ between $s$ and $t$. In the
  graph $D_k^{s,t}$, vertices $s$ and $t$ are connected by $2^k$
  edge-disjoint paths of length $2^k$. Hence, the graph has $4^k$ edges in total.}
\end{figure}
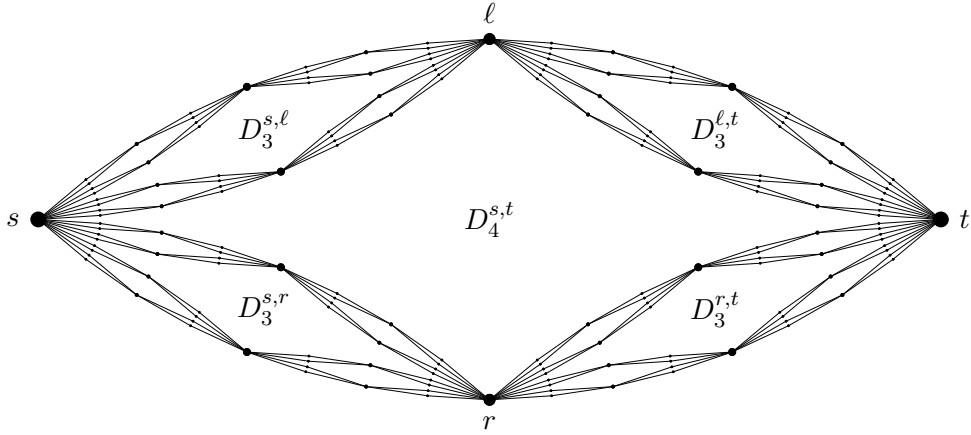

We next show  a lower bound for the competitive ratio for oblivious routing schemes on the diamond graph.
This bound is known, but we provide a proof, as we could not find a formal proof for it in the literature.
\begin{lemma}\label{lem:diamond}
Every oblivious routing scheme for the recursive diamond graph $D_k$ has
competitive ratio at least $\Omega(k) = \Omega(\log n)$, where $n$ is the number of vertices in the graph $D_k$. 
\end{lemma}
\begin{proof}
The recursive diamond graph $D_k^{s,t}$ of order $k$ is defined between a
source $s$ and target $t$. $D_0^{s,t}$ consists of just a single edge
$\{s,t\}$. For $k\ge1$, $D_k^{s,t}$ consists of a \enquote{\emph{left path}}
formed by sub-graphs $D_{k-1}^{s,\ell}\circ D_{k-1}^{\ell,t}$ that only share the vertex $\ell$ and a \enquote{\emph{right path}} formed by graphs
$D_{k-1}^{s,r}\circ D_{k-1}^{r,t}$ that only share the vertex $r$.
Note that (1) $D_k^{s,t}$ contains at most $2\cdot 4^{k}$ vertices;
%and $D_{k+1}^{s,t}$ contains at most 4 times as many vertices as $D_k^{s,t}$;
(2) 
%the left path of $D_{1}^{s,t}$ consists of 2 edges and so does the right path, i.e., $D_1^{s,t}$ has 4 edges and for general $k\ge 1$,  
both the left and the right path of $D_{k}^{s,t}$
consists of $2 \cdot 4^{k-1}$ edges, for a total of $4^{k}$ edges in $D_k^{s,t}$; and (3)   $s$
and $t$ are at distance $2^k$ in $D_{k}^{s,t}$.
The bound on the number of vertices immediately implies that $k=\Theta(\log n)$, where $n$ is the number of nodes in the graph.

In general we call two graphs $G_x=D_{i}^{x,m}$, $G_y=D_{i}^{m,x'}$ that are induced
sub-graphs of $D_k^{s,t}$ for $k$, a \emph{path (of order $i$)}, denoted by $D_{i}^{x,m} \circ D_{i}^{m,x'}$ in $D_k^{s,t}$
if (i) $G_x$ and $G_y$ share exactly one vertex, namely the \emph{middle vertex} $m$ on the path, and (ii) $x$ and $x'$ are the only 
vertices of $G_x \cup G_y$ that are connected to vertices in $D_k^{s,t}-(G_x\cup G_y)$. With this notation $D_k^{s,t}$ consists of two paths of order $k-1$, which each consist of four paths of order $k-2$.

We now recursively construct a bad demand for online routing. We do this in
several iterations, where in the $i$-th iteration we construct a demand on a
\emph{path of order $k-i$} inside $D_k^{s,t}$. For the first iteration we
choose either the left or the right path of $D_k^{s,t}$. This is a path of
order $k-1$ and we denote it with $P$. Let $m$ be the middle vertex on this
path. We issue a demand of $2^{k-1}$ flow units between $s$ and $m$ and between
$t$ and $m$. Observe, that regardless how this demand is routed inside the
whole graph $D_k^{s,t}$, each of its $2^{k-1}$ units of flow has to routed over
at least $2^{k-1}$ edges \emph{inside} $P$, since $m$ is at distance $2^{k-1}$
from both $s$ and $t$. There are $4^{k}/2$ edges in the path $P$, and
consequently the average load of an edge of $P$ must be at least $1/2$ after
routing the demand.

Recall that $P$ consists of four paths of order $k-2$, namely paths
$D_{k-2}^{s,a}\circ D_{k-2}^{a,m}$, $D_{k-2}^{s,b}\circ D_{k-2}^{b,m}$,
$D_{k-2}^{m,c}\circ D_{k-2}^{c,t}$, and $D_{k-2}^{m,d}\circ D_{k-2}^{d,t}$. One
of them must have average load at least $1/2$ due to the first demand.
Wlog.~let $D^{s,a}_{k-2}\circ D^{a,m}_{k-2}$ be this path. Next we will issue a
demand between $s$ and $a$ of value $2^{k-2}$ and also between $a$ and $m$, and
so on for paths of lower order. By recursing always on the path with most load
until $i=k-1$ we obtain a path of order 1 that has average load $(k-1)/2$.
However, a path of order 1 consists of a single edge, which implies that this
edge has load at least $(k-1)/2= \Theta(k)$.

It remains to show that an optimum algorithm could route the demand with
constant congestion. Indeed, it is easy to see that in every recursion 
step we can route
the demand for $P$ \emph{within} $P$ 
in such a way that the order $k-1$ path $P'$ that we
are going to recurse on does not get any load on this level. This can be done
with congestion $2$ as the demand is $2^{k-i}$\,---\,routing all of it along
edges in $P\setminus P'$ is possible and generates congestion of at
most $2$ on these edges. By this routing method an edge can get load from at
most one level of the recursion.
 Thus, all edges
will have congestion at most 2. Hence the competitive ratio is at least
$(k-1)/4=\Omega(k)$.
\end{proof}

\begin{corollary}\label{obs:lowerbound}
Let $k = \lfloor \log_4(n/2) \rfloor$. There is no single-commodity congestion approximator with approximation
guarantee $o(\log n)$ on the recursive diamond graph $D_{k}$.
\end{corollary}
\begin{proof}
    It follows from the construction of the graph $D_{k}$ that it is \emph{series-parallel}. \Cref{lem:diamond} shows that for $k = \lfloor \log_4(n/2) \rfloor$, $D_k$ has at most $n$ and at least $n/4$ vertices. 
    Gupta, Newman, Rabinovich and Sinclair~\cite{GNRS04} showed that the
    flow-cut gap $\lambda$ for series-parallel graphs is constant. If there
    were a congestion approximator with approximation guarantee
    $\alpha=o(\log n)$, we could use \cref{lem:obliviousrouting} to obtain an 
    oblivious routing strategy with competitive ratio $O(\lambda\alpha)=o(\log
    n)$ for the recursive diamond graph contradicting the lower bound in \cref{lem:diamond}.
%
    % Bartal and Leonardi~\cite{BL99} have shown that on a grid no online algorithm can obtain
    % a competitive ratio of $o(\log n)$ for oblivious routing with the goal of
    % minimizing the congestion. The flow-cut gap $\lambda$ for the grid is
    % constant. If there were a congestion approximator on the grid with quality
    % $\omega(1/\log n)$ we could use \cref{lem:obliviousrouting} to obtain an oblivious routing strategy with competitive ratio $O(1/q)=o(\log n)$ contradicting the lower bound in~\cite{BL99}.
\end{proof}

%%% Local Variables:
%%% mode: LaTeX
%%% TeX-master: "main"
%%% End:

\end{document}

%%% Local Variables:
%%% mode: LaTeX
%%% TeX-master: t
%%% End: